\documentclass[12pt]{article}
\usepackage{graphicx}

\usepackage{amsmath, amssymb}
\usepackage{titlesec}
\usepackage{soul}
\usepackage[dvipsnames]{xcolor}
\usepackage{graphicx}
\usepackage{needspace} 
\usepackage{listings}
\usepackage[colorlinks=true,linkcolor=blue]{hyperref}%
\usepackage{longtable} 
\usepackage{lettrine}
\usepackage{ebgaramond}
\usepackage{subcaption}
\usepackage{pdfpages}
\usepackage{wrapfig}

\usepackage{lmodern}

\usepackage[titles]{tocloft}

\usepackage[total={5.5in, 8.5in},top=3cm,bottom=4cm,left=3cm,right=3cm]{geometry}

\usepackage{array}

\usepackage{fancybox}

\usepackage[round]{natbib}

\makeatletter
\renewcommand\maketitle
  {\begin{center}
   {
   \Large\scshape{\@title}
   }
   \end{center}
  }
\makeatother

\titleformat{\section}[display]
    {\clearpage\flushright}
    {\fontsize{96}{50}\selectfont\thesection.}
    {-5pt}
    {\Huge}
    [\vspace{-1.5ex} \hspace{1.3ex} \rule{0.8\textwidth}{0.2pt} ]
\titlespacing*{\section}{0pt}{0pt}{\baselineskip}

\titleformat{\subsection}[block]
    {\Large\scshape}
    {}
    {0pt}
    {\needspace{5\baselineskip}\rule{0.8\textwidth}{0.2pt} \\ \vspace{0.5ex} \thesubsection. \quad }
    [\vspace{-1.5ex} \rule{0.8\textwidth}{0.2pt} ] 
\titlespacing{\subsection}{-1cm}{\baselineskip}{\baselineskip}

\renewcommand{\thesubsubsection}{\arabic{subsubsection}}
\titleformat{\subsubsection}{\scshape}{}{0.2ex}
{\thesubsubsection. \quad}
[ \vspace{-2ex} \rule{\textwidth}{0.15pt} ]

\DeclareCaptionLabelFormat{custom}
{%
      #1 \textbf{(#2)}
}
\DeclareCaptionLabelSeparator{custom}{--}
\DeclareCaptionFormat{custom}
{%
    #1#2\small #3
}
\newenvironment{sectionauthor}
{\begin{flushright}\it}
{\end{flushright}}

\hypersetup{
  colorlinks,
  citecolor=Thistle,
  linkcolor=Orchid,
  urlcolor=Blue}

\definecolor{codegreen}{rgb}{0,0.6,0}
\definecolor{codegray}{rgb}{0.5,0.5,0.5}
\definecolor{codepurple}{rgb}{0.58,0,0.82}
\definecolor{codekeyword}{rgb}{0.5, 0.5, 0.2}
\definecolor{Orchid}{rgb}{0.686,0.447,0.690}

\definecolor{CLIprompt}{rgb}{0.4,0.6,0}
\definecolor{xspeckeyword}{rgb}{0.6, 0.4, 0.3}

\lstdefinestyle{terminal}{    
    commentstyle=\it\color{MidnightBlue},
    keywordstyle=\color{MidnightBlue},
    numberstyle=\tiny\color{codegray},
    stringstyle=\bfseries\color{codepurple},
    basicstyle=\ttfamily\footnotesize,
    breakatwhitespace=false,         
    breaklines=true,                 
    captionpos=b,                    
    keepspaces=true,                 
    numbers=none,                                      
    showspaces=false,                
    showstringspaces=false,
    showtabs=false,                  
    tabsize=2,
    otherkeywords = {user@here:\$, XSPEC12>,PLT>, PyXspec>},
    keywordstyle={\color{CLIprompt}\bfseries},
    xleftmargin=.25in,
    xrightmargin=.25in
}

\lstdefinestyle{file}{    
    commentstyle=\color{codegreen},
    keywordstyle=\color{codekeyword},
    numberstyle=\tiny\color{codegray},
    stringstyle=\bfseries\color{codepurple},
    basicstyle=\ttfamily\footnotesize,
    breakatwhitespace=false,         
    breaklines=true,                 
    captionpos=b,                    
    keepspaces=true,                 
    numbers=left,                    
    numbersep=10pt,                  
    showspaces=false,                
    showstringspaces=false,
    showtabs=false,                  
    tabsize=2,
    xrightmargin=.3in
}
 
\lstdefinelanguage{xspec}{
    basicstyle=\ttfamily\small,
    columns=fullflexible,
    morecomment=[s][\color{Orchid}\bfseries]{[}{]},
    morecomment=[l]{\#},
    morecomment=[l]{;},
    commentstyle=\color{gray}\ttfamily,
    morekeywords = [2]{lmod, cpd, setplot, dummyrsp, dummy, fakeit, model, Model},
    keywordstyle = [2]{\color{xspeckeyword}\bfseries},
    otherkeywords = {XSPEC12>,PLT>},
    keywordstyle = {\color{CLIprompt}\bfseries}
}

\makeatletter \def\lst@gkeywords@sty{\color{CLIprompt}\bfseries} 

\newcommand{\rin}[1][]{%
  \ifx\relax#1\relax
    \ensuremath{r_{\rm in}}%
  \else
    \ensuremath{r_{{\rm in}, #1}}%
  \fi
}

\newcommand{\rout}[1][]{%
  \ifx\relax#1\relax
    \ensuremath{r_{\rm out}}%
  \else
    \ensuremath{r_{{\rm out}, #1}}%
  \fi
}

\newcommand{\rg}[1][]{%
  \ifx\relax#1\relax
    \ensuremath{r_{g}}%
  \else
    \ensuremath{r_{g, #1}}%
  \fi
}

\newcommand{\risco}[1][]{%
  \ifx\relax#1\relax
    \ensuremath{r_{\rm ISCO}}%
  \else
    \ensuremath{r_{#1{\rm (ISCO)}}}%
  \fi
}

\newcommand{\rl}[1][]{%
  \ifx\relax#1\relax
    \ensuremath{r_{\rm L}}%
  \else
    \ensuremath{r_{#1{\rm (L)}}}%
  \fi
}

\newcommand{\rms}[1][]{%
  \ifx\relax#1\relax
    \ensuremath{r_{\rm MS}}%
  \else
    \ensuremath{r_{#1{\rm (MS)}}}%
  \fi
}

\begin{document}

\includepdf{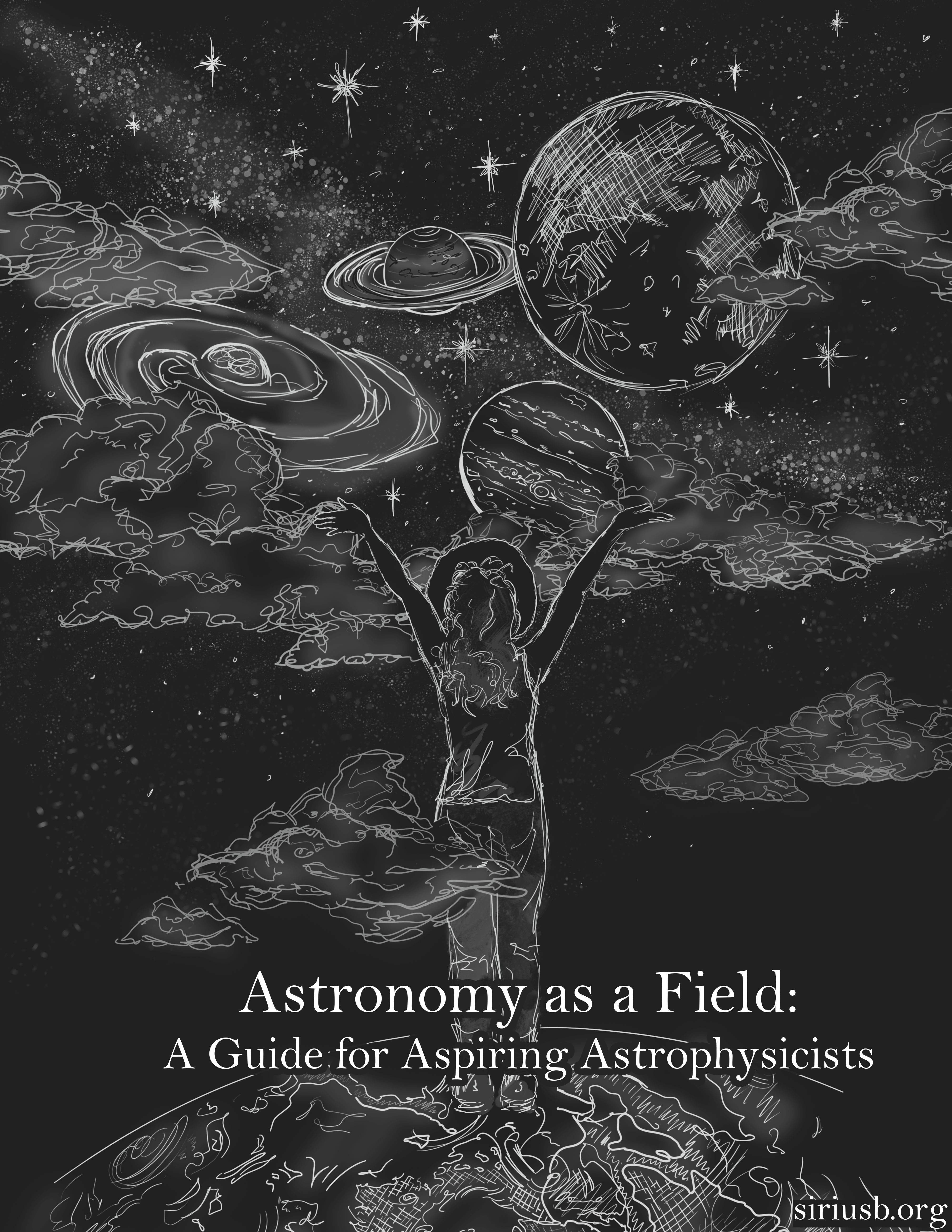}

\pagebreak
\setcounter{page}{1}
\strut \vspace{20pt} 

\Huge{\textsc{Astronomy as a Field:}}\\ 
\Large{\textsc{A Guide for Aspiring Astrophysicists}}\\
\\
\Large{SIRIUS B}\\
\large{\href{https://siriusb.org}{siriusb.org}}

\vspace{40pt}
\Large{\emph{Organizer}}\\
\vspace{-10pt}
\normalsize

\begin{wrapfigure}{l}{0.2\textwidth}
\includegraphics[width=0.9\linewidth]{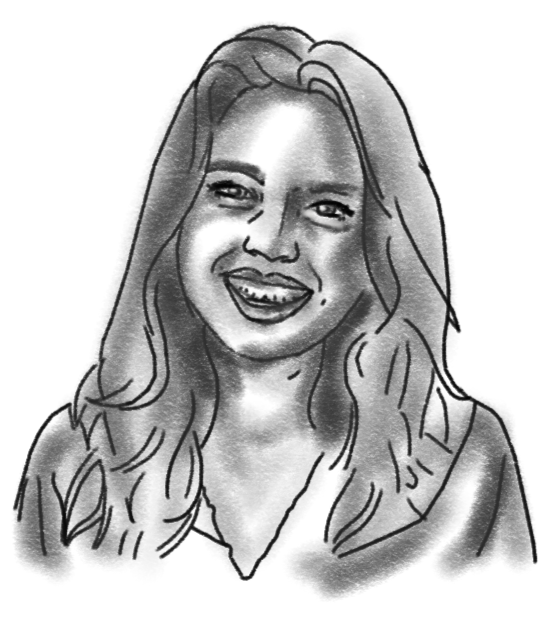}
\end{wrapfigure}
\textbf{Ava Polzin} is a PhD candidate in Astronomy \& Astrophysics at the University of Chicago. Her work focuses on understanding star formation and the baryon cycle in the smallest dwarf galaxies. She lives in Chicago's Hyde Park neighborhood with her two cats, Emma (a mercurial calico) and F\'{e}licette (named for the first cat in space; see last page for the story of the \textit{original} F\'{e}licette). The mini-grant she received from the International Astronomical Union's North American Regional Office of Astronomy for Development and the Heising-Simons Foundation for this work went a long way in printing these booklets thanks to the contributions of more than thirty (!!) women, who are committed to encouraging girls in their pursuit of astrophysics. This effort is the culmination of years of Ava's work toward making science more equitable and accessible (see \href{https://siriusb.org}{siriusb.org} for recent details) and in addition to writing all of the uncredited sections (and some of the credited ones), she will be running the synchronous VERGE program, for which this guide was created, in January 2024 with some of the women who contributed here also giving talks/presentations to the participating girls.

The printing of this work for the 2024 VERGE program is funded by the IAU NA-ROAD and the Heising-Simons Foundation under the Women and Girls in Astronomy Program.
\\

\vfill
\begin{center}
\textit{\small All art by Julie Malewicz.}  
\end{center}

\pagebreak
\LARGE{\emph{Contributors (Listed Alphabetically)}}
\normalsize
\vspace{20pt}

\Large{\emph{Section Authors}}\\
\vspace{-10pt}
\normalsize

\begin{wrapfigure}[5]{l}{0.22\textwidth}
\vspace{-\intextsep}
\includegraphics[width=0.9\linewidth]{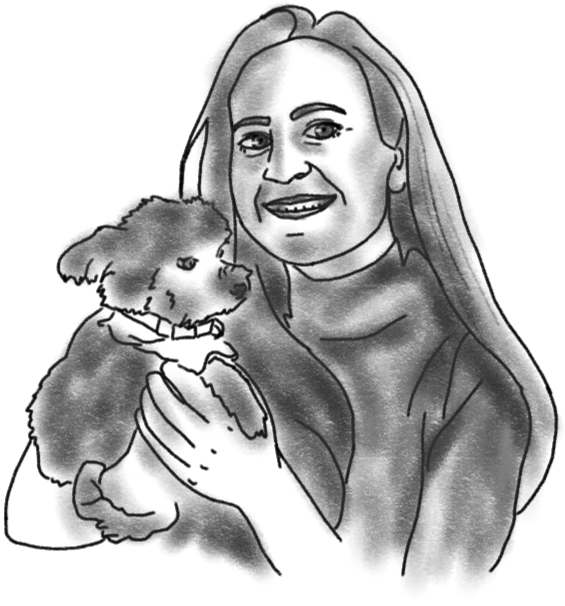}
\end{wrapfigure}
\textbf{Yasmeen Asali} is a PhD candidate in Astronomy at Yale University. She works on understanding how star formation evolves in low mass galaxies and how a galaxy’s environment can impact this evolution. She is interested in the intersection of data science and astronomy, and passionate about teaching.\\
\\

\begin{wrapfigure}[4]{r}{0.22\textwidth}
\vspace{-2.5\intextsep}
\includegraphics[width=0.9\linewidth]{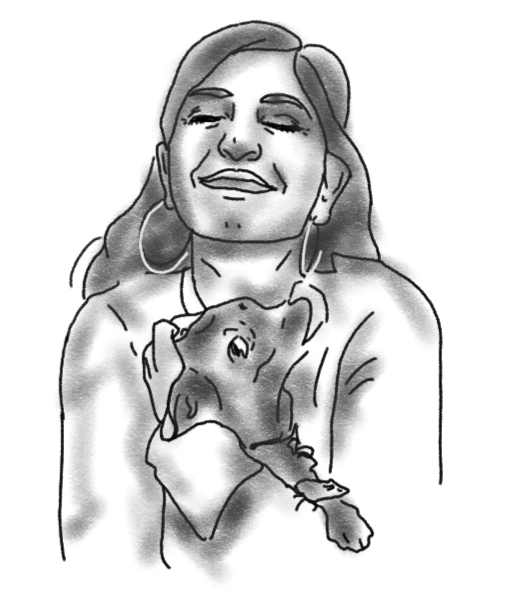}
\end{wrapfigure}
\textbf{Sanah Bhimani} is a PhD candidate in physics at Yale University. She works on CMB experiments like the Simons Observatory. Sanah and her rescue dog Etta enjoy walking around New Haven's East Rock neighborhood.\\
\\

\begin{wrapfigure}{l}{0.22\textwidth}
\vspace{-\intextsep}
\includegraphics[width=0.9\linewidth]{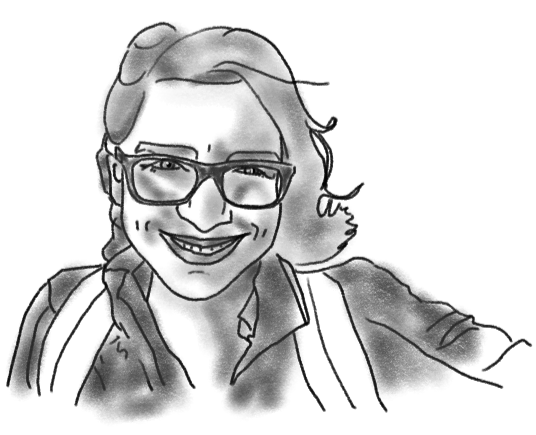}
\end{wrapfigure}
\textbf{Madison Brady} is a PhD candidate in Astronomy \& Astrophysics at the University of Chicago.  She uses high-precision spectroscopic data to measure the masses of nearby exoplanets.  She is interested in studying how rocky, Earth-like planets form and evolve around small stars.  She also wants to figure out whether or not water worlds exist.\\
\\

\begin{wrapfigure}[6]{r}{0.22\textwidth}
\includegraphics[width=0.9\linewidth]{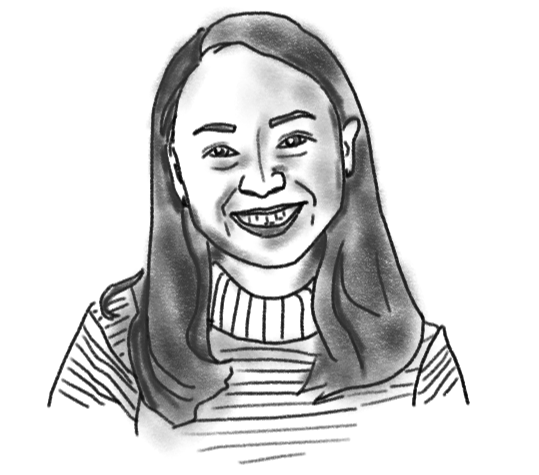}
\end{wrapfigure}
\textbf{Mandy Chen} is a PhD candidate in Astronomy \& Astrophysics at the University of Chicago. She uses integral-unit spectroscopic data to discover and model the motions of diffuse gas surrounding galaxies.  The overarching goal of her research is to understand the interplay between galaxies and their low-density gaseous environments, and how these interactions evolve over cosmic time.\\
\\

\begin{wrapfigure}[9]{l}{0.22\textwidth}
\includegraphics[width=0.9\linewidth]{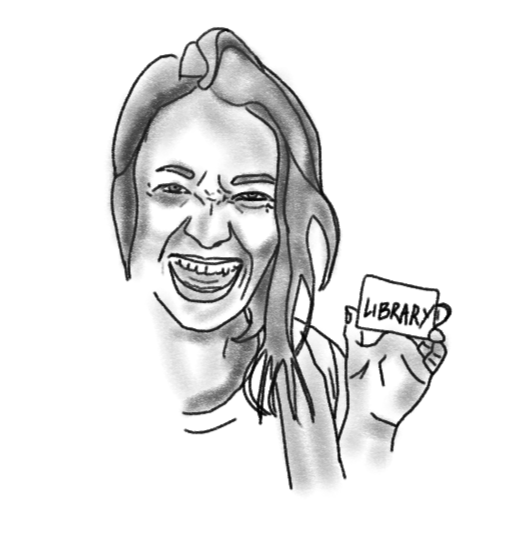}
\end{wrapfigure}
\textbf{Lindsay DeMarchi} is a stellar mortician and space environmentalist. She earned her Ph.D. in Astronomy at Northwestern University by researching multi-messenger, multi-wavelength approaches to dead and dying stars. Her greatest dream is to protect and steward darkness and silence, so she volunteers with the IAU CPS, DarkSky International, and AAS. Her favorite hobby is collaborating with indie filmmakers and video game developers to make their space visions real. \\
\\


\begin{wrapfigure}[5]{r}{0.22\textwidth}
\vspace{-1.5\intextsep}
\includegraphics[width=0.9\linewidth]{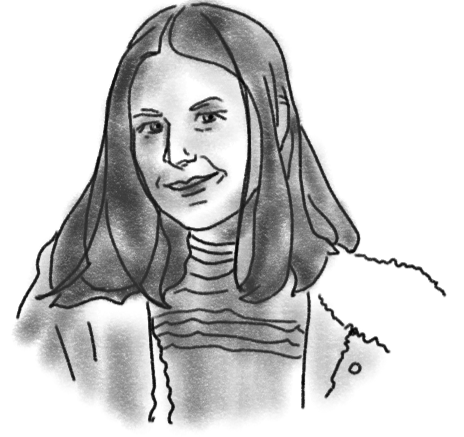}
\end{wrapfigure}
\textbf{Michelle Gurevich} is a PhD candidate in the Theoretical Particle Physics and Cosmology group at King's College London. Her background is in general relativity and gravitational waves. She currently studies the ringdown of binary black hole mergers to test theories of gravity.\\
\\

\vspace{-10pt}

\begin{wrapfigure}[11]{l}{0.22\textwidth}
\vspace{20pt}
\includegraphics[width=0.9\linewidth]{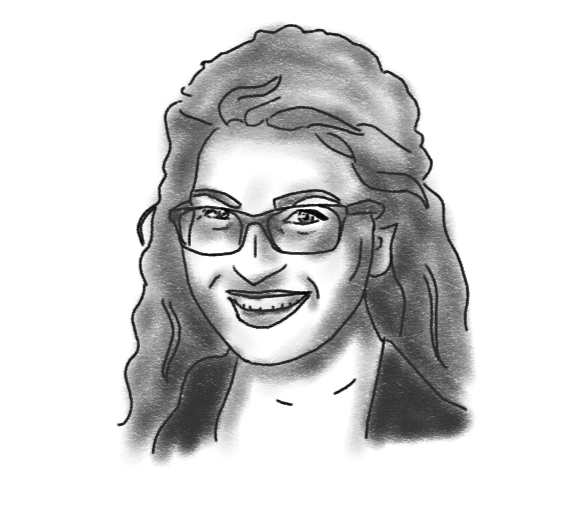}
\end{wrapfigure}
\textbf{Emily Lichko} is a postdoc at the University of Chicago. She received her B.S. in Physics and Applied Mathematics from the University of Michigan in 2013 and her PhD in 2020 from the University of Wisconsin - Madison, working under the supervision of Professor Jan Egedal. Her research focuses on kinetic plasma physics processes in space and astrophysical plasmas, in particular as they relate to questions of particle heating and nonlinear processes that affect the evolution of collisionless, anisotropic plasmas. Outside of her research, she enjoys swimming, biking, running, and failing to replicate recipes from the Great British Bake Off. \\
\\

\vspace{-20pt}

\begin{wrapfigure}[7]{r}{0.22\textwidth}
\vspace{-\intextsep}
\includegraphics[width=0.9\linewidth]{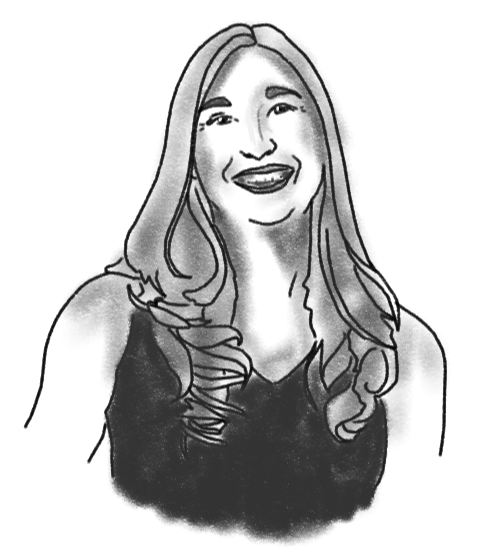}
\end{wrapfigure}
\textbf{Emma Louden} is a PhD candidate at Yale University in New Haven. She studies the geometry of exoplanetary systems. Emma is passionate about a future-focused strategy for astrophysics, engaging the public with space exploration, and applying evidence-based solutions to solve the world’s most pressing problems. She is a 2018 Brooke Owens Fellow and a 2023 Quad Fellow.
\\
\\

\strut \vspace{-50pt}

\begin{wrapfigure}[7]{l}{0.22\textwidth}
\includegraphics[width=0.9\linewidth]{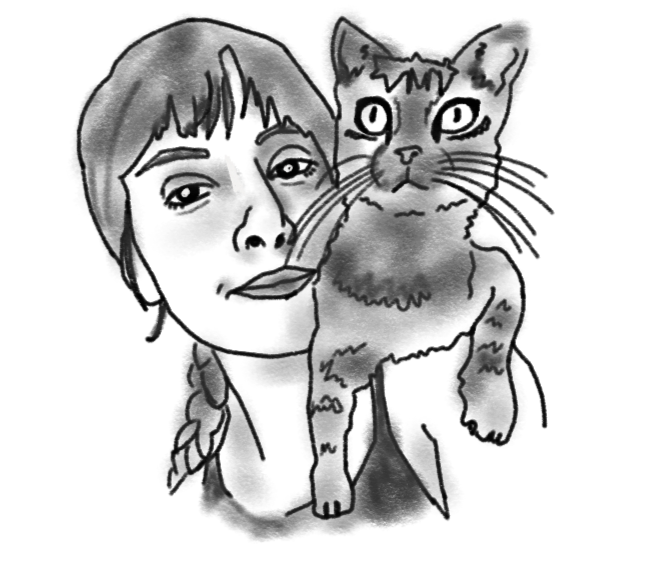}
\end{wrapfigure}
\textbf{Julie Malewicz}  is a PhD candidate down at the Georgia Institute of Technology in Atlanta, where she's working on predicting the X-ray signal from accreting supermassive black hole binaries. She is passionate about public transit, community gardens, taking in stray cats (incl. one very persistent tabby called Ladybug) and advocating for science as a field to be more inclusive and accessible. \\
\\

\begin{wrapfigure}[3]{r}{0.22\textwidth}
\vspace{-2.5\intextsep}
\includegraphics[width=0.9\linewidth]{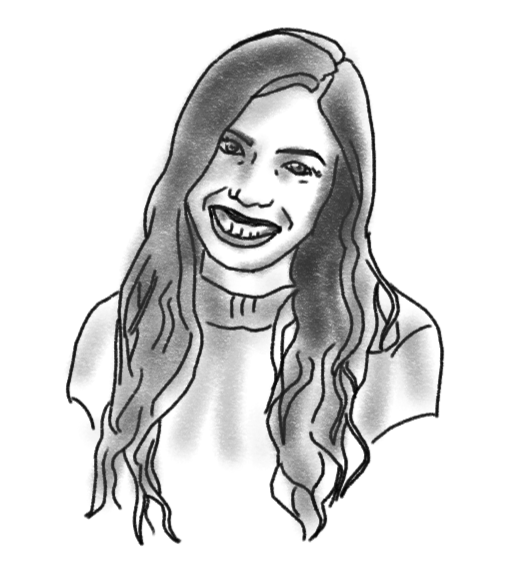}
\end{wrapfigure}
\textbf{Samantha Pagan} is a physics Ph.D. candidate at Yale University. She researches dark matter and neutrinos to explore questions such as why our universe is made of matter.\\
\\

\begin{wrapfigure}[6]{l}{0.22\textwidth}
\vspace{-\intextsep}
\includegraphics[width=0.9\linewidth]{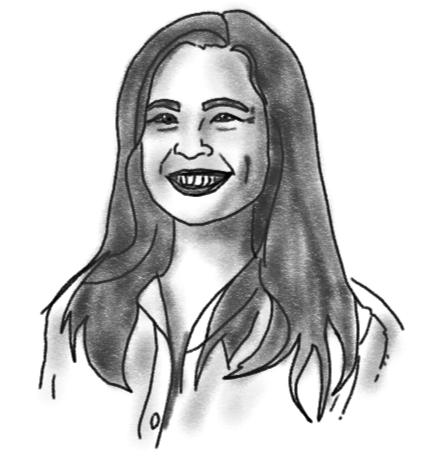}
\end{wrapfigure}
\textbf{Malena Rice} is an Assistant Professor of Astronomy at Yale University. Her research focuses on understanding the diversity of planetary systems, drawing together insights from both exoplanet systems and the solar system. Her recent work has focused on characterizing exoplanet orbital architectures and the distant solar system. \\
\\

\begin{wrapfigure}[4]{r}{0.22\textwidth}
\vspace{-1.75\intextsep}
\includegraphics[width=0.9\linewidth]{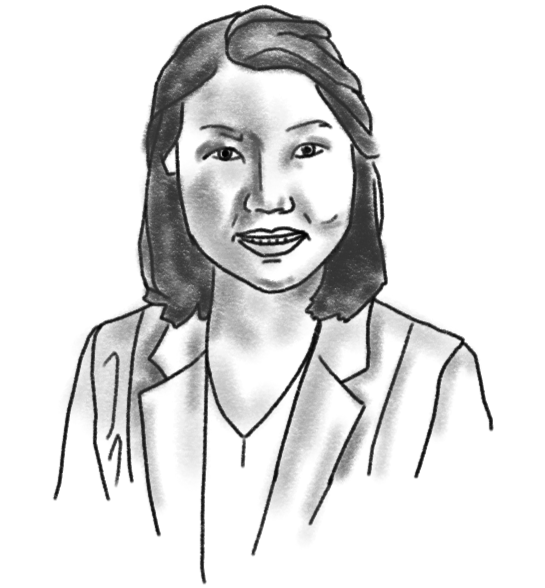}
\end{wrapfigure}
\textbf{Zili Shen} is a PhD candidate in the Yale University Department of Astronomy. She uses the Keck Observatory and the Hubble Space Telescope to observe faint galaxies. She works on discovering the biggest and most diffuse galaxies in our cosmic neighborhood.\\
\\

\begin{wrapfigure}[6]{l}{0.22\textwidth}
\vspace{-0.5\intextsep}
\includegraphics[width=0.9\linewidth]{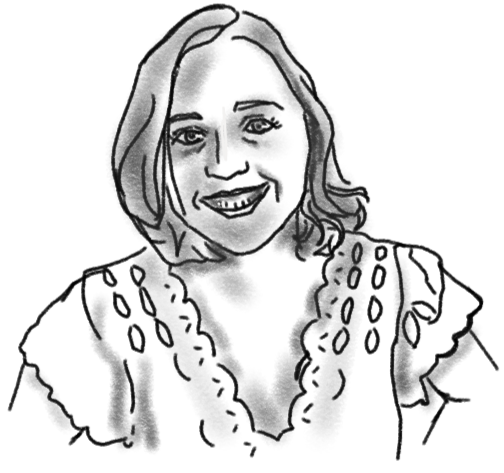}
\end{wrapfigure}
\textbf{Emily Simon} is a PhD candidate in Astronomy \& Astrophysics at the University of Chicago. She is interested in astroparticle physics and hopes to understand the origins and acceleration mechanisms for high energy particles like cosmic rays and neutrinos which likely come from the most extreme environments in the universe like supernova explosions and black hole accretion disks.\\
\\

\begin{wrapfigure}[5]{r}{0.22\textwidth}
\vspace{-1.2\intextsep}
\includegraphics[width=0.9\linewidth]{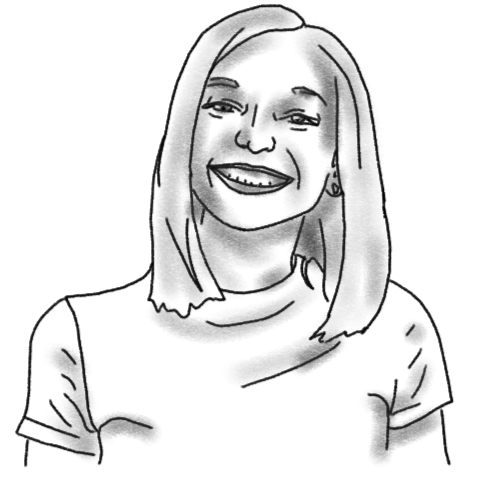}
\end{wrapfigure}
\textbf{Candice Stauffer} received her Ph.D. in Astrophysics from Northwestern University in 2023. She is an expert in using computational and machine learning methods to study stellar explosions, like supernovae, and the multi-wavelength Universe. She currently works as a Data Scientist for the City of Chicago. \\
\\

\begin{wrapfigure}[9]{l}{0.22\textwidth}
\includegraphics[width=0.9\linewidth]{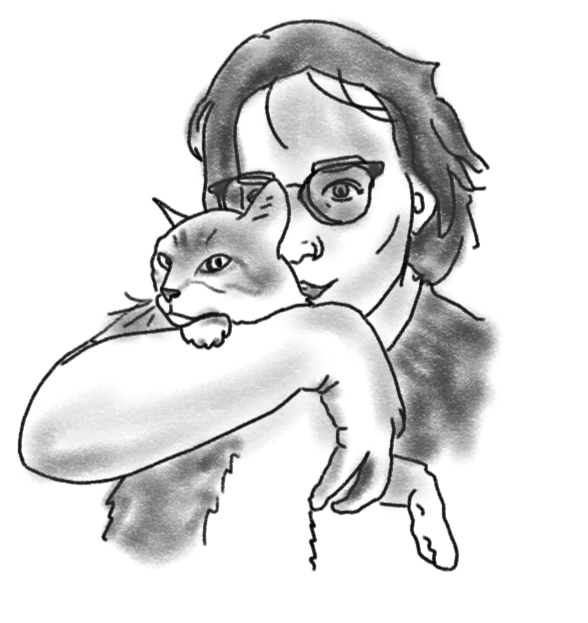}
\end{wrapfigure}
\textbf{Luna Zagorac} is a postdoctoral fellow at the Perimeter Institute for Theoretical Physics in Ontario, Canada. She is a cosmologist through and through: passionate not just about what our silly little Universe is up to, but also about all the ways we as humans interact with it and understand it. When not thinking about dark matter, inflation, and other parts of our invisible Universe, she is using data science to understand astronomical and observational practices in ancient Egypt, roller skating, reading, or playing video games. \\

\pagebreak
\Large{\emph{Advice for Students}}\\
\vspace{-10pt}
\normalsize

\begin{wrapfigure}[18]{r}{0.22\textwidth}
\vspace{50pt}
\includegraphics[width=0.9\linewidth]{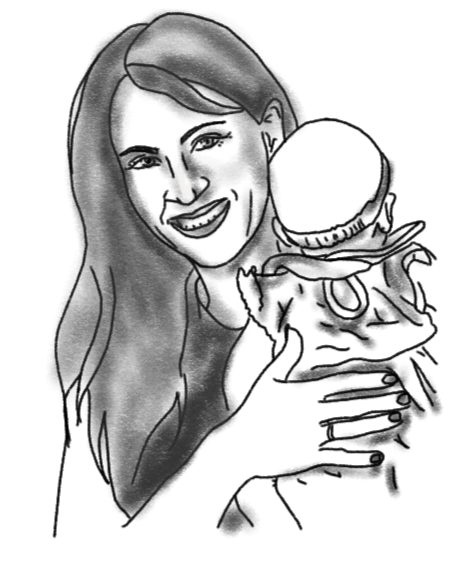}
\end{wrapfigure}
\textbf{Katie Auchettl} is an observational astrophysicist whose research focuses on understanding the physical processes and observational signatures of the extreme death of stars using instruments that observe across the entire electromagnetic spectrum.  Growing up in Melbourne, Australia she was lucky enough to have access to some of the darkest skies anywhere in the world. This naturally led to her excitement about studying the stars in any way she could. As a first generation university student who attended a local public school, it wasn't until university that she realised she could pursue a degree in astronomy. After finishing a PhD at Monash University in Melbourne in 2015, she was a CCAPP Postdoctoral Fellow at the Ohio State University and then an Assistant Professor at the University of Copenhagen before joining the University of Melbourne in 2020, where she is currently an Associate Professor. Apart from studying the stars, Katie is passionate about issues related to diversity, equity and inclusion in the sciences as well as outreach that allows all members of the community to be engaged in science.\\
\\


\begin{wrapfigure}[11]{l}{0.22\textwidth}
\vspace{25pt}
\includegraphics[width=0.9\linewidth]{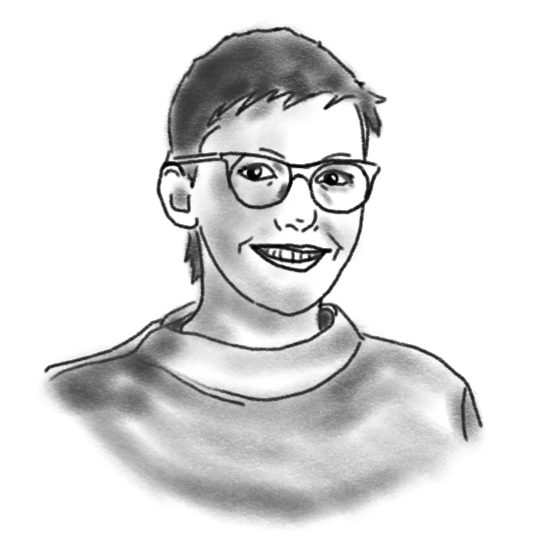}
\end{wrapfigure}
\textbf{Katie Breivik} grew up under the stars of northern Utah and southern Idaho where her family spent almost every weekend building a cabin. During the weekdays, she'd drag her parents across grocery store parking lots to look through telescopes at star parties hosted by the Salt Lake Astronomical Society. She attended Utah State University, and following the advice of her high school physics teacher Ms Amiot, she immediately switched from a Mechanical Engineering major to Physics. After finishing a PhD at Northwestern University and two postdocs, one in Toronto and one in New York, she now lives in Pittsburgh and is an assistant professor at Carnegie Mellon University.\\
\\


\begin{wrapfigure}[8]{r}{0.22\textwidth}
\vspace{13pt}
\includegraphics[width=0.9\linewidth]{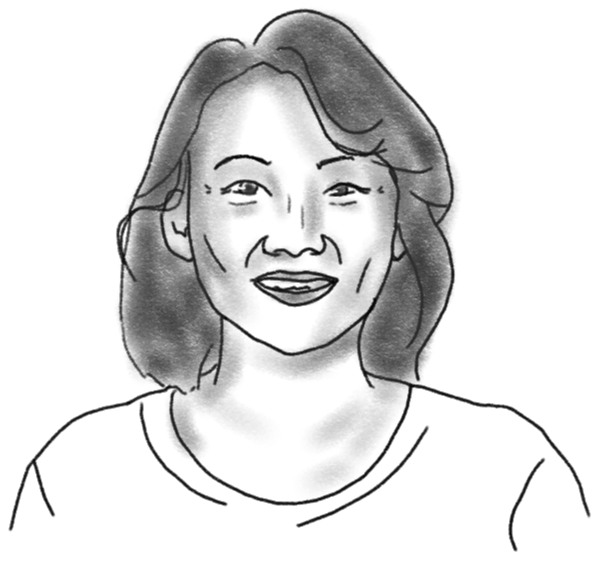}
\end{wrapfigure}
\textbf{Hsiao-Wen Chen} is a Professor of Astronomy and Astrophysics at the University of Chicago.  She grew up in Taipei, Taiwan, and came to the US to pursue PhD in Astrophysics.  Her research interests broadly cover issues concerning the formation and evolution of galaxies across cosmic time. In particular, she is interested in studying the intricate connections between galaxies and the tenuous circumgalactic medium, in order to understand the complex physical processes that drive the gas flows in and out of galaxies.\\
\\

\vspace{-10pt}

\begin{wrapfigure}[13]{l}{0.22\textwidth}
\vspace{35pt}
\includegraphics[width=0.9\linewidth]{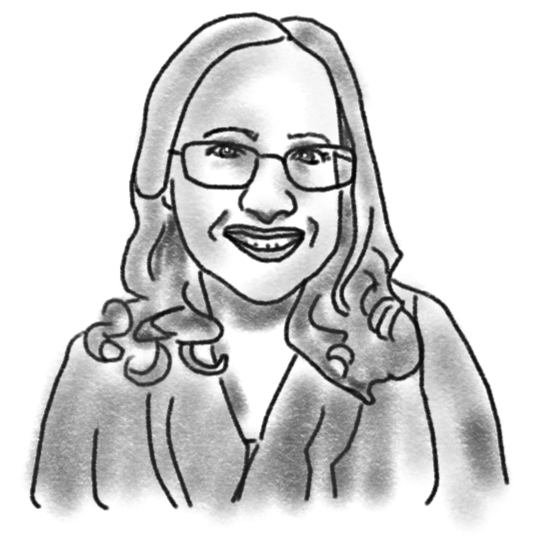}
\end{wrapfigure}
\textbf{Deanne Coppejans} is an assistant professor of astrophysics at the University of Warwick in the United Kingdom (UK). She uses multi-wavelength observations to study the high energy astrophysics of binary stars and stellar explosions. Deanne is from South Africa. During her undergraduate studies, South Africa was competing to host the largest radio telescope on the planet (the Square Kilometre Array, SKA), and this opened her eyes to the fact that she could study the universe as a career. The South African SKA project helped her do this by funding her honours and MSc studies. With the support of the Erasmus Mundus SAPIENT programme, she then went on to do her PhD in the Netherlands. Thereafter she did a 5 year postdoc in the United States, before moving to the UK.\\
\\

\vspace{-10pt}

\begin{wrapfigure}[10]{r}{0.22\textwidth}
\vspace{15pt}
\includegraphics[width=0.9\linewidth]{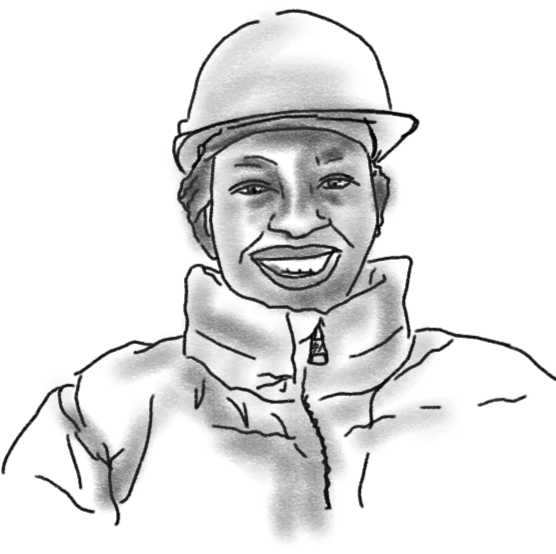}
\end{wrapfigure}
\textbf{Sthabile Kolwa} is obsessed with black holes and has been since first learning about them during primary school in South Africa. After completing her BSc and MSc with funding provided by the SKA project, Sthabile successfully defended her PhD thesis at the Ludwig Maximilian University of Munich in 2019. After this, she became a SARAO postdoctoral research fellow before beginning a lectureship in 2021 at the University of Johannesburg, where she now teaches physics and studies the evolution of active galaxies. Believing in the fundamental right of access to equal opportunities for all, Sthabile advocates for underrepresented groups in astronomy. \\
\\

\begin{wrapfigure}[8]{l}{0.22\textwidth}
\includegraphics[width=0.9\linewidth]{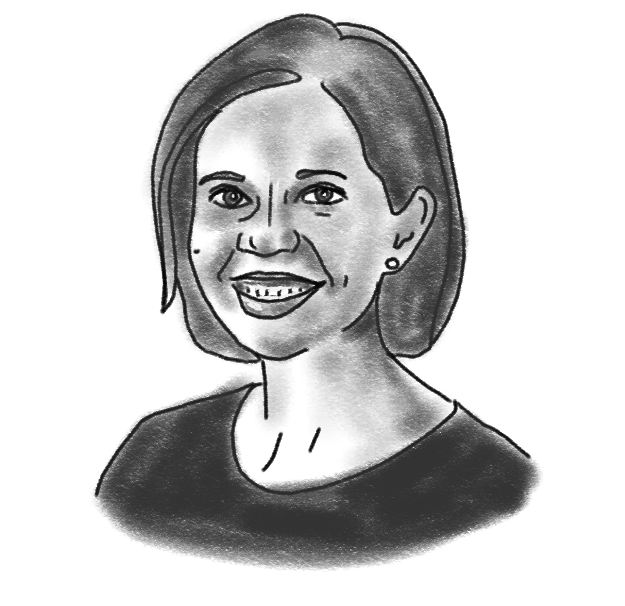}
\end{wrapfigure}
\textbf{Raffaella Margutti} grew up in a small village of $\sim4000$ people in the north of Italy and moved to Harvard as a postdoc after obtaining her PhD in Milano during which she was very fortunate to work for the Swift spacecraft. She is now an Associate Professor at UC Berkeley, where she holds a position in the Astronomy department and in the Physics department. She owes a lot to her PhD advisor Guido Chincarini who has always believed in her ideas.\\
\\

\begin{wrapfigure}[5]{r}{0.22\textwidth}
\vspace{-1.7\intextsep}
\includegraphics[width=0.9\linewidth]{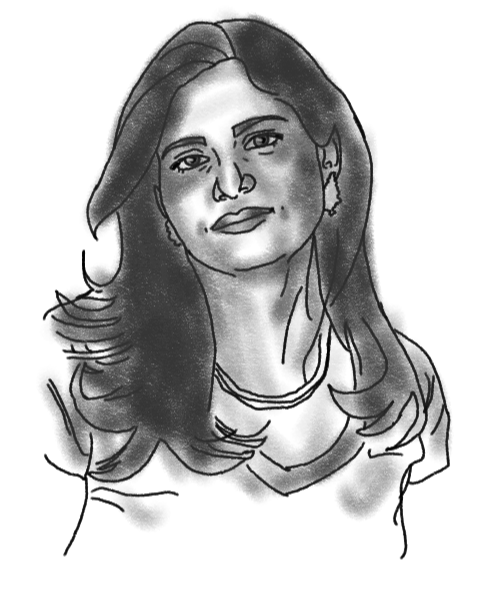}
\end{wrapfigure}
\textbf{Priyamvada Natarajan}  is a theoretical astrophysicist and is the inaugural Joseph S. and Sophia S. Fruton
Professor of Astronomy and Physics at Yale. Her research work is focused on understanding the nature of the invisible Universe, namely, black holes, dark matter and dark energy.\\
\\

\begin{wrapfigure}[5]{l}{0.22\textwidth}
\vspace{-1.7\intextsep}
\includegraphics[width=0.9\linewidth]{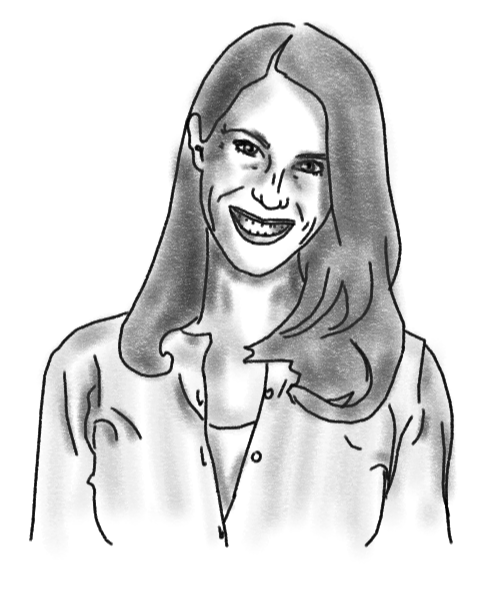}
\end{wrapfigure}
\textbf{Erica Nelson} is an astrophysicist and Assistant Professor in the Department of Astrophysical and Planetary Sciences at the University of Colorado, Boulder. She is obsessed with data on the first galaxies from the newly launched JWST and loves discovering new things about our weird and wonderful universe. \\
\\

\begin{wrapfigure}{r}{0.22\textwidth}
\vspace{-\intextsep}
\includegraphics[width=0.9\linewidth]
{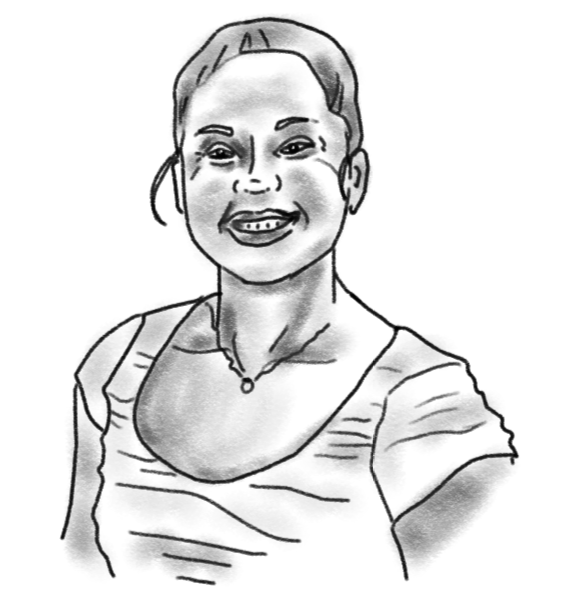}
\end{wrapfigure}
\textbf{Kim Page} obtained both her undergraduate degree in Physics with
Astrophysics, and her PhD in X-ray Astronomy, from the University of
Leicester, UK. She has been a member of the Neil Gehrels Swift Observatory
team since before the satellite was launched in 2004, and helps to run the
UK Swift Science Data Centre based at the University of Leicester, as well
as working on X-ray observations of novae and GRBs.\\
\\

\begin{wrapfigure}[9]{l}{0.22\textwidth}
\vspace{15pt}
\includegraphics[width=0.9\linewidth]{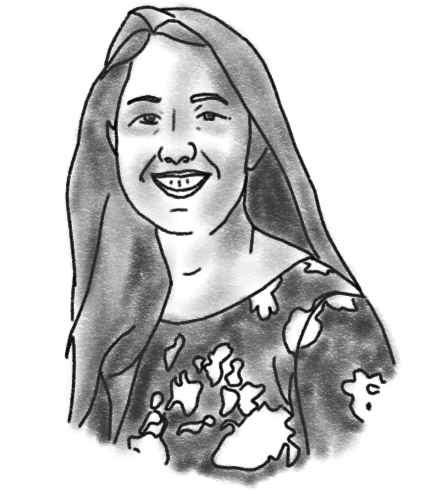}
\end{wrapfigure}
\textbf{Silvia Toonen} is an Assistant Professor in astrophysics at the University of Amsterdam in the Netherlands. She grew up in a small town in a rural part of the country, and was the first of her family to go to university. She is interested in the ways that stars live their lives, and in particular how they interact with other stars which may lead to bright and energetic transients. Her research focuses on numerical simulations of the evolution of binary star systems, and in recent years she has pioneered the field of triple star systems.\\
\\

\begin{wrapfigure}[9]{r}{0.22\textwidth}
\includegraphics[width=0.9\linewidth]{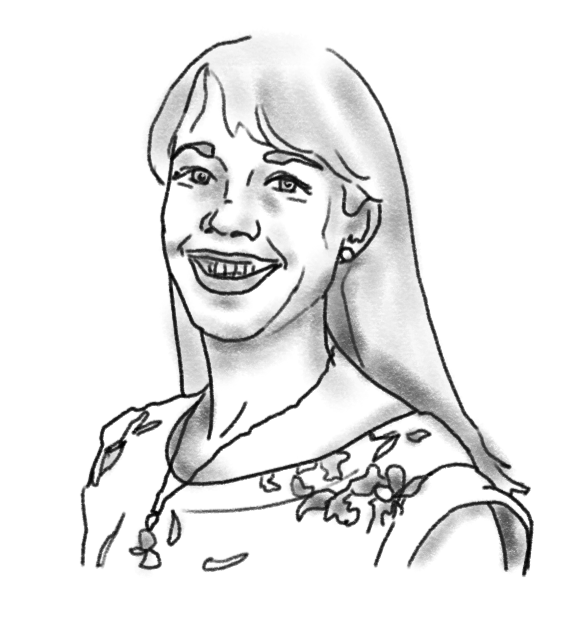}
\end{wrapfigure}
\textbf{Katherine E. Whitaker} obtained her undergraduate degree double majoring in Physics \& Astronomy at the University of Massachusetts Amherst, and later her PhD from Yale University. She is currently an Associate Professor of Astronomy at the University of Massachusetts Amherst, as well as associate faculty at the Cosmic Dawn Center of Excellence in Copenhagen, Denmark. Her research involves understanding how the most massive galaxies evolve over billions of years of cosmic time using big telescopes in space.\\
\\

\begin{wrapfigure}[7]{l}{0.22\textwidth}
\vspace{-5pt}
\includegraphics[width=0.9\linewidth]{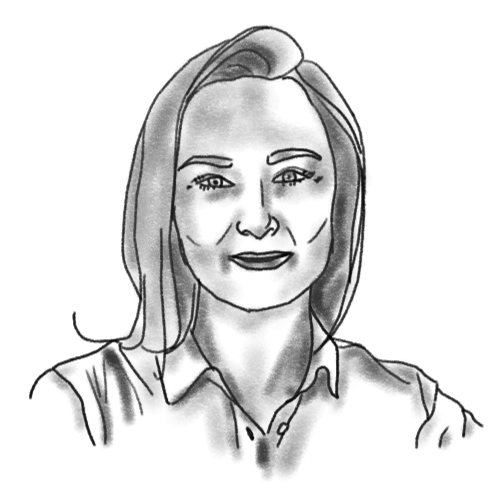}
\end{wrapfigure}
\textbf{Irina Zhuravleva} is an Assistant Professor at the University of Chicago. She is interested in the hottest and most energetic phenomena in the Universe, including interactions of supermassive black holes with galaxies and the evolution of the largest objects containing thousands of galaxies. Her studies are based on data from X-ray satellites and numerical modeling. She is also a NASA participating scientist for the recently launched XRISM satellite.
\tableofcontents

\pagebreak
\strut \vspace{130pt}

\noindent \textsc{``Astronomy and Astrophysics (and more generally Science) provides an amazing opportunity for discovery, excitement, learning, creativity and fostering a passion in a particular (or many!) topic(s). However, it can feel overwhelming or challenging at some points and all of us have gone through this. But always remember that you absolutely deserve to be here and you have already made the hardest step by taking the very first steps towards cultivating your passion and building your foundation for a career (either in astronomy or in another field entirely!). Continue that passion, motivation, and persistence no matter what you do, and surround yourself with people who provide a welcoming and supportive environment and that lift you up and support you unequivocally in your journey. If you don't have someone locally, try to find your people, as I know there are so many of us that want to see your star shine brighter and brighter. We can't wait to see what you will do next!''}
\\
\\
\strut\hfill \textemdash \textsc{Prof. Katie Auchettl, University of Melbourne}\\
\strut \hfill \footnotesize{\textsc{Associate Professor of Astrophysics}}
\normalsize

\section{Introduction}

One of the hardest parts of starting a career in astrophysics, as in most fields, is getting a foot in the door. The goal of this booklet is to demystify some aspects of research careers in astronomy/astrophysics and offer a jumping off point for further independent investigation of topics that might spark interest. Though it is not possible to comprehensively cover a science tasked with understanding something as large as the universe, we attempt here to offer a primer on many broad (and hopefully representative) topics in astronomy and provide some insights into the current frontiers in those sub-fields. For clarity, each sub-field is presented in its own section, but these research areas are not islands -- instead, much of the knowledge, and many of the questions, presented here are interconnected.

At the same time, modern astrophysics is, by its very nature, interdisciplinary. On top of having a vast working knowledge of physics, astrophysicists generally have strong foundations in mathematics, statistics, chemistry, and computer science. In fact, scientists in general wear many hats. Not only are scientists often charged with engineering and construction, software development, science communication, and more in the course of their science, they are also frequently charged with lobbying and informing matters of public policy. Though there is substantial freedom to study what you want (past a certain career stage), it comes with the institutional and intellectual responsibility to share what you have learned and use your science for social good.

In astronomy, we are very fortunate; the breathtaking pictures produced by ground- and space-based telescopes (across almost the entire electromagnetic spectrum) speak to people of all backgrounds and interests. There is something innately human about wanting to know where we came from, what our place is in the universe, and where we are going, and the larger questions we ask connect to this desire. The women who've contributed here fundamentally believe in the shared human pursuit of these questions.

\pagebreak
\strut \vspace{185pt}

\noindent \textsc{``Your curiosity, love of science, motivation and persistence are the most important things for a successful career in astrophysics. There will be ups and downs in your pursuit of your goals and dreams -- please really celebrate your achievements and also use them to remind yourself that you can do it when you are struggling. If you ever find that you are doubting yourself, please remind yourself how far you've come and that we are all cheering you on.''}
\\
\\
\strut\hfill \textemdash \textsc{Prof. Deanne Coppejans, University of Warwick}\\
\strut \hfill \footnotesize{\textsc{Assistant Professor of Astrophysics}}
\normalsize

\section{Thinking Mathematically}

It's really common to approach equations as intangible, but, in science, they are a shorthand for describing the world around us. There is no part of astrophysics where you can avoid understanding the way math is used to describe physical concepts. In this section, you will walk through two (perhaps already obvious) points -- 1) these physical descriptions can be broken down to be more easily digested and retained and 2) using units to build physical intuition (i.e., \textit{dimensional analysis}).

\subsection{Interpreting Mathematical Expressions}

Math, often from the way it's taught in school, can feel somewhat intimidating. Especially when you're encountering concepts for the first time, it can be a challenge to break down what you're seeing and develop an intuition.

One of the things that can make math more accessible in physical contexts is recognizing that physical intuition applies. If we take a simple example $v = \frac{\Delta x}{\Delta t}$, where $v$ is velocity (or speed in this case, since we can ignore direction), $\Delta x$ represents a change in position, and $\Delta t$ represents a change in time. (\emph{$\Delta$ generally represents a change in whatever variable follows.})

Since $v$ is proportional to $\Delta x$, if Object 1 goes farther than Object 2 in the same amount of time, the velocity of Object 1 must be higher than the velocity of Object 2. Similarly, since $v$ is inversely proportional to $\Delta t$, if Object 1 takes less time to go the same distance as Object 2, then Object 1 must once again be moving faster than Object 2.

This sort of thinking can be extended to much more complicated systems and equations. In essence, it is a matter of ensuring that you engage critically with the math rather than just ``plugging and chugging'', where you use equations to an end without considering what they're telling you. It takes practice to become comfortable reading the math this way, but once you begin doing it regularly, it will substantially deepen your understanding of what you are doing.

Looking toward the next section, we might consider that a change in position is measured in units of length, and a change in time is measured in units of time, which means that velocity must carry units of length/time. Because these units are different, we can multiply or divide them, but cannot add or subtract them. The following expression is fine:
\begin{equation}
   v_1 =  v_0 + \frac{\Delta x}{\Delta t} \notag
\end{equation}
while $v_1 = v_0 + \Delta x$ makes no physical or mathematical sense. Additionally, the units on either side of an expression must be equivalent so the rearranged version, $\Delta x = v_0 + v_1$, would similarly be non-physical and incorrect. Another way to think about this is that units essentially act algebraically -- this is why it is always important to know what dimensions and units your variables carry!

(An example from calculus, using the same expression for velocity and position, is that $x_1 = x_0 + \int v \, dt$ -- here $dt$ already confers dimensions of time, so that this integral is valid.)

Knowing the physical dimensions that numbers carry is really powerful. You can gain a tremendous amount of insight into a problem by balancing these dimensions in a practice called dimensional analysis.

Don't worry if this isn't totally clear yet, it takes time and repeated exposure. In an undergraduate physics major, you will generally be required to take formal courses in differential, integral, vector, and multivariate (or multivariable) calculus, linear algebra, differential equations, and Fourier analysis. (Real and complex analysis are often deemed useful, but are usually not required.) Take it one step at a time, really internalize the math, and eventually you will gain intuition for the physical meaning of expressions employing even these more advanced mathematics.

\subsection{Dimensional Analysis}

Everyone's favorite example of dimensional analysis is the analytic (using only algebra and not plugging all variables in) calculation of the extent of a black hole's event horizon. Say that you're given that exercise without further information or equations, only the value of certain constants: all you need to know is that the event horizon (a term we will use loosely here) is the distance from the black hole at which the escape velocity becomes greater than the speed of light. Because the event horizon is the location at which $v_\mathrm{esc}$ becomes larger than the speed of light, $c$, we can consider the boundary case where $v_\mathrm{esc} = c$. The escape velocity is, physically, the velocity at which the kinetic energy ($K$) of some object is greater than the gravitational potential energy ($P$) of the body from which it is trying to escape, which gives us another boundary $|K| = |P|$ (we take the absolute value here to avoid worrying about the orientation of our coordinate systems, since we only care about the relative strength of $K$ vs. $P$).

We can see that both of these boundaries make sense dimensionally -- we are simply stating that one velocity is equivalent to another and one energy is equivalent to another at this boundary. But what of the actual expressions for $K$ and $P$? Energy has dimensions of mass length$^2$ time$^{-2}$. In the case of kinetic energy, we often see it as [mass] $\times$ [length/time]$^{2}$, which, even without knowing the constant at the beginning, will get you within a factor of two of $K = \frac{1}{2}mv^2$. Given that $G$ (the gravitational constant) carries dimensions of mass$^{-1}$ length$^3$ time$^{-2}$, without worrying about the value of $G$, can you figure out the expression for potential energy knowing that it has to depend on radius (or else there wouldn't be a way to infer the radius of the event horizon!), so that $P$ decreases as $r$ increases? Similarly, $P$ increases with increasing mass of the body $M$ and increasing mass of the object $m$.  Remember that the final expression for $P$, should have units of energy (mass length$^2$ time$^{-2}$).\\
\\
\vfill

\pagebreak

Let's think about this in parts. We know that we want units of mass length$^2$ time$^{-2}$ and already have mass$^{-1}$ length$^3$ time$^{-2}$ from $G$. $M$ and $m$ will each have units of mass by default, and $r$ has units of length.

We already know that $P \propto G$ gives us time$^{-2}$, so this is a good first step. $G$ does give us an extra factor of length, though. To fix this, we can take $P \propto \frac{G}{r}$, which has units of mass$^{-1}$ length$^2$ time$^{-2}$. Now we need two powers of mass and, knowing that the gravitational potential depends on the masses of both the body and the object, we can assume that this comes from $M\times m$, so our gravitational potential energy is $\propto \frac{GMm}{r}$. It turns out there's a factor of $-1$ here relative to $K$, but this can also be set by the orientation of $r$, so $P = -\frac{GMm}{r}$ which we largely recover just by thinking about the units!

Now for the rest of the question -- we are trying to find the radius of the event horizon, so we can set up the following equation:
\begin{equation}
    \frac{1}{2} m v_\mathrm{esc}^2 = \frac{G M_\mathrm{BH} m}{r} \notag
\end{equation}
Immediately, we can see that the mass of the object trying to escape ($m$) does not matter because that term cancels (and good thing, since in the case where $v_\mathrm{esc} = c$, we are discussing light, and photons are massless). This means that we are left with:
\begin{equation}
    \frac{1}{2} v_\mathrm{esc}^2 = \frac{GM_\mathrm{BH}}{r} \notag
\end{equation}
Rearranging things and taking $v_\mathrm{esc} = c$ as in the definition of the event horizon, we find that:
\begin{equation}
    r = \frac{2 G M_\mathrm{BH}}{c^2}
\end{equation}
This tells us that the extent of the event horizon \emph{only} depends on the mass of the black hole, since $G$ and $c$ are constants. In fact, it relates linearly ($r \propto M_\mathrm{BH}$), which means that if there are two black holes with $M_\mathrm{BH,1} = 10 M_\mathrm{BH,2}$, then $r_1 = 10\,r_2$.

\pagebreak
\strut \vspace{175pt}

\noindent \textsc{``Studies show that the best predictor of success when pursuing advanced degrees in STEM is not how well you do on standardized tests, but rather your grit, determination, and motivation.  My advice to you is to have confidence in yourself and your abilities, even when you don’t feel that way on the inside. You are strong and capable, and, with perseverance, you can tackle any problem. It has been said that you should \textsl{fake it until you make it}; instead, try faking it until you become it.''}
\\
\\
\strut\hfill \textemdash \textsc{Prof. Katherine E. Whitaker, University of Massachusetts Amherst}\\
\strut \hfill \footnotesize{\textsc{Associate Professor of Astronomy}}
\normalsize

\section{Modes of Study}

As with anything, there are multiple ways to approach science. In (astro)physics, we usually break this down into two to three categories. In the binary classification, there are experimentalists, who use purpose-built instruments to study the universe, and there are theorists, who rely more heavily on physical first principles to make predictions about the universe. In the ternary classification, there are instrumentationalists, who build, maintain, and operate telescopes and other instruments; there are observers, who use existing telescopes to study the universe with collected data; and there are theorists.

\subsection{Instrumentation}

Scientists who are focused on instrumentation spend much, if not all, of their time, working on the construction, commissioning, and maintenance of an instrument or several. In astronomy, these instruments are generally telescopes, though some folks who work closer to the intersection of biology or geology (primarily planetary scientists and exoplanet-focused astronomers) may work on robotic probes or other projects.

Instrumentationalists (sometimes called experimentalists) will shepherd an instrument from its initial conception through its construction and commissioning, and then remain working on maintaining/running it. Given the long timescales of large-scale astrophysical projects, some people will only ever work on a few of these different phases, but there is still a tremendous variety of activities that fall under the experimentalist umbrella. Ground-based instruments at the commissioning phase, for instance, may require travel to the site -- often in remote locations like the Atacama Desert or the South Pole.

For students who are interested in engineering and in astrophysics, this may be a means of combining those passions. 

\subsection{Observation}

Observers are like the explorers of modern astrophysics -- they use telescopes to answer questions about how the universe operates. Some of these questions are specific and were posed to a telescope time allocation committee to earn dedicated time on that instrument; others come out of looking at data (often from large surveys, but sometimes from more targeted observations) and seeing something warrants follow-up.

With increasingly large datasets available to astronomers, this is sometimes referred to as data-driven astronomy. In the case of archival data, scientists will not have to observe but simply download and analyze the data. On the other hand, when a specific question is asked and bespoke observations are taken, it can go one of two ways. There are queue-based instruments, where observations are taken automatically when they're scheduled and data need only be reduced (i.e., made usable) and analyzed, and there are more classical instruments, like the Keck telescopes, which require that astronomers actually oversee and run observations themselves. Sometimes this involves travelling to the telescope, but, especially after the COVID-19 pandemic, there has been a move toward remote, ``pajama mode'' observations. This mode also has the benefit of helping to lessen our field's carbon footprint by limiting truly unnecessary travel. Those data will also need to be reduced and analyzed after observations are completed.

Because the reduction and analysis of data of variable quality is so necessary to this science, there is a fair bit of creativity and scientific software development that goes into observational research. For students who are interested in such things and/or are night owls, observational astrophysics may be for you.

\subsection{Theory}

There are two types of theorists -- your classic pen-and-paper types and computational scientists who use simulations.

In general, there are fewer pen-and-paper theorists now than there were historically. In some parts of cosmology and particle astrophysics, there are people who still work this way, but most astrophysics theory is done with either numerical simulations or semi-analytic modeling, which take known physical principles and apply them together in a consistent way to make predictions about the world/universe around us. 

Computational astrophysics can accurately reproduce observations and use the tunable model parameters to offer insights into why things are the way they are or the simulations may probe scales/environments that are not accessible with current observations, making predictions that will later be substantiated (or not!) by observations.

Pen-and-paper theorists use pure math and physics, while computational astrophysicists also have to have strong backgrounds in coding/computing. For the would-be-mathematicians who are interested in astrophysics, the former may be most engaging, while would-be-computer scientists may find the work of computational astrophysics satisfies interests in both physics and computing.

\pagebreak
\strut\vspace{175pt}

\noindent \textsc{``Almost everyone suffers from imposter syndrome at some stage, so don't let it discourage you! You don't need to be an expert at everything, and working together as part of a team of people with different skill sets can accomplish a great deal. Enthusiasm, passion for the subject, and a willingness to work hard will take you a long way.\\
\\
Remember - not only the world, but the whole Universe, is your oyster!''}\\
\\
\strut\hfill \textemdash \textsc{Dr. Kim Page, UK Swift Science Data Centre}\\
\strut \hfill \footnotesize{\textsc{Swift Data Centre Scientist}}
\normalsize

\section{Exoplanets \& Planetary Science}

\begin{sectionauthor}
    Madison Brady (The University of Chicago) \\
    Prof. Malena Rice (Yale University)
\end{sectionauthor}
\vspace{20pt}

\noindent Planets that are not in our own solar system are referred to as \textit{exoplanets}.  So far, we know of thousands of exoplanets in our universe.  From our observations, we are pretty confident that most stars host exoplanets.

\subsection{Types of Exoplanets}

There are several different types of exoplanets.  We usually describe exoplanets by comparing them to planets in the solar system.  However, there are some that are unlike anything in our solar system.

Small, rocky planets (also called \textit{terrestrial}) planets are usually compared to Earth.  They are made of similar stuff to the Earth (mostly rocky with some iron), though they can be more than ten times as massive (these larger rocky planets are called \textit{super-Earths}).  Rocky planets are extremely common in the universe, and tend to form close to their host stars, where the heat from the star prevents the buildup of gases or ices that would form less dense planets.  Due to their small sizes, rocky planets are very difficult to study.  This is disappointing, as rocky planets are the most likely places to host extrasolar life.

\begin{figure}[h!]
    \centering
    \includegraphics[width=0.4\linewidth]{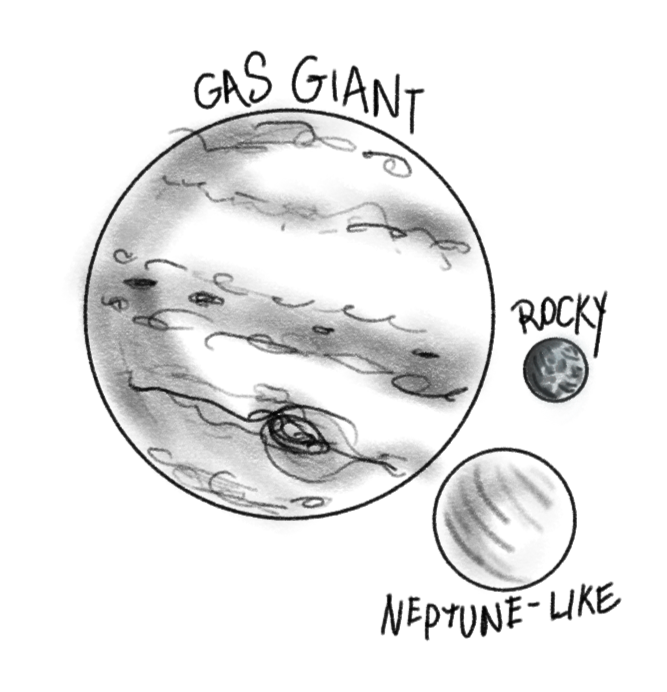}
    \caption{There are many different types of (exo)planets. We often use planets in our Solar System to help define exoplanet classification.}
    \label{fig:planets}
\end{figure}

Another common type of planet is the \textit{sub-Neptune}, named because they usually have sizes between that of Earth and Neptune.  These planets are less dense than rocky planets.  However, scientists don't know what they are made of.  They could be planets with rocky Earthlike cores and thick hydrogen/helium atmospheres, or planets with less dense ice-rich cores and much thinner atmospheres.  One hint as to their compositions comes from the well-known exoplanet radius gap- there are very few planets with radii in between sub-Neptunes and rocky terrestrial planets.  One explanation for this is that sub-Neptunes start out with thick atmospheres and then rapidly lose them (through either their own internal heat or that of the host star) to form rocky planets.  Another explanation is that planets have different compositions based upon where they form.  We need to study many more planets to know which explanation is true.

Giant planets are usually compared to Jupiter.  They are made of mostly hydrogen and helium, and can be up to ten times more massive than Jupiter. They are usually found around stars comparable to the sun or larger, as very low-mass stars frequently don't have enough mass around them to form large planets.  \textit{Hot Jupiters} are interesting because they are unlike anything in our solar system.  They are massive gas planets with orbital periods (or ``years") on the order of ten days or less.  These extreme planets likely formed far from the host star and migrated inwards.  Because hot Jupiters have large observable signatures and are relatively easy to find, hundreds are known in spite of their relatively low occurrence rate (around 1\% of sunlike stars).

\subsection{Observing Exoplanets}

Planets are very small and dim relative to stars.  As an example, the sun is about a hundred times larger in radius than the Earth and ten times larger than Jupiter.  This can make observing exoplanets directly very challenging.  A few common detection techniques are described in the sections below.

\subsubsection{Transits}
Exoplanets can eclipse their host stars, just as how our moon can sometimes eclipse the sun.  We can thus discover planets by studying the brightnesses of stars and looking out for dimming events.  The amount of light blocked by the planet is directly related to its size relative to the host star, so a transit event gives us a measurement of the planet’s size, which can be used to figure out what type of planet it is.

However, not every system has an orbit that causes eclipses, as the planet’s orbits need to line up so that the Earth, the planet, and the star fall on a straight line, which is a rare occurrence.  The further out the planet is from its star, the less likely it is to eclipse its host, so this technique tends to mostly discover planets on very short-period orbits, frequently with periods on the order of one to ten days.  This is much shorter than Mercury’s orbital period, which is 88 days.  Thus, it is difficult to look for planets like those in the solar system with this method.

It is (relatively) easy to design surveys to look for transiting planets with wide-field cameras that monitor many stars at once.  The \textit{Kepler} and \textit{TESS} missions are satellites that have discovered hundreds-to-thousands of exoplanets by studying large regions of the sky.  Transits are thus the most common exoplanet detection method.

With the launch of the James Webb Space Telescope (\textit{JWST}), we can also use exoplanet transits to study their atmospheres.  Molecules in an exoplanet’s atmosphere (such as water, methane, and carbon dioxide) can block different colors of light, so by studying the spectrum of light that passes through an exoplanet’s atmosphere as it transits its host star, we can learn about its contents.  We can use this information to learn about where planets form, how they evolve, and whether or not they are habitable.

\subsubsection{Radial Velocity}

One of the earliest methods of detecting exoplanets was the radial velocity method.  As a planet orbits around its host star, it will pull on the star with its own gravity, causing the star to complete its own (much smaller) orbit.  In general, all planets have some gravitational influence on their star (even the Earth causes the Sun to move), and the strength of this influence will be based both upon the mass ratio of the planet to the star and the distance between the planet and the star.  We can measure the motion of the star directly using the \textit{Doppler effect}.

\begin{figure}[h!]
    \centering
    \includegraphics[width=0.5\linewidth]{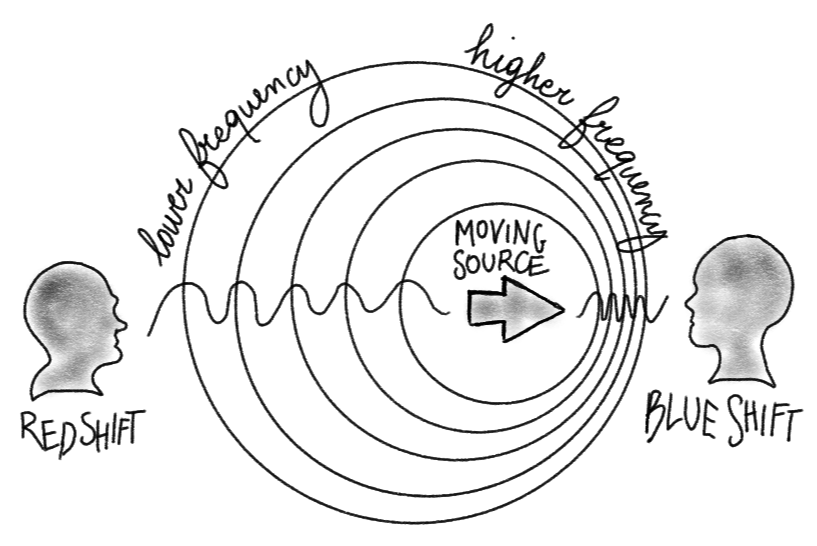}
    \caption{The Doppler effect caused by a moving source with respect to two observers, one \textit{in front} of the source, along the direction of motion, and one \textit{behind it}.}
    \label{fig:Doppler}
\end{figure}

The Doppler effect describes how waves change in frequency as the source moves.  As an example, consider the sound of a car driving past you and honking its horn.  As it is moving towards you, its horn sounds higher-pitched, which happens because the sound waves are compressed.  Meanwhile, after it passes you, the sound waves from the horn are stretched out, making it sound lower-pitched.  Something similar happens with light, which also has wave properties.  As something bright moves towards you, it will appear bluer, while it appears redder as it moves away. These differences in color are usually not something you can spot with the naked eye, but can be easily spotted by our telescopes.  

We can thus use the Doppler effect to study the motions of stars by looking at how they seem to change in color with time, looking bluer when they are moving towards us and redder when they are moving away.  With this information, we can learn about the mass and orbital period of any orbiting planets.  Understanding both the mass and radius of a planet is crucial when trying to figure out what it's made of, so planets are frequently studied with both the radial velocity and transit method (if possible).

While the radial velocity method allows us to study planets that don't transit their host stars, it tends to be more time-consuming.  Measuring the Doppler effect precisely typically requires much higher-powered telescopes than transit methods, and (unlike transits) it is difficult to measure the radial velocities of many stars simultaneously.  Thus, radial velocity surveys are typically used to more precisely characterize systems that were already identified as interesting through some other means (maybe they already have a transiting planet, maybe they are bright and nearby).

\subsubsection{Direct Imaging}
We can attempt to observe an exoplanet by just pointing a telescope at it and taking a picture.  While this may seem very straightforward, it is complicated by the fact that planets can be a million to a billion times dimmer than their host stars.  We thus cannot see the planets easily, much as how it’s difficult to see a firefly when it’s flying past a spotlight. 

One solution to this issue is to use \textit{coronagraphs}, which are astronomical instruments that block out the stellar light to allow us to search for any nearby planets.  However, it is difficult to fully block out all starlight, and these instruments thus have limitations that prevent us from directly imaging very small planets or planets that are very close to their host star.  However, we have directly imaged several large, young, bright planets on long-period orbits. A popular example is the HR 8799 system.

In addition, due to the wave nature of light, the light from a star and its nearby planet can blend together, making it impossible to distinguish them unless the planet has a very long-period orbit or the system is very close.  As this is an issue due to the nature of light (known as the \textit{diffraction limit}) and not due to any sort of instrumental limitations, there is no way to solve this problem besides using a telescope with a bigger mirror.  Even a ten-meter telescope (representative of the largest current-generation instruments) cannot directly image planets on Earthlike orbits outside of the solar neighborhood.

\subsection{Orbital Architectures}
Exoplanets orbit around their host stars in configurations, or ``architectures'', that can be examined to understand what types of planets are common, as well as how planets tend to be grouped into systems. The solar system includes four inner, rocky planets and four outer, gas giant planets, each on nearly circular orbits and with all planets orbiting in roughly the same direction that the Sun spins. In extrasolar systems, planets have been found in a wide range of configurations -- with neighboring planets of very different sizes, like in the WASP-47 system, or with planets orbiting sideways/backwards, like WASP-131 b.

The full orbital architecture of a planetary system is usually only partially known: that is, some planets are easier to find than others. Larger planets tend to produce the largest observable signatures, and planets that orbit in the same plane as their companions are often easier to find than planets that have larger orbital mutual inclinations. This means that the current census of known exoplanets is biased toward those with larger planets. While we don't yet have a conclusive answer regarding how common Earth-like planets are -- particularly since this occurrence rate is heavily dependent on what we would consider to be ``Earth-like'' -- it does appear that Jupiter analogues are relatively common.

One relatively well-characterized category of exoplanets is the hot Jupiters, which include a lone giant planet orbiting close to the host star. Hot Jupiters are often found on orbits that are not well-aligned with their host stars' spin axes, and they are generally not found with nearby planetary companions -- that is, hot Jupiters tend to be relatively lonely. 

\begin{figure}[h!]
    \centering
    \includegraphics[width=0.5\linewidth]{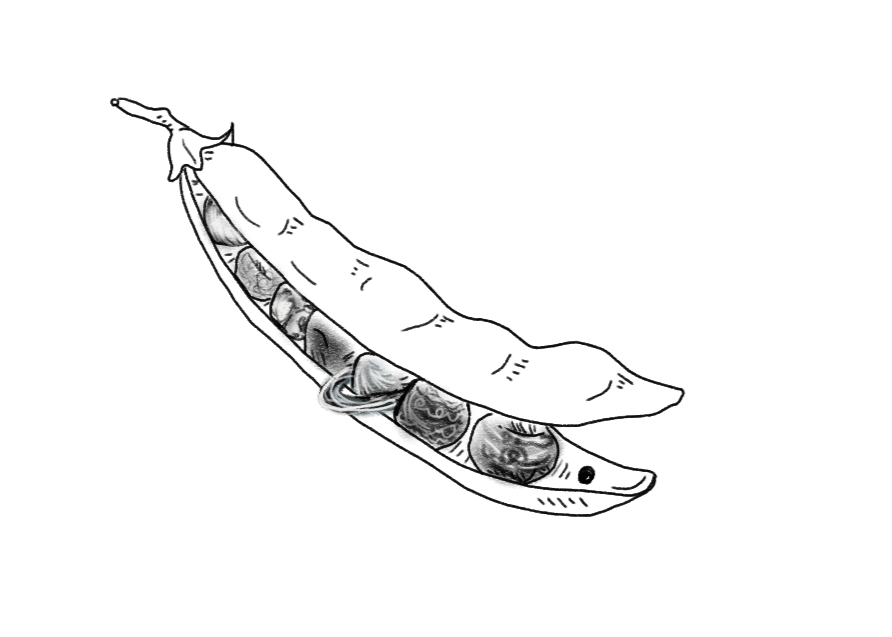}
    \caption{Planets aligned like peas in a pod.}
    \label{fig:peapod_planet}
\end{figure}

Another exciting and surprisingly common orbital architecture is the ``peas in a pod'' configuration, where multiple super-Earth or sub-Neptune sized exoplanets are found together with relatively equal spacings, masses, and radii. Peas-in-a-pod exoplanet systems somewhat resemble the inner solar system, but with much closer orbital spacings and with generally larger and more uniform planets (at least in terms of their masses and radii).

\pagebreak
\strut\vspace{150pt}

\noindent \textsc{``My advice is to follow your genuine interests and don’t be discouraged by what other people may say or think. You will follow your interests passionately, and others will see it too and help you. You may not know everything or enough about a subject, but passion will help you to learn and fill the gaps. Keep your curiosity alive, and not only in your primary field. Often, ideas come when learning about something absolutely different. And don’t be afraid to challenge yourself. It may feel scary and very uncomfortable, yet this is the only way to learn and grow. You’ll get used to the feeling of being outside your comfort zone, but only with regular practice.''}\\
\\
\strut\hfill \textemdash \textsc{Prof. Irina Zhuravleva, the University of Chicago}\\
\strut \hfill \footnotesize{\textsc{Clare Boothe Luce Assistant Professor of Astronomy \& Astrophysics}}
\normalsize

\section{Stars \& Stellar Astrophysics}

\begin{sectionauthor}
   Dr. Candice Stauffer (Northwestern University PhD) 
\end{sectionauthor}
\vspace{20pt}

Stars are born, they change, grow and evolve over time, and perhaps most surprisingly to a lot of us -- stars die.
However, for some at least, death is not the end but rather a new chapter in the great book of stellar astrophysics. 
Let me take you back to the very start.

\subsection{Stellar Evolution}
 
\subsubsection{Birth}

Stars are born in a complex process that begins deep in the cosmos from vast, cold clouds of gas and dust called nebulae. Over time, something very interesting begins to happen in these nebulae. Due to the gravitational pull, the particles of these clouds begin to clump together, forming \textbf{giant molecular clouds}. As they gather, they create denser regions. As the density in those regions increases, the gravity becomes stronger, pulling even more gas and dust, making the clumps larger. Typical giant molecular clouds are roughly \(9.5 \times 10^{14}\) km across, a distance so vast that light, the fastest thing in the universe, would take a century to traverse it. In terms of mass, these clouds typically contain about \(10^4\) to \(10^6\) times the mass of our Sun\footnote{Mass is typically compared to the mass of the Sun: $1.0\,\mathrm{M_{\odot}}$ ($2.0\times 10^{30}\;\mathrm{kg}$) means 1 solar mass.}.

As the very core of these clumps grow even hotter and denser, the star formation process begins. The core of the giant molecular cloud begins to collapse and as this occurs, it breaks into smaller and smaller pieces. In each of these fragments, the collapsing gas releases gravitational potential energy as heat. As its temperature and pressure increase, a fragment condenses into a rotating ball of superhot gas known as a \textbf{protostar}. At this stage, the protostar is a hot, dense ball of gas and dust, glowing with heat but not yet shining with the steady light of a mature star. It's like a star in the making, going through its early developmental stages.

As the protostar matures, its core reaches a critical juncture of density and temperature — a pivotal moment in the birth of a star. During this transformative phase, the protostar continues to draw in mass from the surrounding nebula, growing larger and more massive. With this growth, the pressure and temperature within its core escalate steadily, setting the stage for a stellar spectacle.

This period is crucial in a star's life cycle. If the protostar acquires sufficient mass, the core's temperature will eventually soar to such extremes that it ignites a remarkable process: hydrogen atoms, under intense heat and pressure, start to merge, forming helium. This process, known as \textbf{nuclear fusion}, is akin to striking a match on a vast cosmic scale, sparking a brilliant and sustained glow. Nuclear fusion unleashes an immense surge of energy, which is the secret behind a star's radiance.

\begin{figure}[h!]
    \centering
    \includegraphics[width=0.4\linewidth]{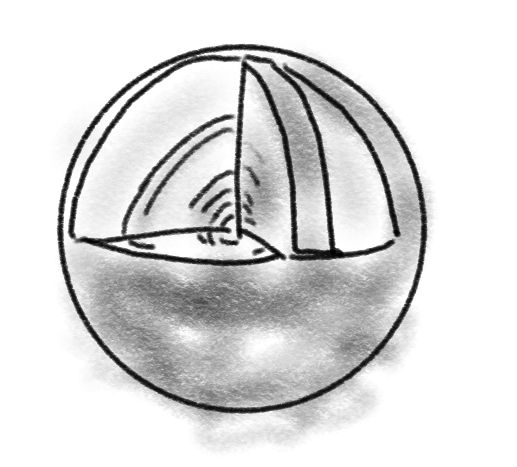}
    \caption{Stars are structured in radial layers defined by a gradient in temperature and density. For post-main sequence stars particularly, this also manifests as a chemical gradient, with the heavier elements in the star's fusing core and the lightest elements in the star's outer layers.}
    \label{fig:star}
\end{figure}

Imagine lighting a colossal furnace that can blaze for millions, even billions of years. That's what happens when a star is born. This fusion process not only lights up the star but also marks the commencement of its extensive journey, a journey spent predominantly in converting hydrogen into helium, fueling its luminous presence in the universe.

This is just the beginning of a star's life. Depending on its size, a star can live for a few million years if it's really big, or for tens of billions of years if it's more like our Sun. Over its lifetime, a star will undergo various changes, eventually leading to its death, which is another remarkable story in itself.

\subsubsection{Life}
Stars are classified into different types based on their temperature, color, and size. A common mnemonic to remember these types is `OBAFGKM,' where each letter represents a class of stars — O, B, A, F, G, K, and M.\footnote{An easy way to remember the order of the `OBAFGKM' mnemonic is with the phrase: `Only Boring Astronomers Fight Green Killer Martians'.} These classes range from the hottest, most massive O-type stars to the cooler, smaller M-type stars, essentially forming a temperature scale for stars.

In the heart of every star, a remarkable process occurs: nuclear fusion. This is the key to a star's life, akin to the burning flame in a fireplace; it's what makes them shine and prevents gravity from collapsing them into a mere point. During their main sequence phase, stars spend most of their lifetime fusing hydrogen into helium in their cores. This phase is stable and can last from a few million years for massive O-type stars to tens of billions of years for smaller M-type stars.

As stars exhaust their hydrogen fuel, they evolve off the main sequence. Their composition subtly shifts, with the core, now rich in helium, moving on to fuse this new element. Meanwhile, hydrogen fusion continues in a shell surrounding the core. For instance, a star like our Sun (a G-type star) will eventually swell into a red giant, fusing helium into heavier elements. Larger O and B-type stars, due to their massive size and higher core pressure and temperature, can fuse elements all the way up to iron.

However, there is a limit. The fusion of iron in the most massive stars marks a turning point. Iron fusion consumes more energy than it releases, effectively halting the fusion process. This is a critical juncture in a star's life. Stars like our Sun, with about a solar mass, exhaust their nuclear fuel and shed their outer layers in a planetary nebula. The remaining dense core collapses into a white dwarf.

In contrast, stars more than ten times the mass of the Sun meet a dramatic end. They explode in \textbf{supernovae}, events so powerful they briefly outshine entire galaxies. The aftermath of such an explosion leaves behind either a neutron star — an incredibly dense object composed of neutrons — or, for the most massive stars, a black hole, an entity with gravity so strong that not even light can escape it.

This stellar lifecycle is a testament to the dynamic and ever-changing nature of the universe, reflecting the delicate balance of forces and intricate processes that govern the life and death of the stars illuminating our night sky.

\subsection{Stellar Graveyards}
\subsubsection{Death}

As stars reach the end of their lives, they undergo remarkable transformations, leading to various final stages depending on their initial mass. Two notable outcomes are white dwarfs and neutron stars, each a fascinating testament to the laws of physics operating under extreme conditions.

\subsubsection{White Dwarfs}

A white dwarf represents the final evolutionary stage of medium-sized stars, like our Sun. When such stars exhaust their nuclear fuel, they shed their outer layers, leaving behind a hot, dense core. This core becomes a white dwarf. Despite being small in size, white dwarfs are incredibly dense. A white dwarf's mass is comparable to that of the Sun, but its volume is similar to that of Earth.

What keeps a white dwarf from collapsing under its own gravity is a quantum mechanical phenomenon known as \textbf{electron degeneracy pressure}. In this state, electrons are packed so tightly that the Pauli exclusion principle (which states that no two electrons can occupy the same quantum state) provides a pressure strong enough to counteract gravitational collapse.

Interestingly, not all stars have had the chance to evolve into white dwarfs. The lightest stars, with lifespans much longer than the current age of the Universe, have not yet reached this stage of their life cycle. They will continue to burn hydrogen for trillions of years, far longer than the more massive stars.

\subsubsection{Neutron Stars and Pulsars}

For stars significantly more massive than the Sun, the end of life is even more dramatic. After exploding in a supernova, the core that remains can form a neutron star. Neutron stars are among the densest objects in the universe; their mass is typically 1.4 times that of the Sun, but they are only about 20 kilometers in diameter.

In neutron stars, it is \textbf{neutron degeneracy pressure} that counteracts gravitational collapse. This pressure arises from the Pauli exclusion principle, similar to electron degeneracy pressure, but it involves neutrons. Due to their immense gravity, the protons and electrons in the star's core merge to form neutrons.

Some neutron stars are observed as pulsars. Pulsars are highly magnetized, rotating neutron stars that emit beams of electromagnetic radiation from their magnetic poles. As the pulsar rotates, these beams sweep across space, and if aligned with Earth, they appear as regular pulses of radiation, hence the name `pulsar.'

\begin{figure}[h!]
    \centering
    \includegraphics[width=\linewidth]{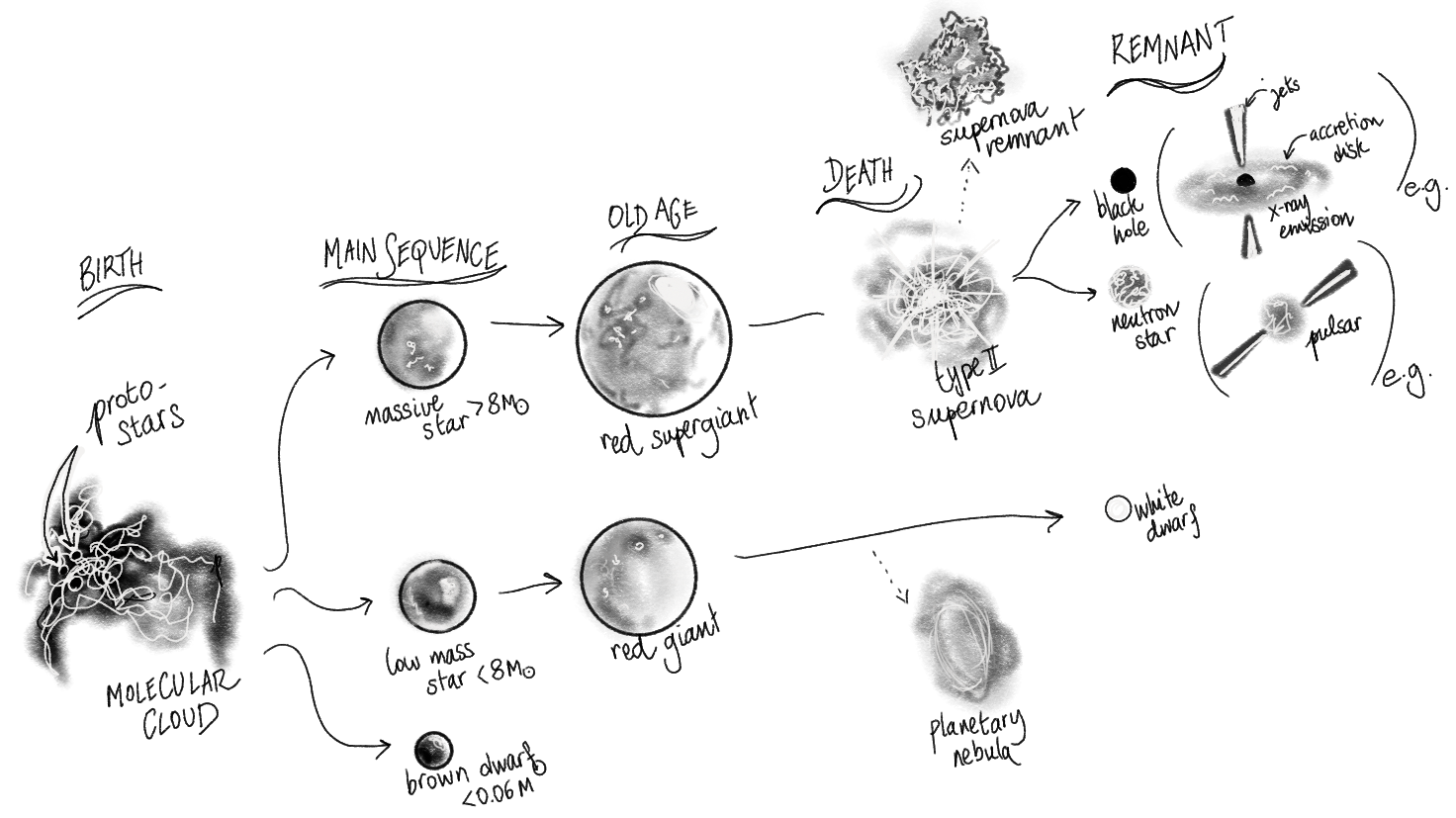}
    \caption{The different stages of stellar evolution as a function of the main sequence star's mass.}
    \label{fig:stellar_cycle}
\end{figure}

\subsection{Astrophysical Transients}

Astrophysical transients are cosmic events characterized by their changing brightness over relatively short time periods. These phenomena stem from a variety of astrophysical sources and each type of transient showcases distinct characteristics. In this section, we will delve into several types of transients. It's important to note, however, that the array of transients we discuss here represents only a portion of the myriad transient phenomena observable in our universe.

\subsubsection{Supernovae (SNe)}

Supernovae (SNe) rank as some of the universe's most powerful and luminous events, signaling the explosive and dramatic demise of stars. These cosmic explosions can occur in one of two primary ways. The first involves massive stars at the end of their life cycle. When such a star exhausts its nuclear fuel, its core collapses under the force of gravity, triggering a violent explosion. This type of supernova is known as a core-collapse supernova.

The second type, known as a Type Ia supernova, occurs in binary star systems where a white dwarf star accretes matter from its companion. Once the white dwarf reaches a critical mass, a runaway nuclear reaction ensues, leading to a spectacular explosion.

Supernovae are essential to the cosmic cycle of matter. In these stellar explosions, elements heavier than iron are synthesized and then scattered across the universe, contributing to the formation of new stars, planets, and even the building blocks of life. In terms of brightness, a supernova can briefly outshine an entire galaxy, emitting as much energy in a few months as our Sun will in its entire lifetime.

The remnants of supernovae are equally fascinating. They can leave behind dense \textbf{neutron stars} or, in the case of the most massive stars, \textbf{black holes}. The expanding shock waves from supernovae also help to shape the interstellar medium, triggering the formation of new stars.

Observations of supernovae have been pivotal in astrophysics, providing insights into stellar evolution, the interstellar medium, and cosmology. Notably, the study of distant Type Ia supernovae has been instrumental in understanding the expansion of the universe and the mysterious dark energy driving this expansion.

\subsubsection{Gamma-Ray Bursts (GRBs)}

Gamma-Ray Bursts (GRBs) stand as some of the most intense and energetic phenomena in the cosmos, observable in distant galaxies. They manifest as sudden, extremely bright flashes of gamma rays, the most energetic form of light. The luminosity of these bursts is so immense that, for their brief duration, they can become the brightest electromagnetic events in the visible universe.

GRBs are typically divided into two categories based on their duration: short-duration GRBs and long-duration GRBs. Short-duration bursts, lasting less than two seconds, are thought to originate from the merger of binary systems containing neutron stars. This collision results in a violent burst of energy and, often, the formation of a black hole.

On the other hand, long-duration GRBs, which last more than two seconds and can continue for several minutes, are generally associated with the deaths of massive stars. In these cases, the core of the star collapses, leading to a supernova or a hypernova, and subsequently, the formation of a black hole. The intense gamma rays are believed to be released as highly concentrated jets of energy and matter, moving at nearly the speed of light.

Despite their transient nature, GRBs leave behind afterglows in other wavelengths like X-ray, optical, and radio, offering astronomers valuable clues about their origins and the characteristics of the intervening space.

\subsubsection{Active Galactic Nuclei (AGN)}

Active Galactic Nuclei (AGN) are located at the centers of some galaxies. These compact regions significantly outshine the rest of their host galaxies, with their extraordinary luminosity originating from supermassive black holes at their cores.

The immense gravitational pull of these black holes draws in surrounding gas and dust, forming an accretion disk. As matter spirals into the black hole, it heats up to extremely high temperatures, emitting vast amounts of energy across the electromagnetic spectrum, from radio waves to visible light to X-rays. This process not only illuminates the AGN but also can impact the host galaxy's evolution.

AGNs exhibit a wide variety in their characteristics and are classified into different types based on their properties. For instance,\textbf{ quasars} are the most luminous and distant AGNs, visible across the universe, while \textbf{Seyfert galaxies} are closer and slightly less luminous. \textbf{Blazars} are another type of AGN known for their rapid and highly variable emissions, believed to be due to their jets being aligned with our line of sight.

One of the defining features of AGNs is their variability. They can change in brightness over a range of timescales, from days to years. This variability is a crucial tool for astronomers to understand the size and structure of the regions close to the black hole, as well as the dynamics of the accretion process.

\subsubsection{Tidal Disruption Events (TDEs)} 

Tidal Disruption Events (TDEs) are spectacular and violent occurrences in the cosmos, happening when a star strays too close to a supermassive black hole, typically situated at the center of a galaxy. The intense gravitational forces of the black hole exert different strengths of tidal forces on different parts of the star, ultimately pulling the star apart in a process often referred to as ``spaghettification,'' due to the elongation effect on the star's material.

As the star disintegrates, its remains — a stream of gas and stellar debris — spiral inward towards the black hole, forming a hot, rotating accretion disk. The friction and intense gravitational forces within this disk heat the material to extremely high temperatures, causing it to emit bright flares of energy, particularly in the ultraviolet and X-ray regions of the electromagnetic spectrum.

TDEs offer astronomers a rare opportunity to study the properties and environments of supermassive black holes. Since these black holes are typically quiescent, the sudden influx of material from a TDE can illuminate the black hole’s surroundings, providing valuable information about the black hole's mass, spin, and the nature of accretion processes. Additionally, TDEs can shed light on the population and distribution of stars around supermassive black holes.
\vspace{-5pt}
\subsubsection{Fast Radio Bursts (FRBs)}
Fast Radio Bursts (FRBs) are one of the most intriguing astrophysical phenomena discovered in recent years. These are sudden and intense bursts of radio waves originating from distant parts of the universe, each lasting only a few milliseconds. Despite their brief nature, FRBs are incredibly powerful, with the most energetic FRBs potentially releasing as much energy in a few milliseconds as the Sun does in nearly a century. The nature of FRBs is compounded by their seemingly random occurrence in the sky and the wide range of frequencies they cover.

The exact origins of FRBs remain one of the great mysteries in astronomy. Various theories have been proposed, ranging from highly magnetized neutron stars and black hole events to more exotic possibilities like cosmic strings or even \textit{alien technology} (although most astronomers agree this is the least likely possibility)!

Some FRBs have been observed to repeat, coming from the same location multiple times, while others appear as singular events. The repeaters have provided vital clues, suggesting that at least some FRBs are likely linked to highly energetic and dynamic astrophysical objects like neutron stars in extreme environments.
\vspace{-5pt}
\subsubsection{Kilonovae}
Kilonovae represent a relatively new and exciting class of cosmic events, characterized by their luminous outbursts. These spectacular phenomena are the result of the merger of two neutron stars or a neutron star with a black hole. Unlike typical supernovae, which are driven by the collapse of a massive star, kilonovae are powered by the immense gravitational energy released during these compact object mergers.

The collision of neutron stars or a neutron star with a black hole triggers a cascade of nuclear reactions, resulting in the synthesis of heavy elements like gold, platinum, and uranium. This process, known as rapid neutron capture or the r-process, is thought to be a primary source of many heavy elements in the universe. The material ejected in a kilonova explosion is rich in these newly formed elements, and as it expands and cools, it emits a distinct signature of light. This light is typically in the infrared part of the spectrum, different from the optical emissions seen in traditional supernovae.

Kilonovae were largely theoretical until the historic observation of a neutron star merger in 2017, detected both in gravitational waves (\textbf{GW170817}) and electromagnetic radiation. This event marked the first time a cosmic event was observed in both gravitational waves and light, opening a new era of multi-messenger astronomy.

\subsubsection{Novae}

Novae occur in binary star systems, where a white dwarf and a companion star orbit each other closely. They are significantly less energetic compared to the colossal explosions of supernovae, but they are still very notable events in the cosmos. A typical nova can increase in brightness by a factor of thousands to hundreds of thousands over a few days, making them visible to the naked eye and allowing astronomers to study them in detail.

The process leading to a nova begins when the \textbf{white dwarf} starts pulling in material, typically hydrogen, from a very close by neighboring star, called its \textbf{companion star}. This material accumulates on the white dwarf's surface, gradually increasing in pressure and temperature.

As the layer of accreted material grows thicker and hotter, it eventually reaches a critical point where its temperature is high enough to ignite nuclear fusion reactions. Unlike the prolonged fusion process in the core of a regular star, this fusion occurs explosively on the surface of the white dwarf in a burst of energy. The result is a thermonuclear explosion that causes the white dwarf to suddenly brighten dramatically, emitting a vast amount of light and energy. This explosive event is what we observe as a nova.

The aftermath of a nova explosion often leaves behind an expanding shell of gas, which is ejected at high speeds from the white dwarf's surface. Over time, this shell disperses into the interstellar medium, contributing to the enrichment of the galaxy with heavier elements.

\pagebreak 
\strut\vspace{200pt}

\noindent \textsc{``My advice is ... a self-affirmation phrase that I think really helps push past obstacles - \textsl{I am capable of anything and everything that I put my mind to and put sincere effort into.}\\
\\
It's a mantra that will help build a sense of self-assurance about one's capabilities.''}\\
\\
\strut\hfill \textemdash \textsc{Prof. Priyamvada Natarajan, Yale University}\\
\strut \hfill \footnotesize{\textsc{Chair, Department of Astronomy}}\\
\strut \hfill \footnotesize{\textsc{Joseph S. and Sophia S. Fruton Professor of Physics, Astronomy}}
\normalsize

\section{High Energy Astrophysics}

\begin{sectionauthor}
    Yasmeen Asali (Yale University) \\
    Michelle Gurevich (King's College London) \\
    Dr. Emily Lichko (The University of Chicago) \\
    Julie Malewicz (Georgia Institute of Technology) \\
    Samantha Pagan (Yale University) \\
    Emily Simon (The University of Chicago)
\end{sectionauthor}
\vspace{20pt}

\subsection{Astrophysical Plasmas}

While we are mainly familiar with three states of matter in our daily lives-- liquids, solids, and gasses-- there exists a fourth state of matter called a plasma. Plasmas are electrically charged gasses, meaning that the molecules of the gas have been pulled apart into positively and negatively charged components (for example: removing one negatively-charged electron from a hydrogen atom will leave behind a positively charged proton). Even though most gasses we encounter on Earth are neutral, in outer space where things are very hot and energetic, most gasses are electrically charged and are therefore called plasmas. 

It's important to understand how plasmas work in order to study highly energetic astrophysical phenomena such as solar winds, supernovae, and the disks of superheated matter that are captured by black holes (accretion disks). The key concept is that plasmas are full of electrically charged particles which are susceptible to electricity and magnetism. In the presence of a strong electric or magnetic field, particles can get huge energy boosts, have their trajectories through space reoriented, or even collide with one another and form other particles. 

On Earth, we can create plasmas in the lab to study their behaviors, but it can be challenging or impossible to inject enough energy or to make large enough magnets to match the scales that we see in space. We can also study what happens when plasmas from space interact with the Earth's atmosphere. Periodically the sun will release large bursts of energized particles (called the solar wind) towards the Earth which are deflected by the Earth's magnetic field and funneled into the North and South poles. When these particles collide with atoms in the air like hydrogen, nitrogen, or carbon dioxide, they transfer their energy and cause the atmospheric molecules to fluoresce in bright colors. This is known as the aurora borealis in the Northern hemisphere and the aurora australis in the Southern hemisphere.

\subsection{Cosmic Rays} 

The most energetic particles in the universe are called ``cosmic rays''. While their name may make them sound like a wave, in reality they are made of ions and electrons, the positively and negatively charged particles that work together to make electricity run through our homes. Because they are so energetic, even a small number of cosmic rays, each of which is the size of an atom, roughly $10^{-10}$ m, can play a huge role in the evolution of astrophysical objects like galaxies that are roughly $10^{20}$ m. 

One of the biggest questions that we still don't understand about cosmic rays is how they get to be so fast. One of the most likely theories involves supernova remnants, the remains of stars that have exploded. The cosmic rays are ejected from the explosion, and then bounce back and forth over the boundary between the inside and the outside of the supernova remnant, picking up energy each time they bounce. One could imagine an air hockey puck, hit harder and harder each time it nears the other player, until it is hit so far it flies off the table. For the cosmic rays, eventually they are moving so fast that they can't be contained near the boundary of the supernova remnant, and they shoot out into the rest of the universe.

\subsection{Astroparticle Physics}

Astroparticle physics is a relatively new branch of study that began in the early 1900s with the discovery that at higher latitudes in Earth’s atmosphere, more particles exist that knock off electrons from other particles (called ionization). Scientists famously first observed this by measuring ionization at the top of the Eiffel Tower and later at higher altitudes in the atmosphere during air balloon flights. This discovery demonstrated that high-energy particles called cosmic rays travel to Earth from the cosmos. In astroparticle physics today, scientists study cosmologically created high-energy particles of many types (e.g. neutrinos, high-energy gamma rays, cosmic rays). Studying these particles enables discoveries about fundamental particles, cosmology, astronomy, and astrophysics–––making it a truly interdisciplinary area of study. 

In particle physics, The Standard Model, colloquially called the ``Theory of Everything,'' is an encompassing theory that impressively explains and predicts the interactions of many fundamental particles like electrons and quarks (which make up protons and neutrons). Based on the Standard Model, physicists have even predicted the existence of new particles like the Higgs Boson before it was discovered experimentally. Many particle physics experiments occur at particle colliders, like the Large Hadron Collider (LHC) at CERN in Geneva, Switzerland. Here, particles are accelerated to high energies and then smashed together. Rather than using colliders, astroparticle physicists study fundamental particles that travel to Earth from various sources in the cosmos. These particles can provide information about our universe as well as be used to study the properties of fundamental particles themselves. Some cosmological sources accelerate particles to far greater energies than are currently possible with human-made instruments. For example, the LHC's highest energy protons produced at CERN have 10$^{12}$ eV, while cosmic rays containing protons have 10$^{20}$ eV. (An eV is a unit of energy equal to moving one electron through a 1 Volt potential difference, which is about that of a AA battery.)

Some of the main unanswered questions in astroparticle physics are: What are dark energy and matter? Why is there more matter than antimatter in our universe? How did the universe evolve? What are the properties of fundamental particles like neutrinos? How are cosmic rays created?

Particles studied in the field include:
\begin{itemize}
    \item High-energy gamma rays: Gamma rays are the highest-energy form of light and, therefore also have the shortest wavelength. Astronomical objects like black holes, neutron stars, and supernovae create gamma rays. 
    \item Neutrinos: Neutrinos are extremely low-mass particles that rarely interact with other types of matter. They are often described as "ghost-like" particles and often pass through the Earth without interacting. Particle astrophysics experiments studying neutrinos emitted from the Sun were essential to the Nobel Prize-winning discovery that neutrinos strangely oscillate between flavors, a property of particles, which contradicted the Standard Model of particle physics.
    \item Cosmic rays: Energetic particles made of ions, electrons, protons, and gamma rays. 
    \item Dark matter candidates: What dark matter is (e.g., a particle, wave, black holes) is still unknown and undiscovered. However, multiple theories of what dark matter could be predict that dark matter particles could be created in the Sun and the universe or exist throughout our galaxy and commonly travel through Earth. 
\end{itemize}

The scientific instruments used in astroparticle physics are diverse and varying, too. The particle of interest determines the detectors and materials best suited for an experiment or observatory. Some astroparticle searches are built in underground laboratories like Xenon1T in Laboratori Nazionali del Gran Sasso (LNGS) in Italy. LNGS is located under a mountain and covered by 1,400 m of rock. The rock above underground laboratories shields experiments from some cosmological-originating particles like low-energy cosmic rays so scientists can more sensitively study particles like high-energy neutrinos and dark matter, which commonly pass through mountains. 

IceCube, the largest neutrino observatory in the world, is built inside the ice at the South Pole in Antarctica, at depths of almost 2,500 meters. Some of IceCube's searches use the entire Earth as a shield for particles, so particles like high energy neutrinos (up to 10$^{21}$ eV in energy) that rarely interact with matter and commonly pass through the Earth can be carefully studied. In June 2023, IceCube released a first image of our galaxy, the Milky Way, from high-energy neutrinos\footnote{\href{https://icecube.wisc.edu/news/press-releases/2023/06/our-galaxy-seen-through-a-new-lens-neutrinos-detected-by-icecube/}{https://icecube.wisc.edu/news/press-releases/2023/06/our-galaxy-seen-through-a-new-lens-neutrinos-detected-by-icecube}}.

\begin{figure}[h!]
    \centering
    \includegraphics[width=0.5\linewidth]{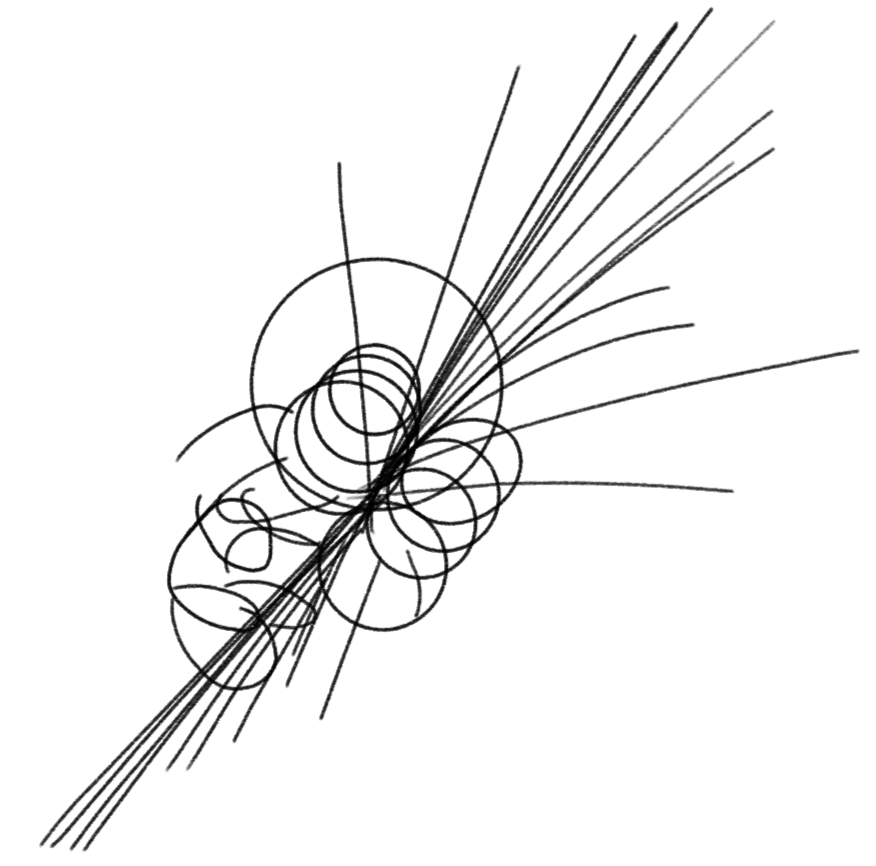}
    \caption{Particle tracks as from a bubble chamber experiment.}
    \label{fig:particle_detector}
\end{figure}

Other efforts, such as VERITAS (Very Energetic Radiation Imaging Telescope Array System), are large telescope observatories located at the surface of the Earth. VERITAS detects the high energy gamma rays (in the 10$^9$-10$^{12}$ eV energy range) to further study astronomical objects like supernovae, pulsars, and black holes, which emit these highest energy forms of light.

\subsection{Black Holes}

Black holes are fascinating objects to study. Though marvelously simple in their construction, they are beaming with some of the most exotic physical phenomena the Universe has to offer. They defy our common understanding of many fundamentals, like space, time, mass and energy (to name a few). There are many ways to go about defining exactly what a black hole is. I will try to offer you a few different ways of conceptualizing what I am talking about.

Colloquially, it is sometimes said that black holes are where all known laws of physics go to die. Though I will concede that indeed, classical physics (think Newtonian, i.e. $\vec{F}=m\vec{a}$) becomes wholly inadequate to describe what is going on there, overall, black holes are very simple. Astrophysical black holes can be condensed down to 2 fundamental properties: how heavy they are, and how fast they spin. That's it! Now, lots of weird stuff can and does happen around them, but mass and spin is all there is to it! 
Black holes do not have what we could consider to be a ``proper'' surface, but as you get closer and closer to their center, you can reach a point where their gravitational pull is so strong that not even light can escape! This forms an apparent surface called the event horizon.

\begin{figure}[h!]
    \centering
    \includegraphics[width=0.6\linewidth]{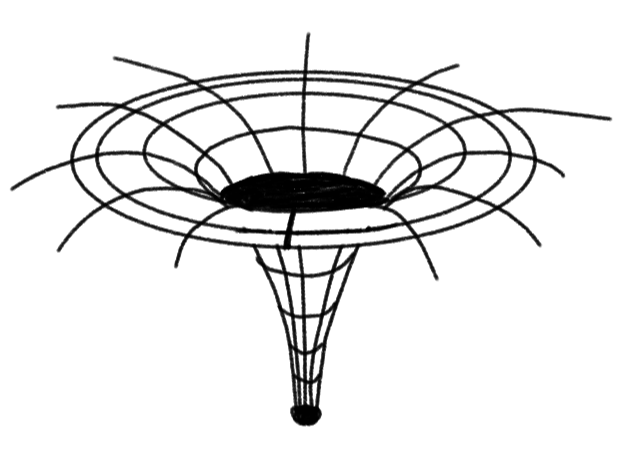}
    \caption{Black holes warp the very fabric of spacetime.}
    \label{fig:bh}
\end{figure}

Now, we know that one way to end up with a black hole is to have a star so massive that it ends up collapsing under its own weight. And we also know that the most massive objects in the Universe cannot be anything other than black holes (not counting composite structures, like galaxies, which are a collection of thousands to millions of stars). Supermassive black holes weigh somewhere in the neighborhood of millions to billions of times the mass of our Sun! But! A high mass is not a requirement to produce a black hole. Instead, what defines them is how prodigiously concentrated their matter content is. Nothing precludes us from finding a black hole that weighs as little as you: it would just be incredibly small! There is a common misconception of a black hole as some sort of cosmic vacuum cleaner, when in reality, black holes follow the exact same rules as every other thing in the Universe! Were we to replace the Sun with a similarly heavy black hole, nothing would meaningfully change for us -- but we would start to miss the warmth of sunlight pretty quickly.

\subsubsection{An Introduction to General Relativity}
The very idea of a black hole first arose in 1916 as a straightforward solution to Einstein's equations for gravity and spacetime. You might have heard of those equations referred to as the theory of General Relativity. Its foundational tenets are actually quite simple in essence.

Space and time are intertwined into a fabric aptly referred to as ``spacetime.'' In concrete terms, this simply means that everything that ever was or ever will be can be placed on a 4-dimensional grid, described by 4 coordinates: time $t$, and position $(x, y, z)$. 
Though this grid is flat by default, matter distorts spacetime and makes it curve inwards. Straight lines start bending towards the places where matter is most concentrated. The physical manifestation of this idea is what we call gravity. General relativity tells us that we humans exist on the surface of the Earth, not because it acts as a giant magnet on us, but because its mass is bending space and time around it -- and we just happen to be there to feel it! Now, luckily for us, the Earth isn't too heavy compared to other cosmic objects, so the surface forces under our feet are enough to keep us aground. However, the gravitational field of objects like black holes is such that there exists no other force in the entirety of the Universe that can counteract it! That's what we mean when we talk about \textit{gravitational collapse}.

\subsubsection{Supermassive Black Holes}

A fun fact is that most sufficiently massive galaxies are hosts to supermassive black holes, snugly located at their centers: gargantuan monsters whose masses can exceed millions to tens of billions times that of our Sun. They act as superengines, powering entire galaxies through prodigiously energetic processes. A certain fraction of them are actually so luminous that astronomers have taken to calling them AGNs, for Active Galactic Nuclei, as they have sometimes been observed to outshine entire galaxies!

Now, you heard me right, I just called a black hole \textit{luminous}. That is not a misnomer, or a mistake, or even some vague metaphor. Though you might remember from above that, indeed, black holes themselves might not allow light to escape from their surface and therefore appear completely dark, all you need to ``see'' them is to place them in gas-rich environments, where you can trigger a process called accretion. Through accretion, surrounding matter gets gravitationally locked onto tight orbits around the black hole, and is slowly stripped of its energy as it falls onto the black hole -- releasing it in the form of light! Now, matter does not plunge directly into the black hole, because of conservation of angular momentum. What happens is that gas particles usually come in with some initial tangential velocity (perpendicular to the line between the point object and black hole), which is conserved even as their radial velocity (directly towards the black hole) precipitates them towards annihilation. Those particles will form spiral trajectories through what we call an \textit{accretion disk}. Now, through friction and a number of other more complicated processes, energy is released as radiation. It begins with thermal radiation, which, for accretion disks, will usually peak in the UV, or ultraviolet. Any matter with any temperature at all emits thermal radiation, even you! Though, people are quite a bit colder than the matter surrounding a black hole, so the light we emit is much less energetic, and instead peaks in the IR, or infrared, which we cannot see with our naked eyes. Some of those UV photons are going to collide with very, \textit{very} hot electrons (with a temperature around $10^9$ $^\circ \mathrm{F}$) and get kicked into the even more energetic domain (this time, into the X-ray!). This X-ray light signal is very particular and is what we can observe with our space-based telescopes, like Chandra or (in the near future, hopefully) XRISM, STROBE-X, Athena, HEX-P, etc.

\subsection{Numerical Relativity \& Gravitational Waves}

Historically, our understanding of the universe has heavily relied on what we can learn from light (formally known as electromagnetic radiation). We can see some of this light when we look up to the skies on a dark night, but we can also observe the universe in types of light that human eyes can't detect. X-rays, microwaves, and radio waves are all forms of electromagnetic radiation just like visible light, and we can use these types of radiation to learn about the universe. In addition to light, another way we have learned about the universe is by detecting astrophysical particles like we've described in the sections above. 

In 2015, we opened an entirely new window to the universe by detecting a completely separate form of radiation -- gravitational radiation, more commonly called gravitational waves. This form of radiation was predicted by Albert Einstein in 1916, and it was finally detected 99 years later by an experiment called LIGO, the Laser Interferometer Gravitational-Wave Observatory.

So, what are gravitational waves? They are ripples in the fabric of space and time, just like the waves you see in a pond when you toss a pebble. But instead of traveling through water, the waves travel through space itself and are created by some of the most powerful events in the universe. Let's think about this analogy in a bit more detail. The surface of a pond reacts differently when we move different objects across it - bigger objects create bigger ripples. The same thing is true in space! Heavier objects will cause space to distort and bend more, meaning the gravitational waves they produce will be larger.

\subsubsection{Gravitational Wave Sources}

Every object moving in space can produce gravitational waves. However, you need a super massive object or energetic event to distort the fabric of space time enough to create a detectable gravitational wave signal. So when we detect astrophysical gravitational waves on Earth, they must be coming from cataclysmic events. The first signal we detected in 2015 came from two black holes that had been orbiting around each other for millions of years. In the final seconds before they collided and merged with each other, the two black holes were moving so quickly that they produced huge gravitational wave ripples that stretched and squeezed space itself. These ripples propagated across the universe, passing through Earth and stretching and squeezing our planet and the LIGO detectors on it. These distortions happened to us as well, so on that day everyone and everything on our planet was stretched and squeezed! However, since the colliding black holes were so far away, the gravitational wave ripples that passed through Earth were so small that the planet was only distorted by 1/1000th the size of a proton! 

It is important to note that we can study these sources by measuring them with detectors, or we can consider them in a theoretical and computational context. Numerical relativity is particularly useful for studying the merger of two astrophysical objects (such as black holes or, in more exotic scenarios, boson stars). This involves running a simulation on a powerful computer which, as the name implies, simulates the two objects on their trajectories. This is done by solving the field equations - the equations which describe the system - iteratively. One way to set up this simulation is called a 3+1 formalism, which separates the 3 spatial dimensions and iterates over the 1 time dimension. The output of a numerical relativity simulation can then be used to model real systems or to inform our understanding of as-of-yet undetected objects.

Merging black holes are not the only sources that emit detectable gravitational waves. Binary neutron star (BNS) mergers, and potentially nearby supernovae, can also produce gravitational waves within the frequency range detectable by instruments like LIGO. In addition, other astrophysical sources can create gravitational wave signals in frequencies that LIGO cannot detect. For these sources, we can turn to the Laser Interferometer Space Antenna (LISA), as discussed in the next section. LISA is designed to observe supermassive black hole binaries in different frequency bands, providing a new window to the universe's gravitational wave landscape. This range of frequencies for different GW sources is called the gravitational wave spectrum, and it is a parallel to the electromagnetic spectrum that we introduced earlier.

\begin{figure}
    \centering
    \includegraphics[width=0.8\linewidth]{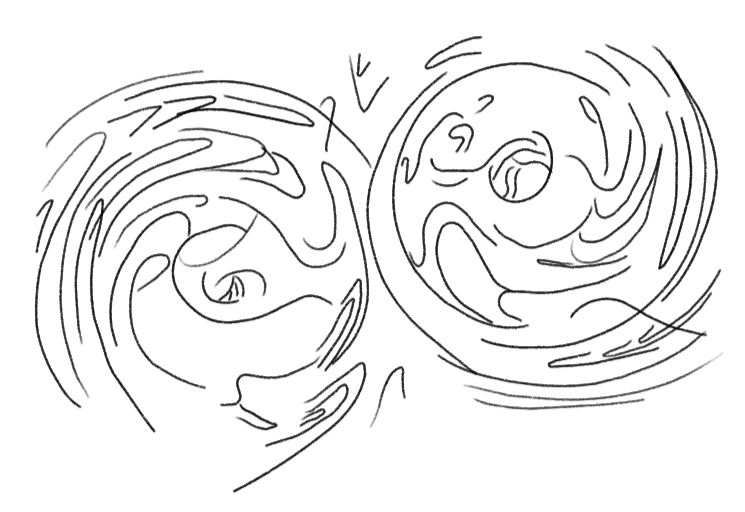}
    \caption{A binary pair of black holes -- such configurations may give rise to gravitational waves that are detectable by instruments like LIGO.}
    \label{fig:stylized_bbh}
\end{figure}

\subsubsection{GW Detectors}

For a long time, scientists were unsure whether or not gravitational waves were a truly physical phenomena. Some believed they were the result of a choice of coordinates, and others were troubled by the implication of a singularity in the relevant equations. Feynman famously argued gravitational waves were indeed physical because they would do ``work'' and thus would be expected to transmit energy. This not only convinced his colleagues it would make sense to think of gravitational waves physically, but suggested a means of detecting them.

Current gravitational wave detectors (observatories such as LIGO-Kagra-Virgo) measure the distortion of spacetime that results from a gravitational wave passing through. The distortions correspond to two types of polarization, called ``plus'' and ``cross'' which describe how the wave moves the spacetime as it passes through. If you imagine a flat circle of many points, you can think of the plus polarization as making these points oscillated up/down and left/right, where cross would produce oscillates along the diagonals. From these distortions, we are able to calculate how much energy is transferred by the gravitational wave, and infer information about the systems which produced it.

Detections are made within a range of frequencies over which the detector is sensitive, that is, the frequencies fall within ``sensitivity curves'' which differ between detectors. The primary constraint in measuring a gravitational wave using a particular detector is whether the measured signal can be isolated from the noise the data contains. This ratio of signal to noise is what determines the detector sensitivity band. You may wonder whether you can hear such a signal, and indeed you can (disclaimer: once it has been scaled to human hearing ranges). If you search online for LIGO chirp audio, you can hear the sound of a binary black hole merger!

You may know from reading about more traditional observatories that signals can be distorted by things such as the atmosphere or nearby light pollution. In the case of gravitational wave detectors, we are susceptible to many auditory distortions, such as from earthquakes or even small animals rummaging near the tunnels. In order to get better data, scientists and engineers are working on launching a space-based observatory, called LISA for ``Laser Interferometer Space Antenna''. Based on current projections, LISA should launch around 2037. By comparing signals that are measured by LIGO-Kagra-Virgo and LISA, as well as the future Einstein Telescope and Cosmic Explorer, scientists will be able to develop a better understanding of our universe.

\begin{figure}
    \centering
    \includegraphics[width=1\linewidth]{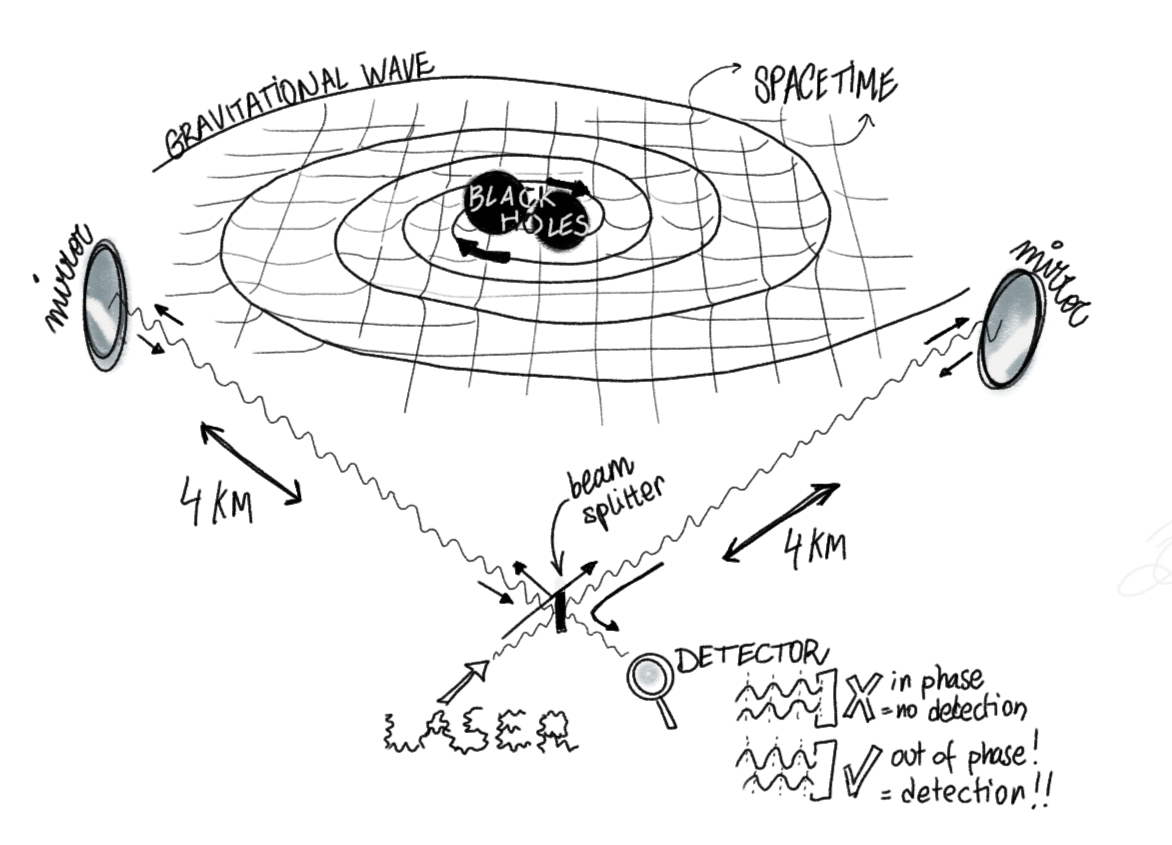}
    \caption{Schematic of the LIGO-Kagra-Virgo gravitational wave interferometer showing detection scenario from a black hole binary system.}
    \label{fig:ligo}
\end{figure}

\pagebreak
\strut\vspace{175pt}

\noindent \textsc{``As I reflect on my life, I've recognized that some of the toughest challenges for me have been (1) trusting myself, (2) walking my own path, and (3) knowing when to let go. With these in mind, I would offer guidance to those who may be searching: (i) trust your instincts (while remaining considerate of others), (ii) resist the allure of conformity, especially when things get tough, and (iii) understand that it's okay to change directions once in a while.''}\\
\\
\strut\hfill \textemdash \textsc{Prof. Hsiao-Wen Chen, the University of Chicago}\\
\strut \hfill \footnotesize{\textsc{Professor of Astronomy \& Astrophysics}}
\normalsize

\section{Galaxies}

\begin{sectionauthor}
     Mandy Chen (The University of Chicago) \\
     Ava Polzin (The University of Chicago) \\
     Zili Shen (Yale University)
\end{sectionauthor}
\vspace{20pt}

\subsection{Morphology and Classification}

Galaxies have traditionally been classified based on their morphology (their shape) and size. There are three major types of galaxies (by shape) -- spirals (characterized by their disk-like shape and clear substructure), ellipticals (generally more diffuse with little-to-no substructure), and irregulars (which are a catch-all class that incorporates low mass galaxies, galaxies actively undergoing mergers, etc.). Galaxies can also have various morphological features, like bars or rings, that can play into the characterization.

There is also a mass/size-based distinction. Dwarf galaxies have stellar masses that are (roughly) $10^4 - 10^9 \times$ the mass of the sun (ultra-faint dwarf galaxies -- named for their incredibly low luminosities -- are generally in the $10^4 - 10^5$ solar mass range, while galaxies that are millions to billions of times the mass of the sun are often referred to as ``classical'' dwarf galaxies). Ultra-diffuse galaxies are characterized by their dwarf-like masses and physical sizes that are more consistent with more massive galaxies, making them more diffuse (as the name suggests). Massive galaxies generically refer to galaxies with stellar masses $\gtrsim 10^{10}$ solar masses. L$^*$ galaxies refer to galaxies that exist at the knee of the Schechter luminosity function, which shows the number density of galaxies as a function of their intrinsic brightness, and are approximately the same mass (and so approximately the same brightness) as the Milky Way galaxy. Though L$^*$ galaxies are often considered ``typical'' galaxies, they actually represent an expansive, heterogeneous population, with a variety of physical and chemical characteristics.

Because of the finite speed of light, when we look at very distant (high redshift) galaxies, we are looking at galaxies in earlier stages of their evolution. It is common to assume that the galaxies observed at high redshift are analogs for the progenitors of modern day galaxies, so this is an avenue for us to understand how galaxies grow and change with cosmic time.

There's a strong connection between a galaxy's size and its intrinsic brightness. More massive galaxies generally have more stars, making them more luminous. The type of stellar population also makes a difference -- younger stellar populations with massive and luminous (but very short-lived) stars will be brighter than older stellar populations in which these very bright young stars have died off. This connection is one of the ways astronomers use the light observed from galaxies to infer their masses -- if you know the distance, and so the intrinsic brightness, and can make some assumptions about the population of stars, you can estimate the galaxy's stellar mass. Of course this isn't the only mass component in galaxies. They also include gas (which is not optically luminous, but can emit at other wavelengths when ionized) and dark matter, discussed more in the rest of this section. This also means that low mass galaxies (whether local dwarf galaxies or high redshift galaxies that are still forming) are exceptionally difficult to observe due to their intrinsically low luminosities.

\begin{figure}
    \centering
    \includegraphics[width=0.8\linewidth]{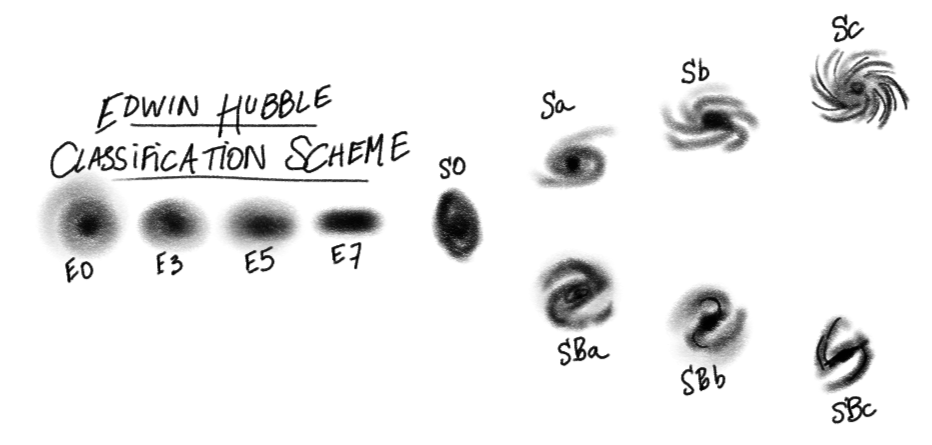}
    \caption{Edwin Hubble's original galaxy classification scheme, sometimes called the ``Hubble tuning fork''. While not all galaxies are well-described by the \textit{elliptical}/\textit{spiral} dichotomy, and it doesn't take all galaxy properties into account and is only based on optical light, the tuning fork still forms the basis of how we describe galaxies in many cases today.}
    \label{fig:galaxyclass}
\end{figure}

\subsection{Distance Estimation}

One of the biggest challenges in discerning a galaxy's properties comes from needing to understand how far away it is to convert between observed properties and intrinsic properties. Without accurate distance estimates, we don't know how bright, how large, or how massive a galaxy is -- we can't even accurately infer anything about a galaxy's immediate environment without constraints on its distance.

Because different distance measurements are useful for objects that range from the very local -- astronomically speaking -- to the very distant, the different measurement methods are referred to as the cosmic distance ladder. The first rung is useful in our immediate neighborhood.

Very locally, when looking at stars in our own Galaxy\footnote{The Milky Way is often denoted Galaxy with a capital ``G'', while galaxy as a generic term is written in lower case.}, we can use the immensely accurate parallax distance. Parallax takes advantage of the changing viewing position of an observer (we can do this on large scales given the Earth's changing position over the course of a year -- two observations six months apart will have the largest baseline displacement) to infer distance from the changed apparent position of the measured object. You can test this yourself. Hold your hand at arm's length and point your finger -- close your non-dominant eye and observe where your finger is pointing. Now, without changing where you are pointing, switch which eye is open; you'll find that, from the perspective of this eye, you are now pointing at a different object. Now bring your finger closer to you and do the same. In which case does your pointed finger end up appearing to move more? The closer one, right? That is because distance ($d$) is proportional to the inverse of the parallax angle ($p$) -- $d \propto 1/p$. This can be worked out directly if $p$ is measured in arcseconds and $d$ is measured in parsecs. The problem is, for large enough distances, you need a sufficiently large baseline to measure any perceptible difference in $p$. This means that extragalactic distances have to be measured differently.

The first extragalactic rung  on the distance ladder is generally the relationship between the light curves of Cepheid variable stars (which show fluctuations in brightness with time as their radii and temperature change) and their intrinsic brightness. Henrietta Swan Leavitt, one of the famous female astronomers that worked at the Harvard Observatory during the early 20th century, discovered that there was a tight correlation between the period of this variability (the period is the time between two peaks or two troughs in the light curve) and the luminosity of the variable star. Since the period can be observed and measured without knowing anything about the distance, this relationship allows us to infer the distance because the luminosity, $L$, is related to the observed brightness $F$ by $L = 4\pi d^2 F$, so $d = [L/(4\pi\, F)]^{1/2}$. In effect, the Leavitt Law, as it is now known, allows Cepheid variable stars to act as ``standard candles''. At slightly greater distances, you can use Type Ia supernovae as standard candles if you're lucky enough to have observed one that is reliably linked to a host galaxy. 

``Tip of the red giant branch'' (TRGB) distance measurements use the same principle -- the intrinsic brightness of the most luminous red giant stars is well understood and can then be used to infer the distance to the galaxy. This requires that there is a \textit{resolved} image of the galaxy, where we can distinguish, and measure the brightness from, individual stars. Unfortunately, even with Hubble and JWST, this places an upper limit on the distance out to which we can reliably infer a TRGB distance. A related effect that we use to determine distance is from ``surface brightness fluctuations'' (SBF). SBF distances rely on the fact that, for a particular telescope, a stellar population will get less and less resolved at greater distances -- we can then use the pixel-to-pixel variation of the images as a proxy for how resolved the stellar population is. A less resolved galaxy will have less pixel-to-pixel variation than a fully resolved galaxy.

Each of those methods relies on imaging, but we can also use spectroscopic measurements to estimate how far away a galaxy is. Taking spectra is very expensive and very time intensive compared to imaging the sky, but ``if a picture is worth one thousand words, a spectrum is worth one thousand pictures.'' If a galaxy is sufficiently far away, so that its velocity is primarily attributable to the Hubble flow (the speed at which the universe is expanding) rather than random motions (the galaxy's peculiar velocity) or interaction with large scale structure, we can infer a distance based on the line-of-sight velocity (also called the radial velocity). Redshift, a term you have probably heard before in this context, is generally just $z = v/c$, where $v$ is the radial velocity and $c$ is the speed of light. We can measure where various spectral lines are relative to where we expect them to be, which gives us $z$. For a single assumed expansion speed of the universe, there is a distinct distance that is related to a galaxy's measured redshift. In addition to getting a distance estimate, spectra give us a tremendous amount of detailed information about the chemical and physical characteristics of the observed galaxy. Spectra are then critical for all sorts of galaxy studies. 

\vspace{-10pt}

\subsection{Star Formation}

Baryons are the ``normal'' (i.e., not dark) matter in galaxies, so when astrophysicists refer to the baryon cycle, we are talking about the way that gas forms stars, is recycled into the galactic environment, and so-on.

Cold, dense gas, when it gets cold and dense enough, coalesces to form stars. These stars can live a very long time (as with low mass stars) or a very short time (as for extremely luminous, massive stars). This gas that formed stars is then partially locked in old stars and partially recycled into the gas in and around the galaxy by the end stages of stellar evolution. When these massive young stars go supernova, the enriched material from their inner layers (replete with heavy elements due to the fusion that takes place in stellar cores) is blown away from the stellar remnant and into the galaxy's gaseous medium. Some of this gas will end up back in a new generation of stars after it has cooled and coalesced and some will end up in the galaxy's halo. Cold gas can also be accreted onto the galaxy (or brought in due to gravity) from the galaxy's surroundings. These processes make up the baryon cycle.

Because you need cold neutral gas to form stars, star formation can shut off when the galaxy's reservoir of gas becomes ionized or heated (from the presence of bright young stars, supernovae, or due to interactions with neighboring galaxies). When star formation shuts off, it's called quenching, and the exact processes that quench some galaxies are poorly understood, particularly for very low mass galaxies and those at high redshifts. A new generation of instruments (including JWST, the Vera C. Rubin Observatory, and some exciting upcoming telescopes like the Nancy Grace Roman Space Telescope) and simulations are enabling us to begin looking into these questions in more detail.

We can infer the amount of active star formation in a galaxy several ways, though perhaps most notably from the light of young stars (generally in the far ultraviolet) or from the presence of gas that has been ionized by young, massive stars (the H$\alpha$ line at 6563\AA~is typically used, though there are other tracers). In addition to understanding how much immediate star formation is occurring (in the last 5 - 100 Myr depending on the selected tracer), we can also access the star formation history of a galaxy since the light in images and spectra is made up of many generations of stars. Using spectral energy distributions (SEDs), which look at the amount of flux across wavelengths, we can better understand when star formation occurred in the galaxy and how it evolved over time -- i.e., were all of the stars formed in a single burst, was there constant star formation across a period, or does the star formation history look different than either of these scenarios? Combining multi-wavelength photometry with spectroscopy gives us a more detailed star formation history than looking at the brightness of an image in one filter or the strength of a single spectral line, since it is anchored on so many individual data points. The calibrations that underlie these methods are generally done empirically, where we have many observations (across the electromagnetic spectrum) that allow us to describe how an observed parameter is connected to the galaxy's properties, while theory can help us pinpoint the exact relationships between tracers. There are still uncertainties, though, particularly as we reach new parts of the observational parameter space, like the extremely high-redshift galaxies being observed for the first time with JWST.

\subsection{Star Clusters \& Stellar Streams}
Most stars in a galaxy are distributed smoothly but not uniformly: the stars are denser in the middle of the galaxy and less dense towards the outskirts. Multiple generations of stars form and disperse in the galaxy, making it hard to disentangle their origins. However, some stars form together and stay together in a cluster because they are gravitationally bound to each other. Star clusters not only look spectacular but also provide key tools for studying galaxies.

We observe two types of star clusters in the present-day universe: open clusters and globular clusters. Pleiades (colloquially known as the Seven Sisters) is an open cluster. These clusters are young, but we know that going forward, they will likely disperse in several million years. Globular clusters, on the other hand, are ancient. They can wander through the host galaxy for ten billion years (except in extreme situations, see stellar streams). They can survive so long because they are dense: millions of stars crowd together inside several parsecs. The progenitors of globular clusters cannot be open clusters because of the age mismatch. JWST has observed “proto-globular-clusters” in high-redshift galaxies, which is the first time that we can study globular cluster formation with observations.

Star clusters and dwarf galaxies can stretch into stellar streams when they fall into a much more massive galaxy. When the external gravitational force from the colliding galaxies exceeds the gravitational pull of the stars inside the cluster, the cluster starts to dissolve into a ribbon-like river of stars called a stellar stream. The stars stay in the stream not due to any physical forces but due to their coherent motion. The motion of these stars is governed by the combined gravity of the massive galaxy and its progenitor, and any close encounters with massive objects can send stars flying off and create a gap in the stream. Thinner streams are more sensitive to these perturbations, making them the ideal tracer of the galactic potential and any substructures within.

\subsection{Dark Matter}
Even though we are most familiar with things composed of atoms and molecules, this type of matter (baryonic matter) only accounts for 15\% of all matter in the Universe. The other 85\% is what we call dark matter. Dark matter doesn’t emit or reflect light, but it does exert gravity on everything else in the Universe. That’s a strange theory about our Universe, you may say. How do we know there exists something that we can’t ever see?

For that, we must thank Vera Rubin for her work on galaxy rotation curves. She was the first female astronomer to observe at the Palomar Observatory in 1965, which had only allowed men until then. She discovered that the outer parts of galaxies rotate as quickly as components closer to the center. That implies that the enclosed mass in the galaxy keeps growing towards the outside, even beyond where stars and gas are observed. The missing mass then became key evidence for the existence of a ``dark matter halo'' in every galaxy. The dark matter would account for over 95\% of the total mass in a galaxy, with only 5\% in stars and gas. Dark matter is a key ingredient in modern galaxy formation theory. Dark matter halos are the sites where galaxies form: their gravity attracts gas which then collapses and forms stars. 

As important as dark matter is in astrophysics, we still don’t have conclusive evidence of what it is. The most widely accepted candidate is Cold Dark Matter, which hypothesizes that dark matter is a particle that moves slowly compared to the speed of light, and that it does not interact with electromagnetic radiation (light). This is the simplest explanation for all the observational evidence, but this particle hasn’t been directly detected despite our best efforts. This leaves room for other types of dark matter, and modified gravity, which are alternative theories to explain the missing mass in galaxies. These theories all make different predictions about properties of galaxies, so astrophysical observations can distinguish between these models in the future.

One example of this is the recent discovery of galaxies that lack dark matter. This is somewhat surprising given the standard theory that galaxies form inside dark matter halos. Thus, these “dark-matter-deficient galaxies” have been extensively observed. There are some astrophysical routes to remove dark matter from existing galaxies (such as tidal stripping, see paragraph on stellar streams), and there are also ways to form exotic galaxies without dark matter (such as a high-speed collision of progenitor galaxies). The recent discoveries support a dark matter particle instead of modified gravity.

\subsection{Gaseous Environments \& Turbulence}
Observations of galaxies, particularly imaging data such as the beautiful pictures from state-of-the-art telescopes (e.g., {\it Hubble Space Telescope} and {\it James Webb Space Telescope}), are dominated by the continuum radiation from stars. While our eyes are naturally drawn to these bright regions of a galaxy, the dominant component that makes up the majority of a galaxy's baryonic mass actually exists in a tenuous gas phase. This low-density gas (with a density as low as $\sim0.0001$ particles per cm$^3$) can extend much beyond the visible stellar disks/bulges and into the circumgalactic and intergalactic space. Generally, people call this diffuse gaseous structure circumgalactic medium (CGM) if it's gravitationally bound to a galaxy or intergalactic medium (IGM) if it's not clear which galaxy the gas belongs to, although in practice the definitions are less well-defined and the boundary between the CGM and IGM is not always clear. 

The IGM and CGM play a critical role in driving galaxy formation, maturation, and eventual quiescence. Star formation consumes the gas inside a galaxy and if left without replenishment, star formation cannot continue and a galaxy will stop from growing further, which is in contradiction to observations that show continuous growth via star formation in many galaxies. Therefore, a key component in modern galaxy evolution models is the presence of gas accretion from the IGM and CGM that serves as the fuel to sustain star formation.  At the same time, activities due to star formation and active galactic nuclei (i.e., active supermassive black holes at the center of galaxies) can inject a large amount of energy and materials back into the CGM and IGM through energetic processes such as galactic outflows. Through depositing energy and materials outside of the star-forming regions, galaxies exhibit suppressed star-formation rates.  Gas accretions/infalls are commonly called {\it feeding} while processes like outflows are called {\it feedback}. As feeding and feedback have the opposite effects on galaxy growth, the intricate interplay between these two processes dictates the cosmic star formation history. Therefore, it is of great interest for us to understand the properties of the CGM and IGM, and in general the gaseous environments both inside and surrounding galaxies, in order to improve our theories of galaxy evolution.

\begin{figure}
    \centering
    \includegraphics[width=0.75\linewidth]{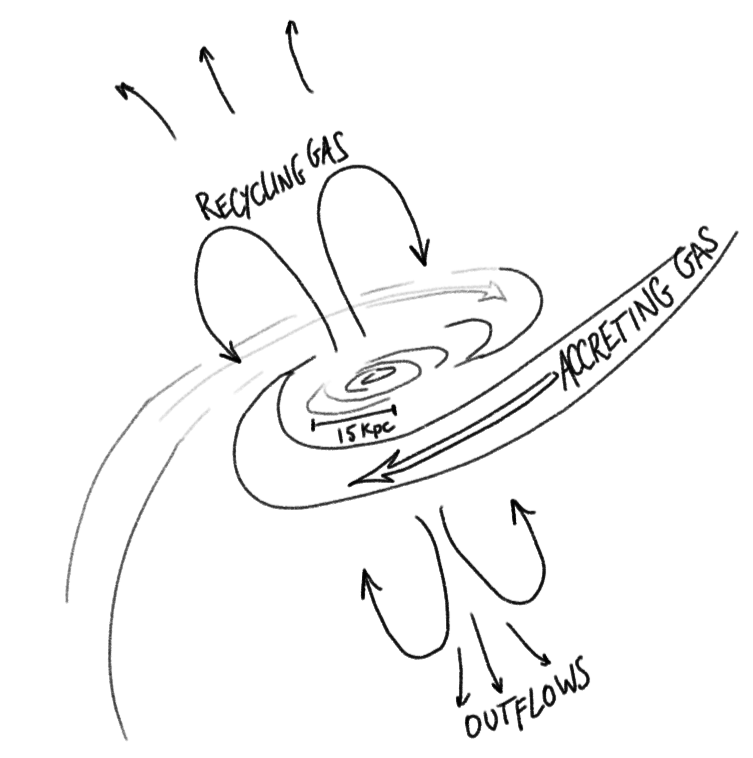}
    \caption{The in/outflow of gas on galaxy scales involves a number of processes internal to the galaxy and is informed by the galaxy's environment.}
    \label{fig:galaxy_outflow}
\end{figure}

As mentioned above, the gas density of CGM and IGM is typically extremely low.  As a result, for the past few decades, observations of this tenuous gas have largely relied on a technique called {\it absorption spectroscopy}. This technique uses the light emitted by a bright background source at large distances (most commonly a supermassive black hole that's actively accreting mass from its surroundings and emitting an enormous amount of energy in radiation) and looks for absorption features in the spectrum of this background source.  The absorption is caused by intervening low-density gas between us, the observer, and the background light source. The shapes and intensity of the absorption features can tell us about many physical properties of the gas, such as density, temperature, kinematics, metallicity, and ionization state. A complementary method of observing the CGM and IGM is to directly detect the light emitted by the gas. However, because of its tenuous nature, such emission is extremely faint and has been mostly impossible to detect until the recent advent of high-throughput integral field spectrographs on large ground-based telescopes, such as the Multi Unit Spectroscopic Explorer (MUSE) on the Very Large Telescopes (VLT) and the Keck Cosmic Web Imager  (KCWI) on the Keck Telescopes. These integral field spectrographs have unrivaled sensitivity to faint emission signals and their data have recently constituted the new frontier for CGM and IGM research. With an IFU, each pixel provides a spectrum in what is called a data cube -- this allows for sensitive, spatially resolved information about the gas composition and velocity.

In addition to its low density, one particularly interesting aspect of the CGM and IGM gas is its turbulent nature.  Because the gas typically moves at a large velocity in space, we expect the gas to be turbulent.  This means that the gas motions are chaotic and have little coherency from one location to the next (in contrast to, for example, the water flowing out of a faucet). The presence of turbulence in the diffuse halo gas and the degree of such turbulence have profound implications for the thermal and dynamic properties of the CGM and IGM. Turbulent energy can be a significant source of heating to offset energy lost due to radiation.  Turbulence can also provide additional pressure support in the gas halos and prevent them from collapsing under gravity. In addition, turbulence produces density fluctuations, triggering and facilitating the formation of gas structures in different temperature phases. Turbulent mixing also provides an efficient transport mechanism for metals from star-forming regions to the CGM and IGM.

\pagebreak
\strut \vspace{150pt}

\noindent \textsc{``In my experience, the best way to combat imposter syndrome is to read more than everyone else -- read the textbook chapters before class, read as many papers as you can on your research topic. Nervousness when channeled into motivation to really know your stuff is actually productive. Ask a million questions; approach everything in this wild ride as an opportunity to learn. \\
\\
There may be really, really bad moments. They may suck a lot. You will get through them. Learn from them to become tougher, wiser, and more empathetic. We are rooting for you! Discovering new things about space is hard, it's also THE COOLEST THING EVER. Try to take a few minutes every once in a while to be in awe of our remarkable universe.''}
\\
\\
\strut\hfill \textemdash \textsc{Prof. Erica Nelson, University of Colorado, Boulder}\\
\strut \hfill \footnotesize{\textsc{Assistant Professor of Astrophysical \& Planetary Sciences}}
\normalsize

\section{Cosmology}
\begin{sectionauthor}
    Sanah Bhimani (Yale University) \\
    Ava Polzin (The University of Chicago) \\
    Dr. Luna Zagorac (Perimeter Institute)
\end{sectionauthor}

\subsection{Theory}

The role of cosmology is to answer ``big-picture" questions about the entirety of the Universe: how and when did it start, what's its shape, how has it been evolving up to today, and what are its contents? In that way it is similar to definitions of ``cosmology" in social sciences and humanities: it is another way we make sense of the Universe around us and how it came to be. Plenty of cosmologists dig into details and smaller questions or certain epochs of the Universe, but for this section we'll take a zoomed-out approach to review our best current understanding of its content, history, and beginning. 

\subsubsection{The Contents of the Universe}

Cosmological observations and experiments tell us at the Universe is expanding, and at an accelerating rate at that. The reason it's expanding has to do with its contents, which are essentially pushing the horizon of the Universe farther and farther away with time. Every single thing described in this guide is found within the Universe, but cosmologists really only divide them into a few types of ``thing:" radiation, matter, and a mysterious ``cosmological constant."

Radiation refers for the most part to light, but can also include very light particles travelling close to the speed of light (like neutrinos). Astronomers spend a lot of time thinking about radiation in the Universe, because a lot of observations are done with light. However, it turns out light is a really tiny budget of the total energy density of the Universe---only about 0.8\%. Most of that budget is the Cosmic Microwave Background (CMB)---the oldest light in the Universe---and cosmologists often don't even include it when doing calculations because it contributes so little. 

Matter is more bulky: it refers to things that have mass and aren't moving close to the speed of light, and makes up about 30\% of the energy density of the Universe. This includes things like planets, stars, galaxies, gasses and dust between them, and more. All of these are examples of ``regular" or ``baryonic" matter, meaning that it's made up of atoms and particles we're familiar with. Then there's dark matter: we know it's dark (meaning, it doesn't interact with light; ``invisible" might have been a better name) and we know that it's matter (meaning, it has mass). We also believe that it's ``cold"---which is cosmologist speak for ``slow-moving"---but we don't know the details. We also know that of the 30\% of matter that exists in the Universe, about 25\% is dark matter and only around 5\% is baryonic matter. As described in the section on galaxies, dark matter is also really important to understanding how galaxies like ours form, making it one of the big mysteries of astronomy and physics today. 

Finally, there is something called a ``cosmological constant." If this sounds uninformative, that's because we know even less about it than dark matter. It doesn't behave like radiation or matter; rather, it seems to be somehow a part of empty space. Each cubic meter of dark matter in the Universe contains the same amount of the cosmological constant, and the more the Universe expands---the more cosmological constant there is. Around 70\% of the energy density of the Universe is in the cosmological constant, which is sometimes also called dark energy or $\Lambda$ (the Greek letter Lambda). This is another huge open question of modern astrophysics. 

Despite not knowing the details of either dark matter or dark energy/$\Lambda$, the big picture view we do know allows us to describe observations of the Universe really well. This is why the ``standard" or ``concordance" model of cosmology is often called $\Lambda$CDM (referring to $\Lambda$, the cosmological constant and cold dark matter -- CDM -- as a generic description of dark matter).   

One ``component" not discussed here is not a component at all, but the shape or curvature of the Universe; nevertheless, curvature sometimes gets added into cosmological equations. Our best data says that we live in a ``flat" Universe, which essentially means that the sum total of all components add up to 100\%. If they added up to more or less than that, we would have reason to believe our Universe is curved, and its cosmology would be very different!

\subsubsection{The History of the Universe}

The picture of the Universe presented in the previous section is valid today; that is to say, $\Lambda$ wasn't always the dominant component in the Universe, and radiation wasn't always such an afterthought. As the Universe expands, its contents get diluted, like pouring water into juice. Its contents dilute at different rates and as a function of redshift $z$, or alternatively the scale factor $a = \frac{1}{1 + z}$; consequently, this dilution is called ``redshifting." Radiation redshifts the fastest, such that its density evolves as $\rho_{\rm{r}} \propto a^{-4}$; this is because there are three spatial dimensions and also the wavelength of the radiation lengthens. Matter is slower: it only redshifts in three dimensions, so $\rho_{\rm{m}} \propto a^{-3}$. The cosmological constant $\Lambda$ doesn't redshift at all; as the name suggests it remains constant, so $\rho_{\rm{\Lambda}} = \rm{const}$.

Knowing this, it's not a surprise that our Universe today is made up mostly of dark energy, some matter, and little radiation: the last two got redshifted away as the Universe evolved, and will continue to do so as it evolves forward. From now until forever, the Universe will expand faster and faster, until all the galaxies grow very distant, the stars run out of gas, and the cosmos grows cold and dark. This scenario is known as the Big Freeze or Heat Death of the Universe, and of several explanations that cosmologists have proposed over the years, our current model favors this one. Thus, the Universe will one day end (a long, long, long time from now): not with a bang, but with a whimper.

Instead of looking on towards the Universe's eventual doom, cosmologists can instead evolve our Universe backwards in time to understand its past. Looking back to higher redshifts doesn't impact the components of our Universe, but it does impact their relative contributions to the total density of the Universe. For instance, at a redshift of about 0.3 or 3.5 billion years ago, the Universe was made up of equal parts matter and dark energy, with some small contribution from radiation (but not as small as today). To find when matter and radiation were present in equal measure, we have to wind the clock even farther back:
\begin{itemize}
    \item farther back than when the first galaxies formed, when the Universe was about 400 million years old...
    \item farther back than the formation of the first stars, when the Universe was around 200 million years of age...
    \item even farther back than the cosmic microwave background, which was released when the Universe was 300 \textit{thousand} years old... 
\end{itemize}
We have to wind the clock back all the way to the Universe's 50,000$^{\rm{th}}$ birthday: the blink of an eye for a Universe that is approaching 14 billion years of age at the time you're reading this text! Still, important moments in our Universe happened even earlier: for instance, the first nuclei of helium and hydrogen formed about 3 minutes after the Universe was born, and the first neutrons, protons, and electrons---only around a microsecond after its birth.  

\begin{figure}[h!]
    \centering
    \includegraphics[width=0.6\linewidth]{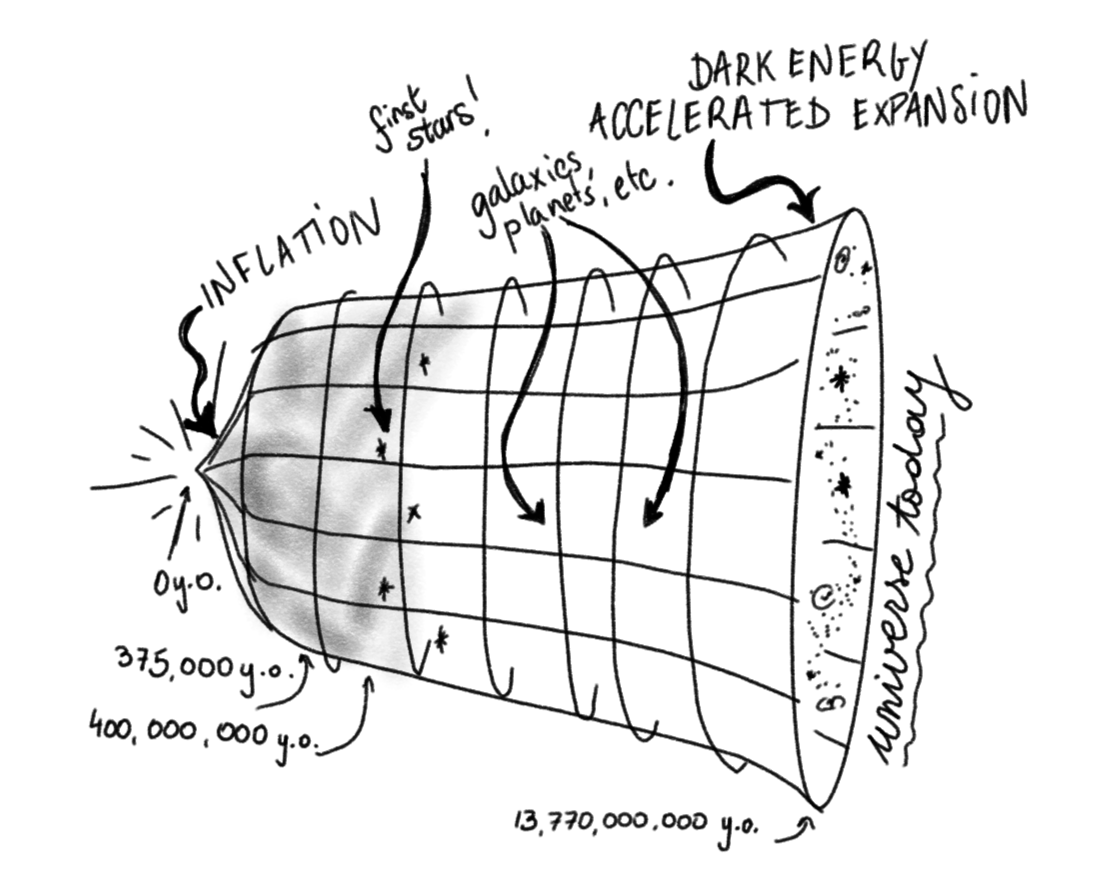}
    \caption{Starting from today and looking back in time, cosmology can help us trace the size and shape of the Universe as well as its contents, all the way back to its (still mysterious) beginnings.}
    \label{fig:cosmology}
\end{figure}

\subsubsection{The Origin of the Universe}

Now that we have an idea of what happened mere microseconds after the birth of our Universe, a natural question to ask is what came before and how it all started. If we continue turning back the clock until the Universe becomes so small it's essentially just one single point---or, in math and physics speak, it reaches a singularity---we reach the event we've come to know as the Big Bang, or the birth of our Universe. What exactly caused our Universe to start expanding and what came before are questions for which nobody has answers right now. 

Some cosmologists believe there must have been a period immediately after the Big Bang---more precisely, around $10^{-32}$ seconds after---when the Universe rapidly expanded from this singularity. This period is called cosmic inflation, and was initially proposed to solve some problems in modern cosmology, such as why our Universe is flat. Imagine slightly inflating a small balloon and drawing a circle on it: the area of the balloon inside the circle will be very curvy to start. Now, if you inflate the balloon way more and look at the same drawn circle, the area inside it will looks pretty flat. Still, other cosmologists think inflation is too weird, and don't think it happened at all. 

All this to say: while there is a general agreement between cosmologists and physicists on referring to the beginning of the Universe as the Big Bang, there is no consensus on how or why it happened, or even what happened immediately after! These are really big and fundamental questions still open for exploration: what do you think happened at the beginning of everything?

\subsection{Experiment}

Thanks to the speed of light acting as a cosmic speed limit, the farther away an object is (on cosmological distances), the farther back we are looking in time. This allows us to look at the universe as it was across different epochs in the 14 Gyr since the Big Bang, which opens doors for us to measure how the universe has evolved over time. (This is the same principle that lets us look at the properties of, say, galaxies over time, except when we discuss cosmology, we are thinking about how the physics behaves more globally.)

\subsubsection{21 cm}

One of the most ubiquitous signals in the universe comes from warm neutral atomic hydrogen, which emits at 21 cm and can be measured by radio telescopes. Because most of this warm HI is connected with large-scale structure (galaxies, galaxy clusters, filaments of the cosmic web, etc.) even at higher redshift, mapping 21 cm emission allows us to map these structures. 

The effective wavelength at which we observe any particular line varies with redshift -- $\lambda_\mathrm{obs} = \lambda_\mathrm{rest}(1 + z)$. Here $\lambda_\mathrm{obs}$ is the observed wavelength, $\lambda_\mathrm{rest}$ is the rest-frame wavelength (or the wavelength at redshift zero), and $z$ is the redshift. This means that we can map a single line (in this case $\lambda_\mathrm{rest} = 21$ cm) at different redshifts by looking at different wavelengths of radiation. At the same time, higher redshift emission is not just more distant, but also corresponds to when the universe was younger (due to the finite speed of light).

As we map the large-scale structure at different redshifts (in different wavelength slices following the equation above), we can actually track the evolution of that structure over time. This, in turn, allows us to understand something called the baryon acoustic oscillation (BAO) scale at different times in cosmic history. Since the BAO scale is a standard ruler (so is well-constrained and comparable at every redshift), we can use it to calibrate our HI mapping to place constraints on the expansion history of the universe!

\begin{figure}[h!]
    \centering
    \includegraphics[width=0.5\linewidth]{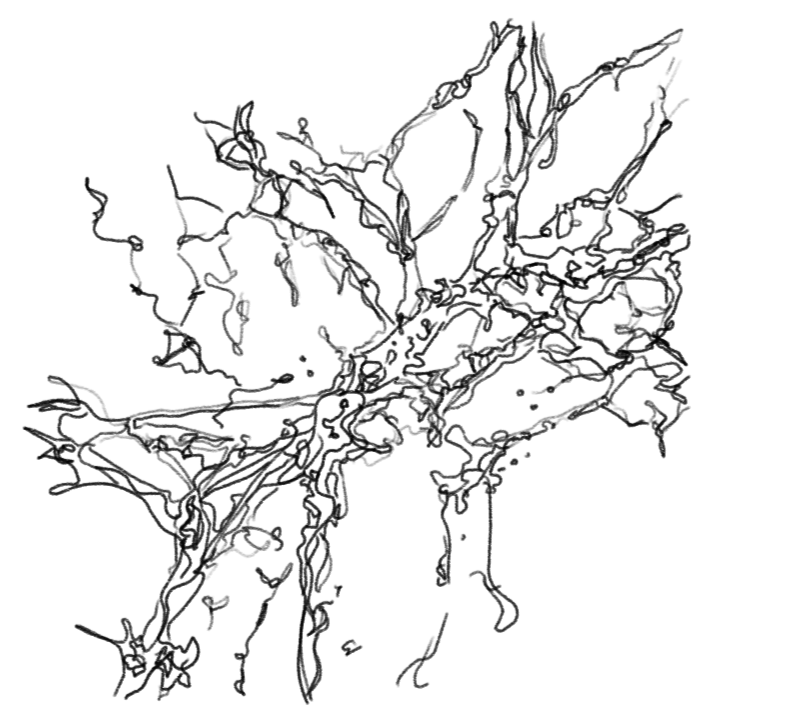}
    \caption{At the largest scales, matter is clustered into \textit{galaxy filaments}, which, as an aggregate, form what we call the \textit{cosmic web}.}
    \label{fig:cosmicweb}
\end{figure}

This expansion history is particularly interesting when dark energy begins to dominate the energy density of the universe at $z \sim 2$. For reference, the first galaxies begin assembling at $z \sim 30$ and the first stars form at $z \sim 15$. With JWST, we've been able to push toward these incredibly high redshifts, so $z \sim 2$ may not seem old, but measurements made at that redshift are of the universe as it was 10 Gyr ago. The Canadian Hydrogen Intensity Mapping Experiment (CHIME) has already detected a cosmological 21 cm signal and is rapidly moving toward being able to recover the BAO scale in their data. It works by mapping the entire Northern sky every night. There is a similar project in the Southern sky, the Hydrogen Intensity Real-time Analysis eXperiment (HIRAX). Other experiments like the Hydrogen Epoch of Reionization Array (HERA) will be able to tell us about the expansion history of the universe at higher redshifts.

\subsubsection{Cosmic Microwave Background}

The Cosmic Microwave Background is remnant radiation from the Big Bang and thus serves as a powerful tool for measuring initial conditions and evolution of the Universe. Measurements of tiny fluctuations in temperature and CMB patterns of light (polarization) have allowed us to extract the age, geometry, and makeup of the Universe, while also providing evidence for dark energy. Future measurements with new, cutting-edge CMB experiments will target their search for inflation, an event predicted to be a rapid, exponential expansion in the very early Universe. Several theories for inflation exist but none have been definitively proven. It's crucial to confirm if inflation occurred because it would provide insight into how small quantum fluctuations in the early, much hotter, and smaller Universe gave rise to the large-scale structures we observe today, like galaxies and galaxy clusters.

We look for inflation in the CMB because, if inflation occured, it would have formed gravitational waves, which would have embedded themselves as a unique signature in our polarization maps of the CMB. If measured, this signal would provide definitive evidence for inflation, eliminate certain inflationary models making it easier to describe the nature of inflation, and provide the only access to physics at Grand Unified Theory (GUT) scales which are inaccessible to particle colliders. Cutting-edge CMB experiments like the Simons Observatory, CMB-S4, BICEP Array will make measurements to detect or constrain the inflationary signal and its energy scale. 

The inflationary signal is extremely faint and this necessitates highly sensitive instrumentation. CMB experiments today are limited by detector count---sensitivity to the CMB, and therefore inflation, can only be achieved with more detectors. The Simons Observatory is currently deploying 4 telescopes with a total of $\sim$60,000 detectors---4$\times$ more than any previous experiment---to the Atacama Desert in Chile. And, CMB-S4, a next-generation experiment will deploy $\sim500,000$ detectors in the 2030s between the South Pole and Chile to measure inflation definitively. 

Increasing the number of detectors allows us to drive down the amount of noise in our data. As we do that, though, ``systematics'' (generally due to the limitations of the instruments/detectors themselves) become the limiting factor in our science results. Since we require precise maps of the CMB to do cosmology, we need to be conscious of these limitations and come up with novel approaches for the mitigation, calibration, and removal of systematic effects, which can be achieved through a combination of hardware and software.

\pagebreak
\strut \vspace{150pt}

\noindent \textsc{``Maybe the best word of advice I have is... to find the topic/field/area that you are truly passionate about -- because we will spend a lot of time and effort -- and \textsl{go all in}. It does not matter what we do, but we need to be the best version of ourselves.\\
\\
Another piece of advice is that you should not follow the advice you receive, and instead you should shape your future the way you like.''}
\\
\\
\strut\hfill \textemdash \textsc{Prof. Raffaella Margutti, University of California, Berkeley}\\
\strut \hfill \footnotesize{\textsc{Associate Professor of Astronomy, Physics}}
\normalsize

\section{Space Policy}

\begin{sectionauthor}
    Dr. Lindsay DeMarchi (U.S. Congressional Fellow) \\
    Emma Louden (Yale University)
\end{sectionauthor}
\vspace{20pt}

\noindent Imagine playing a game without any rules. It would be chaotic, right? Similarly, space policy acts as the ``rulebook'' for space activities. With more countries and even private companies sending satellites, rovers, and astronauts into space, we need to ensure that everyone plays fair and safe.

\subsection{Aerospace/Astronautics}

As humans began to dream of reaching the stars, it became clear that rules were needed. The 1967 Outer Space Treaty, signed by many countries, laid the groundwork. This treaty states some fundamental principles:

\begin{itemize}
    \item Space is for Everyone: No country can claim ownership of the Moon or any other celestial body. Space is the ``common heritage of [hu]mankind.''
    \item Peaceful Purposes: Military activities, like placing nuclear weapons in space, are prohibited. Space should be a zone of peace.
    \item Liability and Responsibility: If a country launches an object into space, they're responsible for any damage it might cause, both in space and on Earth.
\end{itemize}

The United Nations (UN) plays a significant role in setting up these rules. They have a special committee called the ``Committee on the Peaceful Uses of Outer Space'' (or COPUOS for short). Members from different countries come together in this committee to discuss and agree on the best practices for space exploration. 

While the United Nations plays a pivotal role in creating global space rules, individual countries also have their own national space policies. These policies guide their space missions, research, and commercial activities.

Moreover, it's not just countries anymore! Private companies, like SpaceX and Blue Origin, are now major players in space exploration. They, too, must follow international and national space rules.

\vspace{15pt}

\textsc{Challenges and Considerations in Space Policy}\\

\vspace{-1cm}

\strut\hrulefill

\vspace{5pt}

\textbf{Space Traffic Management.} As more countries and private entities launch satellites and spacecraft, space is becoming crowded. Just like cars on a highway, these objects can collide if they're not carefully managed. 

To prevent accidents, we need a system to track all objects in space and predict their paths. This is similar to air traffic control for airplanes but on a much grander scale. Countries and organizations are working on creating ``rules of the road'' for space to ensure that all spacecraft can coexist without crashing into each other.

\textbf{Space Debris.} Over the years, many missions have left behind broken satellites, spent rocket stages, and even lost tools. These pieces of space junk can travel at speeds of up to 28,000 kilometers per hour! At such speeds, even a small piece of debris can cause significant damage to a satellite or space station. In fact, pieces larger than 10\,cm can mean the end of a mission entirely.

There are initiatives to track larger pieces of space debris and predict their orbits. Additionally, new missions are being designed to be ``cleaner,'' either by ensuring they burn up upon re-entry or by building mechanisms to remove them from orbit at the end of their life. Some companies and agencies are even exploring technologies to actively clean up space by capturing and de-orbiting old debris.

The issue of space debris requires top innovative minds, as moving junk to higher ``graveyard" orbits can simply duplicate the problem farther from the Earth, and burning up all debris in Earth's atmosphere can pollute the upper stratosphere with an imbalance of heavy metals. 

\textbf{Planetary Protection.} When we send rovers or astronauts to other planets or moons, there's a risk that Earth microbes could hitch a ride and contaminate these celestial bodies. This could jeopardize the search for alien life by making it hard to distinguish between local and Earth-born organisms. Conversely, if astronauts or probes return from these destinations, they could potentially bring back alien microbes that might be harmful to Earth's ecosystem.

Strict sterilization procedures are in place for spacecraft visiting sensitive destinations. For example, rovers destined for Mars undergo rigorous cleaning to minimize the risk of contamination. Similarly, protocols are being developed for safe return missions to ensure that any samples brought back are securely contained and studied without risk to our environment.

\textbf{Resource Utilization.} Celestial bodies, like asteroids and the Moon, are rich in resources that could be valuable for both space missions and use on Earth. However, unchecked mining could damage these environments and lead to conflicts over who has the right to these resources.

International agreements, like the Outer Space Treaty, state that the exploration and use of outer space shall be carried out for the benefit of all countries. This means that space resources should be used responsibly and equitably. Discussions are ongoing about how to regulate space mining to ensure that it's sustainable, doesn't harm the celestial environment, and benefits humanity as a whole. Most recent in this list is The Artemis Accords, an international partnership as we explore the Moon and other areas of deep space, such as asteroids. The signatories agree to maintain transparency in our science and guide to implement the obligations stated first in the Outer Space Treaty.

\subsection{Dark Sky Conservation}

Dark sky conservation can refer to a number of realms, depending on what you consider to be the ``sky." For some, dark skies hearken to our immediate atmosphere, and the phenomenon of ``noctalgia," or watching our stars disappear from an unnatural atmospheric glow or the scattering of artificial light at night (ALAN). For others, dark sky conservation extends from the surface of Earth to the deep recesses of space, and anything that interrupts our view of nature in between. 

As you've just learned, there are few laws governing outer space and the sky. Instead, the cutting edge of the conversation preceding additional policy is a series of ``best-practices" conversations, documents, and agreements working to set the tone for how we should proceed, regulate, or even \textit{think} about the sky. These conversations are like a wet clay, and \textbf{your voice matters.} 

\textbf{The Atmosphere} You may have noticed that there are some parts of the planet where you can barely see any stars at all, and others where you can see the whole Milky Way! That is because of ALAN and the large number of unshielded, unprotected lights humans use to illuminate the dark. The great thing about this form of light pollution is that it's immediately fixable. Just use proper lighting, et voila! The stars return!

Imagine trying to fill a cup of water using a pipe jetting out of your wall. Water would spurt everywhere, and you'd be lucky to catch any of it in your cup. Instead, we tend to use faucets such that we may easily direct the water into our vessels. Light should be treated the same way! In other words, solving ALAN doesn't mean turning all the lights off; it means directing light like a faucet. The next time you're high in a plane or on a rooftop, ask yourself whether or not it's necessary for photons from the ground to be reaching your eye. Could that be repaired with a simple cap on top of a naked light bulb?

The International Dark Sky Association (recently re-branded as ``Dark Sky International") is one international organization that seeks to educate and promote more conscious uses of proper lighting. Not only does better lighting reclaim our view of the night sky, but we save money, increase safety, save energy, and help critically affected wildlife. Think about it -- animals, plants, trees -- pretty much all life on Earth developed for \textit{millions} of years in a cycle of daylight and natural darkness.

\textbf{From Satellites} One of the top 20 brightest stars in the sky is an artificial satellite. Currently, there are about 5,000 functioning satellites in orbit, but 530,000 are in various stages of planning, and that number is constantly changing! (Seriously, you should check for yourself what the current estimates are!)

Dark Sky International published their own list of five principles to preserve the quiet enjoyment of the night sky and protect the general public from the impacts of ``megaconstellations," or large swarms of satellites:
\begin{enumerate}
    \item Stewardship of the night sky is a shared responsibility that requires participation and consultation with all stakeholders.
    \item The cumulative impact on night sky brightness attributed to satellites does not exceed 10 percent above natural background levels.
    \item Maintained satellite brightness is below the threshold for detection by the unaided eye.
    \item Satellite visibility is an unusual occurrence.
    \item Launch schedules and orbital parameters are publicly available in advance.
\end{enumerate}

In addition to the general public, astronomers who take long exposures of the sky often find their data is tainted or ruined with bright satellite streaks. Though some companies, such as SpaceX, have agreed to paint their satellites darker, the fuselages themselves still block photons from reaching the Earth. This can be particularly harmful to transient astronomy and the world of cutting edge science, where we may not get a second chance to view an event.

The IAU CPS (or, the International Astronomical Union's Centre for the Protection of the Dark and Quiet Sky from Satellite Constellation Interference) is an international group of stakeholders with various ``hubs." These hubs are pillars that include Community Engagement, Policy, Industry, and Science, who all work together to compromise on the issue and suggest regulation and inform members of the public and industry. Their report and recommendations in a series ``Dark and Quiet Skies for Science and Society”  have been elevated to an agenda item in UN meetings on COPUOS. Additionally, NSF’s NOIRLab have held two conferences, summarized in proceedings called SATCON1 and SATCON2 they are invaluable touchstone documents that host recommendations for astronomers and industry alike, such that the negative impacts of satellites on astronomy can be mitigated.

\textbf{(Radio) Quiet Skies} Your Wi-Fi, your smart refrigerator, your car, and your cell phone all communicate via radio waves. In fact, so does everything else that uses satellites: banks, international trading ships, the military, and GPS! Oh, and your television broadcasts, too. The radio spectrum is broken up into wavelengths that can be registered through the ITU (International Telecommunication Union). As industry in space increases, so does the demand for radio wavelengths. Oftentimes, this overcrowding can create unintentional leakage and disrupt protected wavelengths, such as those reserved for astronomy. 

For perspective, the radio observatory called The Very Large Array (VLA) is sensitive enough to detect a flip-phone on Jupiter! Even if cell phones occupy bands adjacent to protected wavelengths and place their satellites on opposite sides of the Earth from the VLA, interference can spill over and be disruptive to science. Oftentimes, radio telescopes that observe low frequencies near the 5\,Ghz band feel their data is too noisy to warrant cleaning or further use. 

\begin{figure}[h!]
    \centering
    \includegraphics[width=\linewidth]{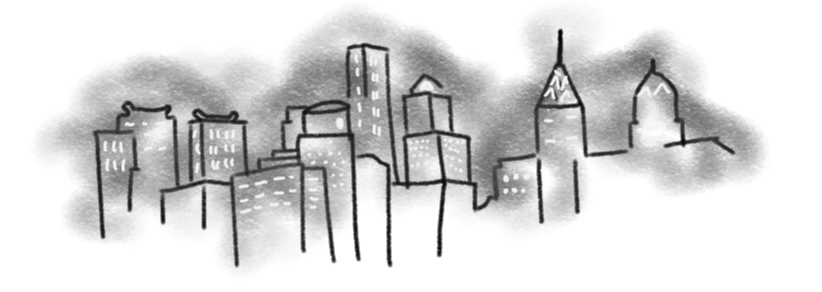}
    \caption{We can bring the Milky Way back to our cities. With proper lighting, even urban areas can see the night sky! }
    \label{fig:lightpollution}
\end{figure}

\pagebreak
\strut \vspace{150pt}

\noindent \textsc{``Recognize that the people who have accomplished amazing things have often done so with the help of mentors who have supported them in different ways; they didn't know all the answers from the start. 
\\
\\
It is important to have multiple people who you can confide in or ask questions. Sometimes your most important mentors will be peers or people who are just one year ahead of you! Lastly, once a mentor, always a mentor -- it's ok to reach out even if you don't talk every week!''}
\\
\\
\strut\hfill \textemdash \textsc{Prof. Katie Breivik, Carnegie Mellon University}\\
\strut \hfill \footnotesize{\textsc{Assistant Professor of Physics}}
\normalsize

\section{Machine Learning \& Artificial Intelligence}
\begin{sectionauthor}
    Yasmeen Asali (Yale University)
\end{sectionauthor}
\vspace{20pt}

\noindent Imagine you have a cat and a dog, and you want to create a simple rule for telling them apart. You might say, ``Cats have pointy ears, and dogs have floppy ears." This is your basic \textit{algorithm}. An algorithm just means a set of clear and specific instructions. So for this example, your algorithm might be: look at the ears, and if they're pointy, you say it's a cat; if they're floppy, you say it's a dog.

But what if you come across a cat with floppy ears or a dog with pointy ears? Your simple rule might not work very well in those cases. That's where machine learning comes to the rescue! Now, instead of relying on a simple rule, you gather lots and lots of pictures of cats and dogs. You show these pictures to a computer and say, ``This is a cat," or ``This is a dog". Machine learning refers to how your computer can learn from looking at lots of examples and get better at recognizing things.

\subsection{How do Computers Learn?}

Let's continue with our cat and dog analogy. As we feed cat and dog pictures to the computer, it will start to come up with rules on its own to decide whether the animal in a picture is a cat or a dog. Here's how it works:
\begin{enumerate}
    \item \textbf{Seeing Examples:} You feed the computer thousands of cat and dog pictures. It doesn't know anything about cats or dogs at first, but as it looks at all these pictures, it starts to notice things on its own. It might see that cats often have whiskers, and dogs have wet noses.
    \item \textbf{Learning Patterns:} The computer starts to learn from these examples. It learns to recognize patterns, like pointy ears, whiskers, and wet noses. It figures out that these are important clues to tell cats and dogs apart.
    \item \textbf{Decision-Making:} Now, when you show the computer a new picture, it looks for these clues. If it sees pointy ears and whiskers, it might confidently say, ``This is a cat." If it sees floppy ears and a wet nose, it might say, ``This is a dog."
\end{enumerate}

Importantly, the computer has come up with its set of rules on its own! It can be really hard for a human to sit down and write a set of rules that can differentiate every single type of dog from every single type of cat. With so many different breeds, the rules need to get increasingly complex to account for all the possible combinations of features. The really special thing about machine learning is that the computer can learn to recognize those patterns on its own, the same way that a baby human will learn to recognize the difference between cats and dogs without having to think through a checklist of features. 

\vspace{15pt}

\textsc{Training}\\

\vspace{-1cm}

\strut\hrulefill

\vspace{5pt}

The process of seeing examples and learning how to make decisions about them is called \textit{training}. Once a machine learning algorithm is trained, we can use it on new examples it's never seen! There are two main ways to train a ML algorithm: \textit{supervised} learning and \textit{unsupervised} learning. The basic difference between these two methods is that supervised learning uses labeled data, while unsupervised using does not require labeled data.  

\textbf{Supervised Learning}: Let's think about our cat and dog sorting example. When the computer is still learning, we are labeling the pictures as cats or dogs so it can figure out the difference between them. This is called supervised learning, because we are providing the computer with labels that inform its learning. Then, for each subsequent or new image we show the computer, we can ask it to label the name image either cat or dog. 

\textbf{Unsupervised Learning}: Now, think of sorting cat and dog pictures without any labels. You put similar images together without someone telling you which group is which. You just notice the similarities and differences. In unsupervised learning, the computer works similarly. It looks for patterns or groups in the data without being given specific labels. It's like exploring and discovering on its own. The computer might group similar pictures without being told what's in each picture, so it might not necessarily create just two groups. Maybe the computer finds patterns based on breed and it creates more than two groups separating out types of cats and types of dogs! Unsupervised learning is really useful for data that we don't already have human labels for. 

\subsection{Machine Learning in Astrophysics}

Astronomy generates an enormous amount of data from telescopes and space missions. Machine learning can help scientists sift through this data, identify patterns, and make sense of it all. For instance, ML algorithms can detect and classify celestial objects, like stars and galaxies, by analyzing the light we detect. They can also recognize transient events, such as supernovae, gravitational wave signals, or exoplanet transits, by picking out patterns in the data that may be otherwise hard to identify. Here are a few examples of the ways in which machine learning is used to aid astronomical research:

\begin{itemize}
    \item \textbf{Galaxy Classification with Galaxy Zoo}: Galaxy Zoo is a citizen science project that uses the collective power of human pattern recognition to generate large labeled training sets for a machine learning classification algorithm. Participants are usually given a set of questions about the observed galaxies. These questions might include whether the galaxy is spiral or elliptical, the presence of distinctive features like bars or spiral arms, and other characteristics relevant to astronomers studying galaxy morphology. The classifications provided by the citizen scientists are aggregated and used by astronomers to further their understanding of galaxy evolution, structure, and other properties. The large-scale classification efforts facilitated by Galaxy Zoo allow researchers to process big datasets much more quickly than would be possible with automated methods alone! 
    \item \textbf{Predicting Solar Flares with the Solar Dynamics Observatory}: Machine learning is used to study day-to-day changes on our Sun based on data from the space-based Solar Dynamics Observatory (SDO). The SDO generates a vast amount of data, including images and videos of the Sun in different wavelengths. Machine learning models can analyze historical data from the SDO to identify patterns and precursors associated with solar flares. These models can then be used for solar flare prediction, providing valuable information for space weather forecasting and mitigating potential impacts on satellite communications and power grids.
    \item \textbf{Hunting for Exoplanets}: The Kepler/K2 observatory has collected an immense amount of data on more than 150,000 stars over several years. Machine learning accelerates the analysis process, allowing us to discover exoplanets that might be difficult for traditional methods to detect. For instance, some exoplanet systems host multiple planets orbiting the same star. The signals from these systems can overlap and become intricate, making them challenging for human observers to differentiate. Machine learning excels in recognizing complex patterns within data, enabling the identification of multiple-planet systems with greater efficiency and accuracy.

\end{itemize}

These examples represent just a glimpse of the many ways ML is transforming astronomy. As ML applications continue to evolve, it's increasingly clear that this powerful tool not only revolutionizes scientific research but also quietly influences various facets of our everyday lives. ML is used by social media companies to sort recommended posts, it's used in the facial recognition software that can unlock your phone, even the auto-complete suggestions in text messaging apps leverage ML. Keep an eye out for the subtle yet impactful presence of machine learning in your own life! 

\pagebreak
\strut \vspace{150pt}

\noindent \textsc{``There are several pieces of advice that I would like to give. First off: Trust yourself. Don't let other people tell you what you are or aren't good in. No one knows better than you what your qualities are. I also want to pass on one of the best pieces of advice I've ever received; \textsl{Some you win, some you lose, but if you let those get to you, you never get anywhere}. Whatever you want from life, chances are you won't get it straight away, but don't give up too easily. Try, try twice, try three times. And if route A doesn't work, try route B, and otherwise route C.''}
\\
\\
\strut\hfill \textemdash \textsc{Prof. Silvia Toonen, University of Amsterdam}\\
\strut \hfill \footnotesize{\textsc{Assistant Professor of Astrophysics}}
\normalsize

\section{FAQ}
\addtocontents{toc}{\protect\setcounter{tocdepth}{0}}

\vspace{-5pt}

\subsection{So what is the difference between an astronomer and an astrophysicist?}

There is none! (At least not anymore ... traditionally, the distinction was whether or not spectroscopy was being used in research.) Now, when someone uses the term \emph{astronomer} or \emph{astrophysicist}, they will use them interchangeably. Colloquially within the field, some people will reserve the term \emph{astronomer} for observers and use \emph{astrophysicist} as the general term regardless of focus on instrumentation/observation/theory. In truth, everyone working in the field is doing astrophysics (i.e., studying the physical mechanisms that govern the universe) and almost no one is doing astronomy in the classical sense of the term, so \emph{astrophysicist} often fits the field better.

A different perspective comes from the ``guidance'' you'll sometimes hear that if asked what you do, \textit{astronomer} will continue the conversation and \textit{astrophysicist} will end it. From experience, people hear these terms the same way, so even this is not a foolproof way to decide which term to use. 

There are also some subfields of astrophysics that like to further differentiate themselves with their own names. Cosmologists are astrophysicists who work on cosmology and, though it is less pervasive as a term, exoplaneteers work on exoplanets/planetary science. Rarely, if ever, will cosmologists introduce themselves as \emph{astrophysicists} -- generally they will say that they are a \emph{cosmologist} or a \emph{physicist}.

\subsection{What is the difference between a physicist and an astrophysicist?}

An astrophysicist is just a type of physicist, so the distinction is one of specificity. Astrophysics is a subfield of physics along with atomic/molecular/optical, condensed matter, high energy, quantum, ... so saying that someone is an astrophysicist is like saying that someone is a condensed matter physicist.

Some confusion arises from the fact that astrophysics is housed in a separate department at some institutions. This is generally due to the historic significance of astronomy (in many universities in the United States, for instance, astronomy was one of the first departments established after the college was founded) and the fact that it is a large and well-funded subfield. To some extent, this is no different than a biophysicist working in a biophysics institute, but it does mean that astronomy departments have more educational autonomy (like running their own classes), which is an advantage that is often not afforded to institutes.

\pagebreak
\strut
\vspace{-3cm}

\subsection{What kind of education/training is necessary to become an astrophysicist?}

Long-term research jobs in astrophysics almost always require a PhD. To work at a national lab or a university as a scientist (or professor in the latter case), the usual track is:
\begin{itemize}
    \item[-]$\sim4$ years for a bachelor's degree (generally in physics, potentially in astronomy/astrophysics)
    \item[-]\textit{Possibly $\sim2$ years for a master's degree -- though not required for PhD programs in the United States, it is increasingly common for PhD applicants to hold a master's}
    \item[-]$\sim 6$ years for a PhD -- in STEM, PhD programs are fully funded, meaning that you will not have to pay tuition and will actually receive a livable (though generally non-competitive compared to industry) wage for the teaching/research you do  
    \item[-] Generally 3 - 6 years in postdoctoral research positions
\end{itemize} 
After those 10 - 15 years, you will be competitive for \textit{permanent} positions like tenure-track professorships at universities or scientist roles at nationally funded facilities.

There are other positions adjacent to astronomy that require much less in the way of preparation. For instance, telescope operators are absolutely integral to the science astronomers do, and may only have a bachelor's degree (ideally in physics/astronomy). Science writing positions are generally open to those with only an undergraduate degree, but may require more of a background in journalism.

Because graduate school admissions are increasingly competitive, it is becoming more and more usual for prospective applicants to do post-baccalaureate research at a university (frequently their undergraduate alma mater) in order to get full-time research experience. These positions often last one or two years and are paid comparably to graduate research positions. Other students may choose to do a master's degree before (re-)applying for PhD programs. Depending upon the institution, these programs may or may not charge tuition and/or pay a stipend.

\subsection{Should I study physics or astrophysics?}

At the undergraduate level, the answer is generally physics, given the expectation that students entering graduate programs will have the same knowledge of advanced math, classical mechanics, statistical mechanics, electrodynamics and electromagnetism, and quantum mechanics that a physics undergraduate affords. If there is the option, you may be able to major in physics with a concentration in astronomy/astrophysics, so that your elective courses are cross-listed grad/undergrad courses in astro, or you may be able to double major in astronomy/astrophysics. Many standalone astronomy majors were established for students with an interest in the science, but no aspirations to pursue it professionally. There are exceptions, though, and you should confirm on an institution-specific level. 

Some people will also apply to astro grad programs from geology/planetary science, math, or computer science majors, but this is a non-standard path and, since graduate admissions are incredibly competitive, there is no guarantee of success in getting into a PhD program this way (or, frankly, even with a physics background).

In graduate school, deciding between a PhD in physics and astronomy/astrophysics is much more dependent on the institution and supervisor with whom you would like to work. Standard advice is that astro-specific programs are generally better resourced than the astro portion of a physics department, and you will be able to focus more heavily on research-relevant coursework, but even within a university with both departments, you may prefer the feel and requirements (for coursework, teaching, etc.) of the physics department. The other deciding factor may be that you prefer to take a physics vs. astronomy qualifying exam (or vice versa), but this is something to think about/decide much further down the line.

\subsection{How do I get involved with undergraduate research?}

There are both funded and unfunded opportunities for undergraduate research. The simplest positions to get are unfunded research assistantships, which are generally with a professor in your department. The tried-and-true approach is to cold contact professors who do research that interests you and ask if they have any opportunities for undergraduates to participate. This may be somewhat easier with faculty you know and have impressed -- i.e., you've already taken a class with them and performed very well -- but not all research groups have roles for undergraduates, so you may have to reach out to people you don't know.

These unfunded positions may become funded after you've proven yourself; there are usually also university-wide undergraduate research grants you can apply for to support work you're doing in a professor's group.

The other avenue for getting a funded position is to apply for more prestigious undergraduate research fellowships, like the NSF REU (only open to US citizens) or the DAAD RISE, which fund summer research on specific projects that are external to your undergraduate institution. These are very competitive, so you will generally need existing research experience for a successful application, though there is some priority given to students from liberal arts colleges or similar schools where research is generally de-prioritized and less available.

\subsection{How do I know what research opportunities exist at a school?}

This is a consideration at both the undergraduate and graduate level. For undergraduates, though, it is somewhat less pressing -- you are not locked in by the research you did in college and can (usually) change your focus in graduate school. (It serves to remember that astrophysics is already a sub-field, so further narrowing is generally not crucial until you begin a PhD.) 

It may be important to you to have lots of options for research direction/modality. In that case, it's best to consider this before you select your undergraduate institution. Schools with a large astro presence (generally due to having a dedicated department and/or an institute of some kind) will often have more diversity in terms of the questions being covered. At the same time, smaller institutions may offer more research opportunities to undergraduates. Certainly, liberal arts colleges and predominantly undergraduate institutions will often have substantially less breadth in available astrophysics research, but \textit{undergraduate} research positions may be more plentiful and come with more responsibility.

If you know in advance that you are going to pursue (astro)physics, check the department website for universities you are considering attending for college. They will often list research areas on which their faculty focus and may discuss specific projects on which students have worked. Some departments will also tout their commitment to undergraduate research and list fellowship opportunities or open positions. Some universities require an undergraduate thesis in order to finish a bachelor's degree. Theses generally signal a department commitment to involving college students in research early, so this can be another positive indicator.

It also does not hurt to ask -- once admitted and at the point of deciding between schools, find the contact information for your prospective department's director of undergraduate studies (or equivalent) and ask explicitly about undergraduate research opportunities. (Note that a non-response is not a reason to worry. Academics are overworked and often receive a deluge of emails -- if you don't hear back, your message was likely buried, not ignored.)

\subsection{What skills will make me more competitive for undergraduate research positions?}

At this point, coding is critical for most aspects of astrophysical research. There are a number of languages that people use -- commonly IDL, Julia, C/C++, and Python, though Python is by far the most used in astro research. Completing free beginner courses from DataCamp or Codeacademy will get you up and running with these skills, but the best way to learn is to write code for a specific application. You can also refer to something like Python for Astronomers (\href{https://prappleizer.github.io/}{prappleizer.github.io}) as a reference or look at other guidebooks/tutorials (some listed at \href{https://astroteaching.github.io/computational/}{astroteaching.github.io/computational}).

Additionally, learning version control and collaborative work with git, via GitHub, Bitbucket, or similar, will give you a leg up. Version control is critical when you are developing software or even individual scripts, and the collaborative aspects of git allow you to contribute to larger projects where you are not necessarily the primary developer, but have reason to access, and potentially edit, source code.

Learning LaTeX will also help -- both for typesetting problem sets/homework assignments (quick way to become teacher's pet!) and working on paper drafts. Instead of Word, Pages, or free/open source alternatives, almost all drafting in (astro)physics -- including for this booklet! -- is done in LaTeX . There are local installations + editors like TeXShop, but if you have LaTeX installed on your computer, you can also use any text editor (some favorites for coding and other applications are Sublime Text and VSCode) to draft and compile your text. Collaborative work is often done in Overleaf, which is a cloud-based option -- even if you're not at the point of collaborating, it can still be very useful to access your documents from multiple devices without worrying about syncing.

For anyone just starting a project, it is also really important to understand the existing literature on the topic. Your supervisor may give you papers to read to start out, but finding others that are relevant (particularly as they are released) is a great way to show interest and initiative \textit{and} learn about where the field is headed.

\vspace{-2cm}

\subsection{Where do I find research papers to read?}

Every astro-relevant paper, at least for the last decades, is indexed in NASA/ADS (\href{https://ui.adsabs.harvard.edu/}{ui.adsabs.harvard.edu}); this is a primary resource for anyone doing astrophysics research. Pre-prints are also ingested by ADS. Those can be accessed directly (as they come out) on arXiv (\href{https://arxiv.org/}{arxiv.org}); pre-prints are released $\sim 9$pm ET (Sunday through Thursday) -- to see what's new, you can click on ``new'' next to astro-ph.

Some pre-prints are summarized in astrobites (\href{https://astrobites.org/}{astrobites.org}), which aims to make the literature accessible to undergraduates. Especially as you're starting out, it can be good practice to read both the paper and the astrobites summary to better digest the content.

\subsection{How do I get actively involved in my university's (astro)physics department?}

In addition to getting involved with research as an undergraduate, which is often critical to full integration in your department, attending seminars and colloquia is a fantastic way to be exposed to new science and meet your future colleagues. Even if you don't understand what is being discussed initially, the more talks you attend (and the further you get in your undergraduate study in general), the more you will take away from each talk. Eventually, you will be giving talks like that yourself.

Additionally, your department may have a chapter of the Society of Physics Students or affinity organizations like Women in Physics. These are wonderful opportunities to take more initiative and be fully involved in pre-professional activities.

\vspace{-2cm}

\subsection{Outside of class/research, what should I do to prepare myself for a career in astronomy?}

Building personal/professional connections and having a network of mentors is really important in the early stages of your career. (That's one of the goals of SIRIUS B's VERGE program!) Involving yourself in pre-professional activities like those listed above does a lot in this respect, and being visible in your department (as from going to talks) can help as well.

Reading broadly and understanding the common language of astrophysics (some of which is subfield specific) will also make for an easier transition to doing science professionally.

\subsection{What can I do now to prepare for a career in astrophysics?}

The best thing you can do to prepare for a career in astrophysics is gain solid footing in mathematics. There are also some formal programs through which you can get advanced pre-college training in astrophysics, namely the Yale Summer Program in Astrophysics (YSPA; \href{https://yspa.yale.edu}{yspa.yale.edu}) and Summer Science Program (SSP; \href{https://summerscience.org}{summerscience.org}). Some universities will also run daytime summer research programming for local high school students -- this will usually be less intensive than YSPA or SSP, but will still amount to an enormous leg up for participating students. 

\vspace{-1cm}

\subsection{I want to play with data now. How do I do that?}

The good news in astronomy is that so much of our software and data (both from observations and simulations) are free and open source by default! 

Observational data is generally stored in FITS -- Flexible Image Transport System -- files. These can be easily opened and perused with Python if you're comfortable coding or you can use SAOImageDS9 (DS9 for short; \href{https://sites.google.com/cfa.harvard.edu/saoimageds9}{sites.google.com/cfa.harvard.edu/saoimageds9}) or similar to go through a graphical user interface.

You can find data of all kinds on different survey and data access websites, for example:
\begin{itemize}
    \item Barbara A. Mikulski Archive for Space Telescopes (MAST) -- \\\href{https://mast.stsci.edu/portal/Mashup/Clients/Mast/Portal.html}{mast.stsci.edu/portal/Mashup/Clients/Mast/Portal.html}
    \item Dark Energy Spectroscopic Instrument (DESI) Legacy Imaging Survey -- \href{https://www.legacysurvey.org}{legacysurvey.org}
    \item Hyper Suprime-Cam Subaru Strategic Program -- \\\href{https://hsc-release.mtk.nao.ac.jp/doc/index.php/tools/}{hsc-release.mtk.nao.ac.jp/doc/index.php/tools}
    \item Sloan Digital Sky Survey -- \href{https://www.sdss4.org/dr17/data_access/}{sdss4.org/dr17/data\_access}
\end{itemize}

If you are comfortable coding, you have even more options -- for instance, you could play with other types of data like those from:
\begin{itemize}
    \item Arecibo Legacy Fast ALFA Survey -- \href{http://egg.astro.cornell.edu/alfalfa/data/}{egg.astro.cornell.edu/alfalfa/data}
    \item Gaia stellar information -- \href{https://gea.esac.esa.int/archive/}{gea.esac.esa.int/archive}
    \item IllustrisTNG (cosmological simulation) -- \href{https://www.illustris-project.org/data/}{illustris-project.org/data}
    \item NASA Exoplanet Archive -- \href{https://exoplanetarchive.ipac.caltech.edu}{exoplanetarchive.ipac.caltech.edu}
    \item X-ray transient light curves -- \href{https://github.com/avapolzin/X-rayLCs}{github.com/avapolzin/X-rayLCs}
\end{itemize}

You can generally find documentation by poking around these pages, but if that isn't sufficiently helpful, don't worry -- these are resources intended for professionals and are not made specifically accessible to the public.

There are also websites where you can engage with data via the site including:
\begin{itemize}
    \item JADES interactive viewer -- \href{https://jades.idies.jhu.edu}{jades.idies.jhu.edu}
    \item Legacy Viewer -- \href{https://www.legacysurvey.org/viewer}{legacysurvey.org/viewer}
    \item Magnetar Outburst Online Catalog analysis tab -- \\\href{http://magnetars.ice.csic.es/#/outbursts}{magnetars.ice.csic.es/\#/outbursts}
    \item NASA Exoplanet Archive confirmed planet plotting tool -- \\\href{https://exoplanetarchive.ipac.caltech.edu/cgi-bin/IcePlotter/nph-icePlotInit?mode=demo&set=confirmed}{exoplanetarchive.ipac.caltech.edu/cgi-bin/IcePlotter/nph-icePlotInit?mode=demo\&set=confirmed}
    
\end{itemize}

If you're looking for a more sanitized experience where you can engage in citizen science, the Zooniverse (\href{https://www.zooniverse.org}{zooniverse.org}) has a number of different projects in astronomy among other topics. This can be a really fun way to spend some time that also contributes to active science!

If you have a telescope, you can always take your own data that can be analyzed! Assuming you have the equipment, you may want to contribute to the monitoring of time-domain events (like exoplanet transits and stellar variability); you can then add to the American Association of Variable Star Observers (AAVSO; \href{https://www.aavso.org/}{aavso.org}) repositories. This will require that any data you take are properly calibrated, which necessitates some additional work, but is a terrific way for an excited amateur to get involved in observational astronomy.

\vspace{-1cm}

\subsection{How do I connect with other women in astrophysics?}

Universities often have a number of organizations that connect women in STEM and specific scientific fields. Most universities will have some version of a Society of Women in Physics, which can be a supportive discipline-specific community even for undergraduates. There are often also Women in Science groups (sometimes dominated by one field or another) that you can get involved with at that level. At the undergraduate level, the American Physical Society holds the Conference for Undergraduate Women in Physics (CUWiP), which is hosted simultaneously at a number of regional hubs throughout the United States. CUWiP generally focuses as much on pre-professional development as it does on science, so this is an opportunity to talk about your research \textit{and} learn best practices for navigating graduate applications, ignoring/mitigating imposter syndrome, engaging in self-advocacy, etc.

There are initiatives that look to aggregate lists of women in astrophysics in order to showcase their work and generate a professional network. For instance 1400 Degrees (\href{https://1400degrees.org}{1400degrees.org}) is a directory of women and gender minorities in (astro)physics. There is also If/Then (\href{https://www.ifthenshecan.org}{ifthenshecan.org}), which showcases a somewhat smaller cohort across all of STEM.

Additionally, as you progress in your career, you will find that individual mentorship -- both informal and formal -- will play a role. Most, if not all, of the women who contributed to this guide, can point to female scientists who shaped their path. These mentoring situations happen organically over time, so there isn't a framework for approaching them; just trust that you, too, will end up with scientists in your extended network who will advise and support you, and that you too will eventually be able to reach back and support the interested girls behind you.

\pagebreak
\strut \vspace{225pt}

\noindent \textsc{``The gap between us being average and being great often involves limiting beliefs that we have internalized about ourselves. They can block you in a big way.''}
\\
\\
\strut\hfill \textemdash \textsc{Dr. Sthabile Kolwa, University of Johannesburg}\\
\strut \hfill \footnotesize{\textsc{Lecturer in Physics}}
\normalsize

\pagebreak
\strut \vspace{20pt}

\begin{figure}
    \centering
    \includegraphics[width=1\linewidth]{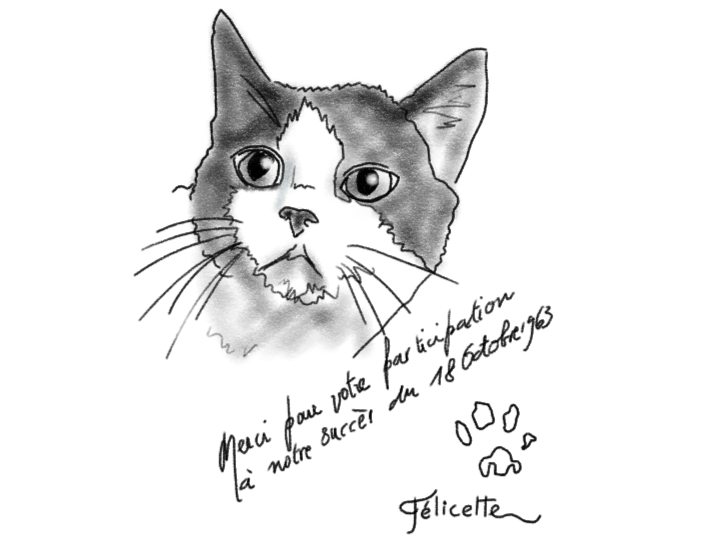}
    \label{fig:felicette}
\end{figure}

In case you don't know her story, it's worth learning about F\'{e}licette, the first cat in space, who was, perhaps apocryphally, nabbed off the streets of Paris, where she was living happily enough as a stray. She became France's first astronomical envoy and was the only cat to ever cross the K\'{a}rm\'{a}n line. Despite surviving her ordeal and returning to Earth unharmed, she eventually gave her life for science, when she was euthanized to allow French scientists to better study the effects of her trip to space. She is now memorialized with a statue at the International Space University in Strasbourg, France.

The phrase under her portrait translates to ``Thank you for your participation in our success on October 18, 1963.''

\pagebreak
\includepdf{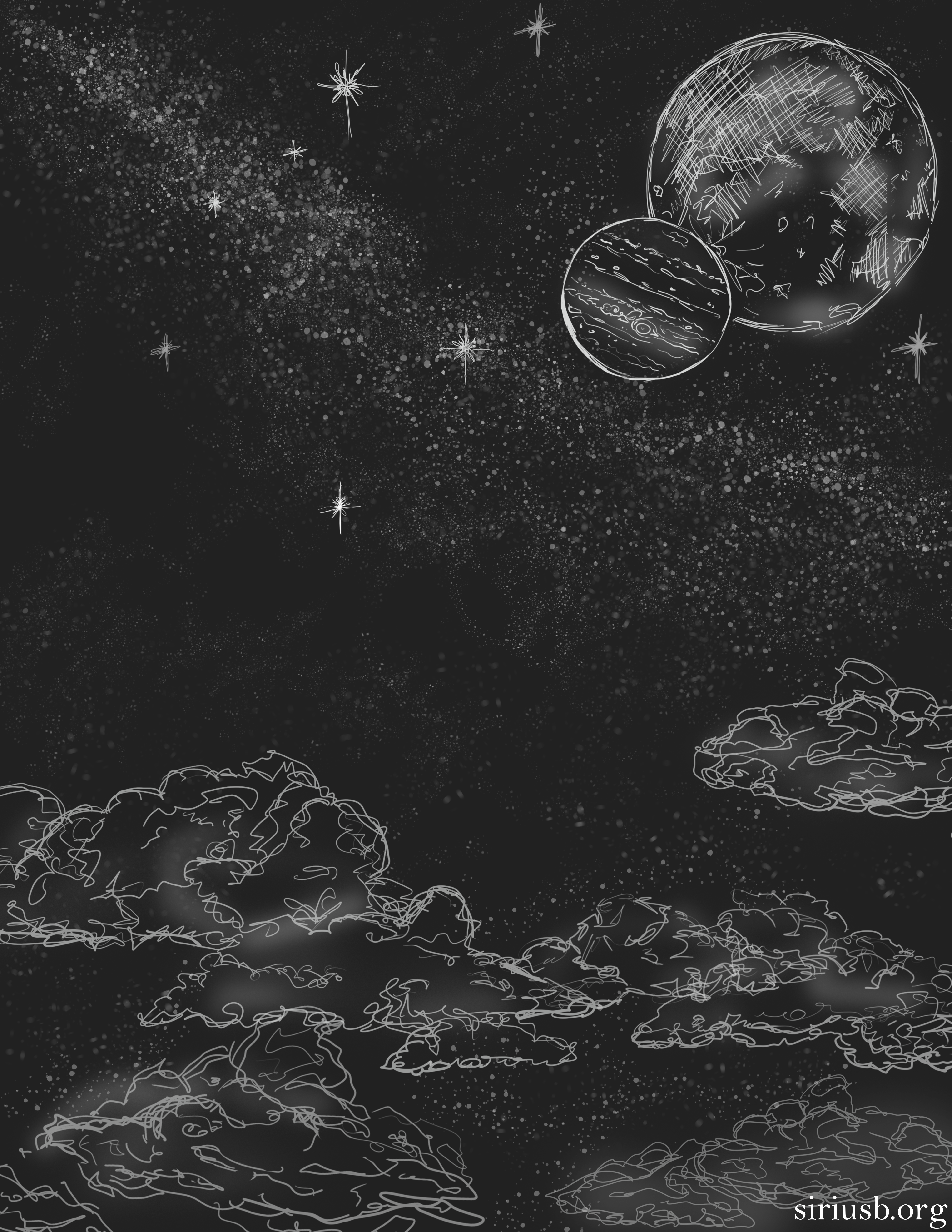}

\end{document}